\documentclass[prd,twocolumn,superscriptaddress,nofootinbib,floatfix,amsmath,amssymb]{revtex4-2}
\usepackage[utf8]{inputenc}
\usepackage[T1]{fontenc}
\usepackage{bm}
\usepackage{amsfonts,amsmath,amssymb,amsthm}
\usepackage{graphicx}
\usepackage{comment}
\usepackage[usenames,dvipsnames]{xcolor}
\usepackage{hyperref}
\usepackage{braket}
\usepackage{mathtools}
\usepackage{xcolor}
\usepackage{enumitem}
\usepackage{hyperref}

\hypersetup{
    colorlinks=true,
    linkcolor=blue,
    filecolor=magenta,      
    urlcolor=cyan,
}

\newcommand{\ii}{\mathrm{i}}
\renewcommand{\a}[1]{\hat{a}_{\bm{#1}}}
\newcommand{\ad}[1]{\hat{a}_{\bm{#1}}^\dagger}
\newcommand{\cre}{\hat{a}^\dagger}
\newcommand{\ann}{\hat{a}}
\newcommand{\bx}{\bm{x}}
\newcommand{\sx}{\mathsf{x}}
\newcommand{\bk}{{\bm{k}}}
\newcommand{\dd}{\textrm{d}}
\newcommand{\abs}[1]{\left|#1\right|}
\newcommand{\spec}{C_{\ba\bb}}
\newcommand{\normal}[1]{:\mathrel{#1}:}
\DeclareMathOperator{\tr}{\text{tr}}
\newcommand{\NN}{\mathcal{N}}

\newcommand{\rr}[1]{\left(#1\right)}
\newcommand{\ba}{{\bm{\eta}_1}}
\newcommand{\bb}{{\bm{\eta}_2}}

\newcommand{\norm}[1]{||#1||}

\newcommand{\pdag}{{\phantom{\dagger}}}
\newcommand{\vac}{\text{vac}}

\newcommand{\bc}{{\bm{\eta}}}

\usepackage[normalem]{ulem}
\usepackage{soul}

\begin{document}



\title{What makes a particle detector click}

\author{Erickson Tjoa}
\email{e2tjoa@uwaterloo.ca}
\affiliation{Department of Physics and Astronomy, University of Waterloo, Waterloo, Ontario, N2L 3G1, Canada}
\affiliation{Institute for Quantum Computing, University of Waterloo, Waterloo, Ontario, N2L 3G1, Canada}

\author{Irene L\'opez-Guti\'errez}
\email{irene.lopez@tum.de}
\affiliation{Technische Universit\"at M\"unchen, Department of Informatics, Boltzmannstra{\ss}e 3, 85748 Garching, Germany}
\author{Allison Sachs}
\email{asachs@uwaterloo.ca}
\affiliation{Department of Physics and Astronomy, University of Waterloo, Waterloo, Ontario, N2L 3G1, Canada}
\affiliation{Institute for Quantum Computing, University of Waterloo, Waterloo, Ontario, N2L 3G1, Canada}

\author{Eduardo Mart\'in-Mart\'inez}
\email{emartinmartinez@uwaterloo.ca}
\affiliation{Department of Applied Mathematics, University of Waterloo, Waterloo, Ontario, N2L 3G1, Canada}
\affiliation{Institute for Quantum Computing, University of Waterloo, Waterloo, Ontario, N2L 3G1, Canada}
\affiliation{Perimeter Institute for Theoretical Physics, 31 Caroline St N, Waterloo, Ontario, N2L 2Y5, Canada}

\date{\today}

\begin{abstract}
   
    We highlight fundamental differences in the models of  light-matter interaction between the behaviour of Fock state detection in free space versus optical cavities. To do so, we study the phenomenon of resonance of detectors with Fock wavepackets as a function of their degree of monochromaticity, the number of spatial dimensions, the linear or quadratic nature of the light-matter coupling, and the presence (or absence) of cavity walls in space. In doing so we show that intuition coming from quantum optics in cavities does not straightforwardly carry to the free space case. For example, in $(3+1)$ dimensions the detector response to a Fock wavepacket will go to \textit{zero} as the wavepacket is made more and more monochromatic and in coincidence with the detector's resonant frequency. This is so even though the energy of the free-space wavepacket goes to the expected finite value of $\hbar\Omega$ in the monochromatic limit. This is in contrast to the behaviour of the light-matter interaction in a cavity (even a large one) where the probability of absorbing a Fock quantum is \textit{maximized} when the quantum is more monochromatic at the detector's resonance frequency.  We trace this crucial difference to the fact that monochromatic Fock states are not normalizable in the continuum, thus physical Fock states need to be constructed out of normalizable wavepackets whose energy density goes to zero in the monochromatic limit as they get spatially delocalized.

\end{abstract}

\maketitle

\section{Introduction}

    Particle detector models may be thought of as non-relativistic, controllable quantum systems that couple locally in space and time to quantum fields.  They provide a way to extract localized spatio-temporal information from the fields while avoiding some of the problems with causality that may appear with the use of projective measurements \cite{sorkin1956,Dowker2011UselessQ,Benincasa2014projective}. Furthermore, as the name particle detectors suggests, the very definition of the notion of particle operationally has been given in recent times through these particle detector models~\cite{Wald1984accel,Earman2011}, in light of the drawbacks of the more traditional, `particle physics' inspired, notion of particle (see e.g., \cite{Lamb1995}). One of the best-known results using particle detector models is the operational formulation of the Hawking and Unruh effects (see, e.g., \cite{Unruh1979evaporation,Candelas1977irreversible,Crispino2008review}), and they are ubiquitous as models for experimental setups in quantum optics \cite{Scully1997book,boyd2008nonlinear} and in superconducting circuits \cite{Wallraff2004superconducting}. 
    
    A particularly simple particle detector model is the so-called Unruh-DeWitt (UDW) model \cite{Unruh1979evaporation,DeWitt1979}. It consists of a two-level quantum system linearly coupled to a scalar field. UDW detectors have been proven to be good models for the light-matter interaction in quantum optics for processes not involving  exchange of orbital angular momentum (see, e.g.,~\cite{Pablo2018rqo,Pozas2016}). Although most of the studies involving the UDW model thus far have considered a linear coupling between the detector and field, models that couple them quadratically have also been used \cite{Takagi1985detector,Takagi1986noise,Iyer1980dirac,Hummer2016,Jorma2016fermionic,Allison2017a,Allison:2018multi}. Quadratic couplings are not only useful to model non-linear processes in optics, but also, quadratic couplings are fundamental to modelling the coupling of a detector to a charged bosonic field or a fermionic field without violating the $U(1)$ symmetry of the theory \cite{Hummer2016}.

    {Perhaps surprisingly, the possible fundamental distinction between quantum field theory in a cavity and in free space has not been investigated much within the particle detector framework. The standard folklore is that one should be able to think of free space as being an extremely large cavity. Indeed this is how one can avoid certain infrared (IR) difficulties associated to infinite extent of free space when performing canonical quantization of the field. However, the Hilbert spaces in these two cases can be really different, and this manifests for instance in the normalizability of one-particle Fock states $\hat a^\dagger_\bk\ket{0}$. It is therefore not obviously clear whether in presence of an external probe (detector), the distinction between them can always be swept under the rug so long as ``the cavity is large enough''. 
    
    Furthermore, non-linear coupling between the probe and the field has also been  mostly investigated only for vacuum states, see e.g., \cite{Takagi1986noise,Allison2017a,Allison:2018multi,Hummer2016bosonfermionZM}.
    However, in addition to trying to understand the cavity vs free space problem, it is already known that even in non-relativistic quantum optics a plethora of interesting phenomena can emerge when non-linear medium and non-vacuum states are involved. Two such examples are \textit{sum-frequency generation} (SFG) and \textit{difference-frequency generation} (DFG)  \cite{boyd2008nonlinear,Mandel1985parametric,Kleinman1962secondharmonic}. As such, the possibility of modelling these non-linear phenomena using a relativistic, non-linear variant of the Unruh-DeWitt model merits further study. }
    
    The above considerations naturally lead us to investigate in this paper the behaviour of an UDW detector interacting with two kinds of non-vacuum state of massless scalar field, namely one-particle and two-particle Fock wavepackets. Firstly, although the UDW model is a monopole-scalar approximation of the usual light-matter interactions given by the atom-electromagnetic field dipole interaction $\hat{\bm{d}}\cdot\hat{\bm{E}}$ \cite{Lopp2021Rontgen}), the two physical questions we would like to address are likely to be present regardless of the multipoles of the detector and the spin of the field. Indeed, for the response functions of the model it has been repeatedly established that the UDW model captures the fundamental features of the light-matter interaction~\cite{Pablo2018rqo,Pozas2016}, to the point that the models typically used in quantum optics such as spin-boson, Rabi and Jaynes-Cummings models can be seen as further approximations performed on the UDW model. The UDW paradigm will therefore provide the same insights without having to deal with subtleties involving gauge choices and exchange of angular momentum. Secondly, one-particle and two-particle Fock wavepackets have very clear physical interpretation in terms of resonances with the detector's energy gap, and they are naturally suited to see if non-linear phenomena such as multiple harmonic-generation can arise in the scalar UDW model. Finally, as we will see, the fact that Fock wavepackets in a cavity and in free space are fundamentally different in the ``monochromatic limit'' (when the wavepacket is very peaked around some frequencies) is the root cause of the fundamental distinction between free space and a very large cavity, a fact that is present for both the scalar and the electromagnetic field.

    More specifically, in this paper we will study the response of linearly and quadratically coupled detectors to one-particle and two-particle Fock wavepacket excitations with a frequency spread, paying special attention to the limits when the wavepacket becomes monochromatic and the interaction time becomes very long. We will see that intuition that can be extracted from quantum optics in cavities will not carry straightforwardly to the free space case: in free space, if we make the wavepacket narrower so that most of its energy is concentrated in the modes near resonance with the detector, the response of the detector for long times actually \textit{decreases}. Furthermore, the detector becomes fully transparent to a monochromatic Fock state with energy exactly equal to its energy gap. This phenomenon occurs for linearly coupled detectors in $(3+1)$ dimensions and higher, while in lower dimensions the behaviour is remarkably different. For quadratic coupling this phenomenon occurs in $(2+1)$ dimensions and higher. This result reveals that there are fundamental differences between a very large cavity and the continuum in the context of the light-matter interaction when resonance is involved. We also show that indeed the standard nonlinear optical phenomena --- {sum-frequency generation} and {difference-frequency generation} arise naturally in the fully relativistic detector model when the detector-field coupling is quadratic, thus effectively mimicking the presence of a non-linear medium.
    
    Throughout our analysis we will study several other interesting aspects of the light-matter interaction comparing linear with quadratic coupling and cavity with free-space scenarios. In particular, we will analyze the spacetime dimension dependence of the energy content of a finite-width wavepacket and whether it is possible to take the monochromatic limit keeping the energy expectation of the Fock state constant. 
    
    We will also study the complementary view: if a detector starts in the excited state and we let it spontaneously decay, in what modes of the field is the energy of the detector deposited? We discuss in detail the differences and similarities between the linear and quadratic models and build intuition about spontaneous decay processes with the quadratic detector. We will see that during spontaneous decay, a quadratically coupled detector preferentially imparts the energy to the field mode with frequency equal to \textit{half} the energy gap of the detector, thus effectively splitting the excitation into two parts. This is in contrast to linearly coupled detector, where the detector's energy excitation is deposited to the field modes that have frequency matching the energy gap of the detector.
    
    This paper is organized as follows. In Section~\ref{sec:setup} we review the UDW model and find the expression for the Wightman two-point function for an arbitrary state of the field in both linear and quadratic coupling. In Section~\ref{sec: oneparticle} we analyze both couplings for the field prepared in a one-particle Fock wavepacket peaked at a given frequency. We also analyze the energy content of the wavepacket and discuss our results in concert with the standard intuition from optical cavities. In Section~\ref{sec:twoparticle} we analyze the two-particle case, whose excitations are possibly peaked at distinct frequencies. In Section~\ref{sec:energydeposit} we discuss how the excited detector deposits energy in a quantum field initialized to the vacuum state. Throughout this paper we adopt natural units $c= \hbar = 1$, and use the notation $\sx\equiv (t,\bx)$ to remove clutter when necessary. We present our conclusions in Section \ref{sec:conclusions}.

    \section{Setup}
    \label{sec:setup}
    
    In this section we introduce the particle detector models that we will analyze in the paper. We will then provide the general expression for the excitation probability of a detector starting in its ground state for both linear and quadratic detector-field coupling. 

    \subsection{Linear interaction: The Unruh-DeWitt model}
    \label{subsec: linear-interaction-setup}

    For convenience, let us consider as our detector model a two-level system comoving with the quantization frame $(t,\bx)$ and energy gap $\Omega$, whose centre of mass is at the origin $\bm{x}=0$ of this frame. For the linear model in \mbox{$(n+1)$-dimensional} flat spacetime, the interaction Hamiltonian that describes the system is given in the interaction picture as \cite{Louko_2006,Pablo2018rqo}
    \begin{equation}
    \label{eq: linear-hamiltonian}
    	\hat{H}_I(t) = \lambda \chi(t) \hat{\mu}(t) \int {\dd}^n\bm{x}\, F(\bx) \hat{\phi}(t,\bx)\,,
    \end{equation}
    where $ n $ is the number of spatial dimensions; $\lambda$ is the coupling strength of the detector with the field which has dimension $[\text{Length}]^{(n-3)/2}$; $\chi(t)$ is the switching function controlling when the interaction takes place and how its intensity varies in time; $F(\bx)$ is the spatial smearing of the detector that in the light-matter interaction would be determined by the wavefunctions of the excited and the ground state \cite{EMM2013wavepacket,Pozas2016,Pablo2018rqo}; $\hat{\mu}(t)$ is the monopole moment operator of the detector which plays the role in a scalar model that the dipole moment plays in the vector version of light-matter interaction. It is given by
    \begin{equation}
    \label{eq:moment}
        \hat{\mu}(t) = \hat\sigma^+e^{\ii \Omega t}+\hat\sigma^- e^{-\ii \Omega t}\,,
	\end{equation}
    where $\hat\sigma^\pm$ are the $\mathfrak{su}(2)$ algebra ladder operators. In the  basis $\{\ket{g},\ket{e}\}$, we can write $\hat\sigma^+=\ket{e}\!\bra{g}$ and $\hat\sigma^-=\ket{g}\!\bra{e}$. The massless scalar field operator $\hat{\phi}(t,\bx)$ can be expanded in terms of plane-wave modes as 
    \begin{equation}
    \label{eq:field}
	    \hat{\phi}(t,\bx) = \int \frac{\dd^n \bk}{\sqrt{2 (2\pi)^n |\bm{k}|}} \left[\ann_{\bk}    e^{-\ii\abs{\bk}t+\ii\bk\cdot \bx} + \cre_{\bk}    e^{\ii\abs{\bk}t-\ii\bk\cdot \bx}\right]\,,
    \end{equation}
    where $\hat a_\bk^\pdag$ and $\hat a^\dagger_\bk$ are ladder operators satisfying canonical commutation relations
    \begin{equation}
        [\hat a^\pdag_{\bk},\hat a_{\bk'}^\dagger]
        =\delta^{(n)}(\bk-\bk')\openone\,.
    \label{eq:canon}
    \end{equation}
    The Hamiltonian in Eq.~\eqref{eq: linear-hamiltonian} is known as the (spatially smeared) Unruh-DeWitt model, and has been shown to capture the fundamental features of the light-matter interaction when angular momentum exchange does not play a fundamental role in the detector dynamics \cite{Pablo2018rqo,Pozas2016}.

    The interaction Hamiltonian \eqref{eq: linear-hamiltonian} generates the time evolution operator
    \begin{equation}
        \hat{U} = \mathcal{T}\exp\Bigg[-\ii \int_{-\infty}^{\infty}{\dd}t\,\hat{H}_I(t)\Bigg],
    \end{equation}
    where $\mathcal T$ denotes time ordering.
    For small enough $\lambda$, we can use the perturbative Dyson series expansion up to second order:
    \begin{align}
        \hat{U} &\coloneqq \openone + \hat U^{(1)} + \hat U^{(2)} + O(\lambda^3)\,,\\
    	U^{(1)}&= - \ii \int_{-\infty}^{\infty}{\dd}t\, \hat H_I(t)\,,\\
    	U^{(2)} &= -\int_{-\infty}^{\infty}{\dd}t \int_{-\infty}^t{\dd}t' \hat H_I(t)\hat H_I(t')\,,
    \end{align}
    where $\hat U^{(j)}$ is of order $\lambda^j$ in the Dyson series. If the full density matrix is initially given by $\hat\rho_0$, then the time-evolved density matrix reads
    \begin{align}
        \hat{\rho} = \hat U\hat{\rho}_0 \hat U^\dagger\,.
    \end{align}
    The time-evolved density matrix of the detector $\hat{\rho}_\text{d}$ can be obtained from tracing out the field's degrees of freedom, 
    \begin{equation}
    	\label{eq:density}
        \hat{\rho}_\text{d} = \tr_\phi  \rr{\hat U\hat{\rho}_0 \hat U^\dagger}\,.
    \end{equation}
    Substituting the Dyson expansion into Eq.~\eqref{eq:density}, we obtain
    \begin{equation}
        \begin{split}
        \hat{\rho}_\text{d} &= \sum_{i,j=0}^\infty \hat{\rho}^{(i,j)}_\text{d}\,,\hspace{0.5cm}
   	    \hat{\rho}_\text{d}^{(i,j)} \coloneqq  \tr_\phi 
   	    \rr{\hat U^{(i)}\hat{\rho}_0 \hat{U}^{(j)\dagger}}\,,
   	    \end{split}
    \end{equation}
    where the terms of order $\lambda^k$ are those with $i+j=k$. 
    
    In this paper we are working up to second order in perturbation theory and hence $k\leq 2$. In particular, if we assume that the full density matrix is initially a product state $\hat{\rho}_0 = \ket{g}\!\bra{g} \otimes\hat{\rho}_\phi$, then the excitation probability $P^\phi$ of the detector from its ground state is encoded in the matrix element $\bra{e}\hat\rho_\text{d}^{(1,1)}\ket{e}$, which reads
    \begin{align}
        P^{\phi} &=
        \lambda^2 \int {\dd}t\int {\dd}t'\int {\dd}^n\bm{x}\int{\dd}^n\bm{x'}\chi(t) \chi(t')F(\bm{x}) F(\bm{x'})\notag\\
        &\hspace{0.5cm}\times  e^{-\ii\Omega(t-t')} W^\phi(t,\bx,t',\bx')\,,
        \label{eq: general-prob-linear}
    \end{align}
    where $W^\phi(t,\bx,t',\bx')$ denotes the Wightman two-point function for the arbitrary field state $\hat\rho_{\phi}$:
    \begin{align}
   	    W^\phi(t,\bx,t',\bx') 
   	    \coloneqq 
   	    \tr_\phi\left[\hat{\rho}_{_\phi} \hat{\phi}(t,\bx) \hat{\phi}(t',\bx')\right]\,.
   	    \label{eq: wightman-linear-gen}
    \end{align}

	\subsection{Quadratic interaction}
	\label{subsec: quadratic-interaction-setup}
    	
    The quadratic coupling is a modification of the Hamiltonian Eq.~\eqref{eq: linear-hamiltonian} where the monopole moment of the detector couples to field quadratically: 
    \begin{align}
        \hat{H}_I = \lambda\chi(t)\hat\mu(t)\int \dd^n\bx \,F(\bx)\left(\normal{\hat\phi(t,\bx)^2}\right)\,,
        \label{eq: quadratic-hamiltonian}
    \end{align}
    where $\normal{\hat O}$ denotes normal ordering of an operator $\hat O$ to remove tadpole divergences \cite{Hummer2016}. This model is useful because 1) it represents the simplest $U(1)$-invariant way to couple a charged scalar field to a particle detector, 2) it algebraically mimics the coupling of a detector to a fermion field as described in \cite{Takagi1986noise, Hummer2016}, 3) it has commonly been employed as a particle detector model in different scenarios \cite{Hinton1984a, Takagi1985detector, Bessa2012, Allison2017a}, and 4) it is a scalar analog of models in which light couples to the square of the electric field amplitude in non-linear media~\cite{boyd2008nonlinear}.
    	
    Repeating the calculation analogous to the one in Section~\ref{subsec: linear-interaction-setup}, the excitation probability of the detector $P^{\phi^2}$ is given by
    \begin{align}
    \label{eq: general-prob-quad}
        P^{\phi^2} &=
        \lambda^2 \int {\dd}t\int {\dd}t'\int {\dd}^n\bm{x}\int{\dd}^n\bm{x'}\chi(t) \chi(t') F(\bm{x}) F(\bm{x'})\notag\\
        &\hspace{0.4cm}\times e^{-\ii\Omega(t-t')}W^{\phi^2}(t,\bx,t',\bx')  \,.
    \end{align}
    where $W^{\phi^2}(t,\bx,t',\bx')$ denotes a Wightman-like two-point function for quadratic coupling and for the arbitrary field state $\hat\rho_\phi$:
    \begin{align}
        W^{\phi^2}(t,\bx,t',\bx') \coloneqq \tr_\phi\left[\hat{\rho}_{_\phi} \normal{\hat{\phi}(t,\bx)^2} \normal{ \hat{\phi}(t',\bx')^2}\right]\,.
        \label{eq: wightman-quadratic-gen}
    \end{align}
    
    In this paper we will make a convenient abuse of terminology and call  Eq.~\eqref{eq: wightman-linear-gen} and \eqref{eq: wightman-quadratic-gen} respectively linear and quadratic Wightman two-point functions. Note that rigorously speaking only the linear one is {a proper Wightman function~\cite{Wightman1956VEV}}. In order to distinguish the Wightman two-point functions for linear and quadratic interactions, we have used the superscript $\phi$ for the linear case in Eq.~\eqref{eq: wightman-linear-gen} and superscript $\phi^2$ for the quadratic case in Eq.~\eqref{eq: wightman-quadratic-gen}.

    \section{one-particle detection}
    \label{sec: oneparticle}
    
    
    In this section we will investigate how a detector responds to a one-particle excitation of the field. We will first define what we mean by one-particle Fock state in free space, and then we will consider how the detector response depends on the properties of the field state. We will see that the ability of detectors to resonate with field quanta strongly depends on both the choice of detector-field coupling and spacetime dimensions.
    
    
    \subsection{One-particle Fock state}
    \label{subsec: energy-wavepacket}

    It is well-known that in free space, the na\"ive one-particle Fock state $\ket{1_{\bk}}$ with momentum $\bk$ is not normalizable since $\braket{1_\bk|1_{\bk'}} = \delta^{(n)}(\bk-\bk')$. Therefore, we cannot take $\ket{1_\bk}$ as a physical one-particle excitation state. We can rectify this by considering instead a Fock \textit{wavepacket} of the form 
    \begin{align}
        \ket{1_f}\coloneqq \int \dd^n\bk\,f(\bk)\hat a_\bk^\dagger\ket{0}\,,
        \label{eq: one-particle-state}
    \end{align}
    where $f$ prescribes a weight on each momentum $\bk$. For convenience we will call $f$ the \textit{spectrum} of $\ket{1_f}$.
    For this state to be physically reasonable, it must be normalizable to unity and this implies that the $L^2$-norm of $f$ is also unity:
    \begin{align}
        \braket{1_f|1_f} = \int \dd^n\bk\, |f(\bk)|^2 = \norm{f}^2 = 1\,,
        \label{eq: one-particle-norm}
    \end{align}
	where we have denoted the $L^2$-norm of $f$ by $\norm{f}$. 
	This state can be regarded as a normalizable version of one-particle Fock state: it is an eigenstate of the total number operator $\hat N\coloneqq \int \dd^n\bk\, \hat a_\bk^\dagger\hat a_\bk^\pdag$ with eigenvalue 1:
	\begin{align}
	    \hat N\ket{1_f} &= \int \dd^n\bk\,\dd^n\bk'\,f(\bk') \hat a_\bk^\dagger\hat a_\bk^\pdag \hat a_{\bk'}^\dagger\ket{0} = \ket{1_f}\,.
	    \label{eq: one-particle-eigenstate}
	\end{align}
	Note that we have not chosen the specific form of $f$ apart from demanding that its $L^2$-norm is unity.

    We can determine how much energy is contained in this one-particle wavepacket. The energy expectation value will depend on the profile of the spectrum $f$, since the weight of each frequency influences the total energy of the state. Furthermore, in general $f$ can have a highly complicated profile\footnote{In particular, $f$ need not have a single peak in order to describe a one-particle excitation: Eq.~\eqref{eq: one-particle-norm} and \eqref{eq: one-particle-eigenstate} only demand the $L^2$-norm of $f$ be unity.}. However, we can obtain an intuitive picture by focusing on a specific class of one-particle Fock states, namely those whose spectrum $f$ is real and has a single peak at $\bk=\bk_0$ with some frequency width (bandwidth) $\sigma$ around the peak. We denote this choice of spectrum by $f_{\bk_0,\sigma}$, so that the state now reads
    \begin{align}
        \ket{1_f}\coloneqq \int \dd^n\bk\,f_{\bk_0,\sigma}(\bk)\hat a_\bk^\dagger\ket{0}\,.
        \label{eq: one-particle-state-peaked}
    \end{align}
    We can then think of a highly \textit{monochromatic}
    one-particle state $\ket{1_f}$ as a normalizable version of monochromatic excitation $\ket{1_{\bk_0}} = \hat a_{\bk_0}^\dagger\ket{0}$ when we take the limit $\sigma\to 0$. 
    
    In order for this \textit{monochromatic limit} to work, we require $|f_{\bk,\sigma}|^2$ to be a family of nascent delta functions, i.e. the following distributional limit holds:
    \begin{align}
        \lim_{\sigma\to 0} |f_{\bk_0,\sigma}|^2 = \delta^{(n)}(\bk-\bk_0)\,.
        \label{eq: distributional-limit}
    \end{align}
    Given the free Hamiltonian of the scalar field
	\begin{align}
	    \hat H_{0,\phi} = \int \dd^n\bk\, |\bk|\hat a_\bk^\dagger\hat a^{\phantom{\dagger}}_\bk\,,
	    \label{eq:free-field-hamiltonian}
	\end{align}
    the energy expectation for $\ket{1_f}$ with spectrum $f=f_{\bk_0,\sigma}$ is then given by
    \begin{align}
        \bra{1_f}\hat H_{0,\phi}\ket{1_f} = \int \dd^n\bk\,|\bk|\,|f_{\bk_0,\sigma}(\bk)|^2\,.
        \label{eq: energy-average-before-nascent}
    \end{align}
    In the {monochromatic limit} $\sigma\to 0$ where $f_{\bk_0,\sigma}$ becomes very sharply peaked around $\bk_0$, the distributional limit gives
    \begin{align}
         \lim_{\sigma\to 0}\bra{1_f}\hat H_{0,\phi}\ket{1_f} &= \int \dd^n\bk\,|\bk|\delta^{(n)}(\bk-\bk_0) = |\bk_0|\,.
         \label{eq: energy-average-nascent}
    \end{align}
    This agrees with the  energy expectation value formally evaluated for the non-normalizable monochromatic state $\ket{1_{\bk_0}}$. For example, if we set $f_{\bk_0,\sigma}$ to be an $L^2$-normalized Gaussian\footnote{The problem of defining localized one-particle states has a long history (see, e.g., \cite{Hegerfeldt1974causality,Iwo1998localization,Palmer2011localizedqubits} for related discussions).} \cite{Benincasa2014projective,Kohlrus2015:1511.04256v6, bruschi2020self}
	\begin{equation}
	\label{eq: GaussianFrequency}
	    f_{\bk_0,\sigma}(\bk) = \frac{1}{(\pi\sigma^2)^{n/4}}\exp\left(-\frac{(\bk-\bk_0)^2}{2\sigma^2}\right)\,,
	\end{equation}
	the energy expectation value can be calculated explicitly in terms of hypergeometric functions (see Appendix~\ref{appendix: energyexpectation}) which indeed yields $\lim_{\sigma\to 0}\braket{1_f|\hat H_{0,\phi}|1_f} = |\bk_0|$. 

    In the following subsections we will analyze the response of linearly and quadratically coupled detectors to a one-particle state with a general single-peaked spectrum $f_{\bk_0,\sigma}$.

	\subsection{Linear coupling: transition probability in arbitrary dimensions}
    \label{subsec: linear-probability-oneparticle}

    	
    Let us obtain the explicit expression for the transition probability when the detector is linearly coupled to the field. We begin by substituting our definition of one-particle Fock state~\eqref{eq: one-particle-state-peaked} into the Wightman two-point function~\eqref{eq: wightman-linear-gen}, then apply the field expansion~\eqref{eq:field} and the canonical commutation relations \eqref{eq:canon}. The Wightman two-point function reads
    
    \begin{align}
        W^{\phi}(\sx,\sx') = 
        W_{\vac}^\phi(\sx,\sx') 
        + \rr{K_{\bk_0}^*(\sx)K_{\bk_0}(\sx')+\text{c.c.}}
       \label{eq: wightman-linear-one},
    \end{align}
    where ``c.c.'' denotes complex conjugation and we define 
    \begin{align}
        K_{\bk_0}(\sx) &\coloneqq \int \frac{{\dd}^n \bm{k}}{\sqrt{2(2\pi)^n|\bm{k}|}} f_{\bk_0,\sigma}(\bk) e^{\ii(|\bm{k}|t-\bm{k}\cdot\bm{x})}
        \label{eq: K-integral},
    \end{align}
    which depends on the shape of the spectrum $f_{\bk_0,\sigma}(\bk)$.
    $W_{\vac}^\phi(\sx,\sx') \coloneqq \bra{0}\hat \phi(t,\bx) \hat \phi(t',\bx') \ket{0}$ is the vacuum Wightman two-point function which has the form
    \begin{align}
   	    W_{\vac}^\phi(\sx,\sx') 
   	    &= \int \frac{{\dd}^n \bm{k}_2}{2(2\pi)^n |\bm{k}_2|}	    e^{-\ii|\bm{k}_2|(t-t')+\ii\bm{k}_2\cdot(\bm{x}-\bm{x'})}\,,
   	    \label{eq:wvac}
    \end{align}
    where this expression is understood as a (bi)distribution.

    After substituting the explicit expression for the Wightman two-point function in Eq.~\eqref{eq: wightman-linear-one} into the general expression for the excitation probability $P^\phi$ in Eq.~\eqref{eq: general-prob-linear}, it is useful to express  $P^\phi$ as 
    \begin{equation}
        P^\phi = P^\phi_{\vac}+P^\phi_K\,.
        \label{eq: prob-linear-schematic}
    \end{equation}
    The first term is the vacuum contribution,
    \begin{align}
        P_{\vac}^{\phi} &:= \frac{\lambda^2}{2(2\pi)^n}\int\frac{{\dd}^n \bm{k}_2}{|\bm{k}_2|} \bigr|\tilde{F}(\bm{k}_2)\bigr|^2  \bigr|\tilde{\chi}\left(\Omega+|\bm{k}_2|\right)\bigr|^2,
        \label{eq:linearvacuum}
    \end{align}
    where $\tilde\chi$ and $\tilde{F}$ are the Fourier transforms of the switching and smearing functions respectively:
    \begin{equation}
    \begin{split}
   	    \tilde \chi(\Omega)&\coloneqq\int_\mathbb{R}\!\dd t\, \chi(t)e^{\ii \Omega t}\,,\\
        \tilde F(\bm k) &\coloneqq\int_{\mathbb{R}^n}\!\!\dd^n \bm k\, F(\bm x)e^{\ii \bm k\cdot\bm x}\,.
   	    \label{eq: Fourier-transform}
    	\end{split}
    \end{equation}
    In order to simplify subsequent calculations, we assume that both $\chi$ and $f$ are real, and further that the switching function is even, i.e. $\chi(t) = \chi(-t)$, and the smearing function is  rotationally invariant, i.e. $F(\bx) = F(|\bx|)$. These restrictions still capture the fundamental phenomenology we will analyze and  guarantee that $\tilde{\chi}(\Omega) = \tilde\chi^*(\Omega) $ and $\tilde{F}(\bk) = \tilde{F}^*(\bk)$. These are satisfied for commonly used switching and smearing functions such as Gaussian switching/smearing and also for pointlike detectors ($F(\bx) = \delta^{(n)}(\bx)$).

    The second term $P^\phi_K$ reads
    \begin{align}
        P_{K}^{\phi} &= \frac{\lambda^2}{2(2\pi)^n}\iint\frac{{\dd}^n \bm{k}_1}{\sqrt{|\bm{k}_1|}}\frac{{\dd}^n \bm{k}_2}{\sqrt{|\bm{k}_2|}} {f_{\bk_0,\sigma}(\bm{k}_1)f_{\bk_0,\sigma}(\bm{k}_2)}\notag\\
        & \hspace{0.5cm}\times  \tilde{F}(\bm{k}_1)\tilde{F}(\bm{k}_2) \Big(\tilde{\chi}(\Omega-|\bm{k}_1|)\tilde{\chi}(\Omega-|\bm{k}_2|)\,+\notag\\
        &\hspace{3cm}\tilde{\chi}(\Omega+|\bm{k}_1|) \tilde{\chi}(\Omega+|\bm{k}_2|)\Big)\notag\\
        &=
        \lambda^2
        \bigr(\mathcal{I}_+^2(\sigma,\bk_0,\tilde{F},\tilde{\chi})+\mathcal{I}_{-}^2(\sigma,\bk_0,\tilde{F},\tilde{\chi})\bigr)\,,
        \label{eq: linear-nonvacuum-probability-general}
    \end{align}
    where we defined (to alleviate notation)
    \begin{align}
        \mathcal{I}_\pm&(\sigma,\bk_0,\tilde{F},\tilde{\chi})\notag\\
        &\coloneqq \int  \frac{\dd^n\bk}{\sqrt{2(2\pi)^n|\bk|}}f_{\bk_0,\sigma}(\bk)\tilde{F}(\bk)\tilde\chi(\Omega\pm|\bk|)\,.
        \label{eq: prob-linear-one-compactify}
	\end{align}
    In order to proceed with the explicit calculation of the transition probability, we will need to make explicit choices for the switching function $\chi(t)$, smearing function $F(\bx)$, and the spectrum $f_{\bk_0,\sigma}(\bk)$.
    	
    First we will choose a spatial profile $F(\bx)$. Pointlike detectors are particularly simple to work with and, furthermore, will be necessary for finding closed form expressions for the non-linear (quadratic) model. For this reason, we set the spatial profile to $ F(\bx) = \delta^{(n)}(\bx)$ so that $ \tilde{F}(\bk) = 1$. This choice corresponds to a pointlike detector located at the origin of the lab coordinate $\bx=\bm{0}$.
        
    Next, let us look at the choice of the switching function $\chi(t)$. For the discussion on the switching it is relevant to note that in quantum optics we often have intuition that comes from applying  single-mode and rotating-wave approximations. Together, these approximations are consistent with taking the limit of long interaction times (for a more nuanced discussion check,  e.g.,~\cite{Funai2019FTL}). It is therefore convenient and useful to compare both linear and quadratic models within this long interaction regime. In our model, the long-interaction limit can be achieved by setting $\chi(t)$ to be constant and without loss of generality we can set $\chi(t) = 1$ 
    The Fourier transform is therefore $\tilde{\chi}(\Omega) = 2\pi\delta(\Omega)$. We will call this the ``long time'' limit.
        
    In the pointlike and long time limit, we can obtain vast simplifications to $P^\phi$ in Eq.~\eqref{eq: prob-linear-schematic}. First, by inspecting Eq.~\eqref{eq: linear-nonvacuum-probability-general}, we see that for the non-vacuum contribution (i.e., $P_K^\phi$) the long time limit commutes with the integral. Therefore, in the long time and pointlike limits we get
    \begin{align}
        P_{K}^{\phi} &= \lambda^2 \bigr(\mathcal{I}_+^2(\sigma,\bk_0,1, 2\pi\delta)+
        \mathcal{I}_{-}^2(\sigma,\bk_0,1, 2\pi\delta)\bigr)\,.
        \label{eq: linear-prob-before-gap}
    \end{align}
    Since we have fixed the smearing and switching functions, we will drop the last two arguments of $\mathcal{I}_\pm$ and simply write $\mathcal{I}_\pm (\sigma,\bk_0)\equiv \mathcal{I}_\pm (\sigma,\bk_0,1, 2\pi\delta)$.
    
    Second, let us suppose that the detector is in the ground state. In this case, it can be shown that taking the \textit{adiabatic limit}\footnote{The way to compute $P^\phi_{\vac}$ in the infinitely long time limit requires us to consider a switching function \mbox{$\chi_{_T}(\tau)\coloneqq \chi(\tau/T)$} whose Fourier transform decays faster than any polynomial and then take the limit $T\to\infty$ at the end. This is known as the \textit{adiabatic limit}. This distributional limit does not commute with the integral. The adiabatic limit represents the physical (UV-safe) way to compute the long time limit~{\cite{Satz_2007,Louko_2008}}.} of long interaction times, $P_{\vac}^{\phi}$ vanishes. Furthermore, the ``counter-rotating'' term $\mathcal{I}_+(\sigma,\bk_0)$ in Eq.~\eqref{eq: linear-prob-before-gap} also vanishes. Namely, since $\Omega>0$, \mbox{$\mathcal{I}_+(\sigma,\bk_0)=0$} since  the argument of the delta $\delta(\Omega+|\bm k |)$ never vanishes in  the integration domain of~\eqref{eq: prob-linear-one-compactify}, so all that remains is $\mathcal{I}_-$.  Thus, the full transition probability in Eq.~\eqref{eq: prob-linear-schematic} only consists of a single ``co-rotating'' term
    \begin{align}
        P^\phi = \lambda^2\mathcal{I}_-^2(\sigma,\bk_0)\,.
        \label{eq: final-linear-probability}
    \end{align}
    This can be evaluated in closed form in arbitrary dimensions for the following judicious choice for the spectrum $f$.

    \begin{figure*}[tp]
        \includegraphics[scale=0.8]{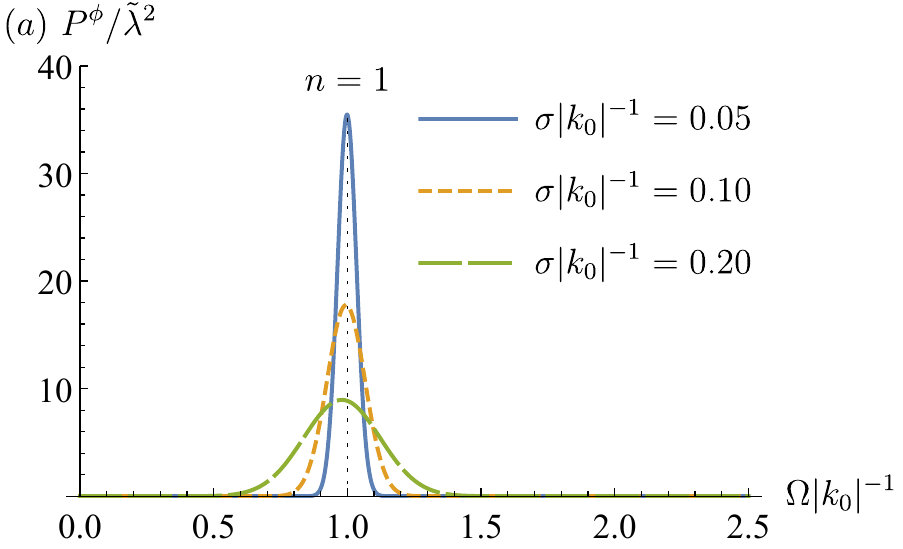}
        \includegraphics[scale=0.8]{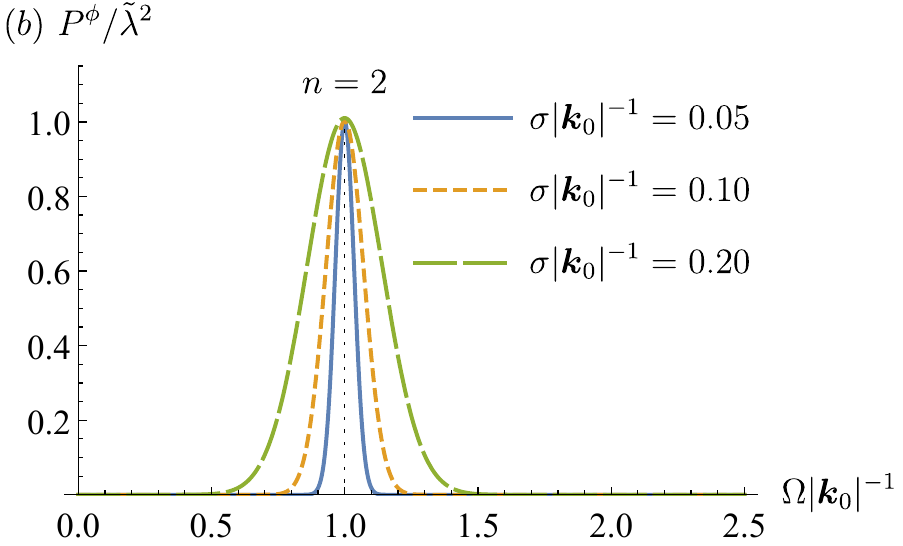}
        \includegraphics[scale=0.8]{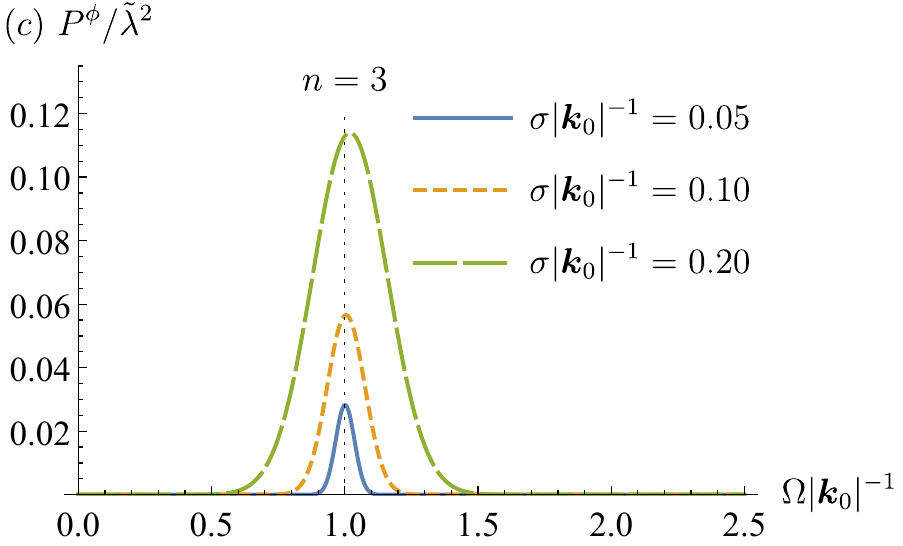}
        \includegraphics[scale=0.8]{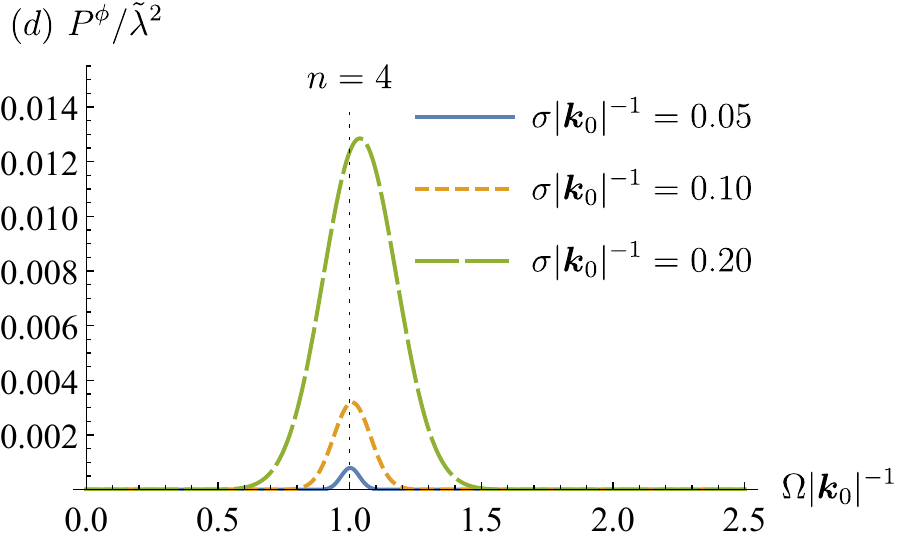}
        \caption{
        Plots of transition probability $P^\phi/\tilde{\lambda}^2$ as a function of detector energy gap for linear coupling and one-particle state in various spatial dimensions $n$, for various  spectral bandwidths $\sigma$. We vary $\Omega$ as we search for resonant peaks while keeping $|\bk_0|$ fixed. Here $\tilde\lambda = \lambda |\bk_0|^{(n-3)/2}$ is the adimensionalized coupling constant. The vertical lines denote the resonant frequency. Note that in the monochromatic limit $\sigma\to 0$, the peak amplitude \textit{diminishes}  for $n\geq 3$, and it approaches a constant value for $n=2$, while it \textit{increases} for $n=1$.}
        \label{fig: linear_resonance1}
    \end{figure*}
    
    Let us now set the the particle excitation spectrum $f$ to be the $L^2$-normalized isotropic Gaussian spectrum given in Eq.~\eqref{eq: GaussianFrequency}. 
    In order to analyze phenomena such as resonance, we need a quasi-monochromatic particle spectrum (i.e. a rapidly decaying spectrum peaking at $\bk_0$ with some bandwidth $\sigma$). Substituting the spectrum in Eq.~\eqref{eq: GaussianFrequency} into $\mathcal{I}_-$, we get
    \begin{align}
        \mathcal{I}_- &= \sqrt{\frac{2\pi ^{\frac{4-n}{2}}}{|\bk_0|^{n-2}} \frac{\Omega ^{n-1}}{\sigma^{4-n}}} e^{-\frac{|\bk_0|^2+\Omega ^2}{2\sigma ^2}} I_{\frac{n-2}{2}}\left(\frac{|\bk_0| \Omega }{\sigma ^2}\right)\,,
    \end{align}
    where $I_\alpha(z)$ is the modified Bessel function of the first kind of order $\alpha$ \cite{NIST:DLMF}. 
    Therefore, the expression for the transition probability \eqref{eq: final-linear-probability} in the long time and pointlike limits now reads (see derivation in Appendix~\ref{appendix: exactprobability})
    \begin{align}
        P^\phi = 
        \lambda^2 \frac{2\pi ^{\frac{4-n}{2}}}{|\bk_0|^{n-2}} \frac{\Omega ^{n-1}}{\sigma^{4-n}} e^{-\frac{|\bk_0|^2+\Omega ^2}{\sigma ^2}} I^2_{\frac{n-2}{2}}\left(\frac{|\bk_0| \Omega }{\sigma ^2}\right)\,.
        \label{eq: prob-linear-gaussian-final}
    \end{align}
    Note that for $n=1$ in Eq.~\eqref{eq: prob-linear-gaussian-final} we require that
    \begin{equation}
        \Omega,|\bk_0| \geq \Lambda > 0\,,
        \label{eq: IR-cutoff}
    \end{equation}
    where $\Lambda$ is an infrared (IR) cutoff to regulate the well-known IR divergence in (1+1)-dimensional massless scalar field (see e.g. \cite{birrell1984quantum,pozas2015harvesting}). Expression~\eqref{eq: prob-linear-gaussian-final}  is only valid for $n=1$  when the IR cutoff is below all relevant scales (see Appendix~\ref{appendix: exactprobability} for details). 

    Let us now plot and interpret Eq.~\eqref{eq: prob-linear-gaussian-final}. We will look at how the resonant peak of the transition probability behaves as a function of spectral width $\sigma$ and detector gap $\Omega$, keeping the wavepacket peak frequency constant (the detector gap is tuned to sweep across the spectral bandwidth of the wavepacket including the `resonance' case $\Omega=|\bk_0|$). The results are shown in Figure~\ref{fig: linear_resonance1}.

    There are three preliminary observations that we can make based on Eq.~\eqref{eq: prob-linear-gaussian-final} and Figure~\ref{fig: linear_resonance1}. The first observation is that for large spectral width $\sigma|\bk_0|^{-1}\gg 1$ corresponding to the wavepacket assigning equal weight to every momentum $\bk$, Eq.~\eqref{eq: prob-linear-gaussian-final} vanishes as fast as $\sigma^n$ in all spatial dimensions. This is in spite of the fact that an infinitely wide spectrum Fock wavepacket also has infinite total energy expectation (as per equation Eq.~\eqref{eq: energy-expectation-dimtwo} in Appendix~\ref{appendix: energyexpectation}, the energy of the wavepacket diverges like as $\sigma\rightarrow\infty$). 

    The second observation is that the maximum of the detector response does not happen at the resonance frequency with the peak of the wavepacket $\Omega=|\bm k_0|$. Only as the wavepacket becomes more and more monochromatic ($\sigma$ decreases) and the resonant peak becomes sharper, the maximum of $P^\phi$ moves towards $|\bk_0|=\Omega$. In other words, only in the limit $\sigma\to 0$  does the largest detector response happen exactly at $|\bk_0|=\Omega$. For $n=1$ this shift is not resolvable in  Figure~\ref{fig: linear_resonance1}. However, it can be seen by solving for the particular value of $|\bk_0|$ that satisfies $\partial P^\phi/\partial |\bk_0| = 0$. 
    	
    The third and perhaps the most important observation is that the amplitude of the resonant peak behaves differently in different dimensions as we take $\sigma|\bk_0|^{-1}\to 0$ keeping $|\bm k_0|$ constant. In particular, as the wavepacket becomes more and more monochromatic, for $n=1$ the peak of $P^\phi$ \textit{increases} in amplitude, for $n=2$ the peak approaches a constant value, and for $n\geq 3$ the peak \textit{decreases} in amplitude.  In the limit $\sigma|\bk_0|^{-1}\to 0$, when $\Omega=|\bk_0|$ we obtain that
    \begin{align}
        P^\phi(\Omega=|\bk_0|) = \tilde\lambda^2 \rr{\frac{\sigma/|\bk_0|}{\sqrt{\pi}}}^{n-2}+O((\sigma/|\bk_0|)^{n})\,,
    \end{align}
    for all $n\geq 1$ and $\tilde\lambda= \lambda|\bk_0|^{(n-3)/2}$ is a dimensionless coupling constant. 

    Note that although the peak vanishes in the monochromatic limit for $n\geq 3$, there is always a resonant peak for finite $\sigma$ because the off-resonant frequencies decay faster than the resonant frequency. Mathematically, it means that in all dimensions we have
    \begin{align}
        \lim_{\sigma\to 0}
        \frac{P^\phi(\Omega\neq |\bk_0|)}{P^\phi(\Omega=|\bk_0|)} = 0\,.
    \end{align}

    We point out that this diminishing probability has nothing to do with the fact that our detector is pointlike. For instance, let us consider a Gaussian smearing function
    \begin{align}
        F(\bx) &= \frac{1}{(\pi \Delta^2)^{\frac{n}{2}}}e^{-|\bx|^2/\Delta^2}\Longrightarrow 
        \Tilde{F}(\bk) = e^{-\frac{1}{4}(\Delta^2|\bk|^2)}\,,
    \end{align}
    where $\Delta$ controls the effective size of the detector. Substituting this into Eq.~\eqref{eq: prob-linear-one-compactify}, we can show that the new excitation probability (denoted $P_{\Delta}^\phi$) is related to the pointlike one by the relation
    \begin{align}
        P^{\phi}_{\Delta} = P^\phi e^{-\frac{1}{2}\Delta^2\Omega^2}\,.
    \end{align}
    We recover the pointlike result when $\Delta\to 0$.
    
    {Observe that  if we increase the size of the detector in proportion to decreasing the wavepacket width (by setting $\Delta = \sigma^{-1}$), the probability actually decreases \textit{faster} than if we were in the pointlike regime. Therefore, one cannot argue that the diminishing resonant probability for $n\geq 3$ is due to the fact that the field quanta is simply large in comparison to the detector and increasing the detector size would help counter this effect.}

    {Instead, we point to our  analysis of the energy density of the wavepacket (see \ref{appendix: energyexpectation}), which approaches $k_0$ (i.e. a finite value) in the monochromatic limit, yet the spread in position space becomes uniform. Thus, the energy density approaches 0 (as we show in Eq.~\eqref{eq: RSET-scaling} in Appendix~\ref{appendix: energyexpectation}). In principle one would think that a detector large enough ($\Delta = \sigma^{-1}$) would have a non-zero excitation probability, since integrating the energy density over the whole of space does give a finite value.} However that is not the case. As a detector is delocalized it has to become more weakly coupled to the field at each point. In the limit of infinite delocalization the response of a detector approaches zero for any state of the field. This is because, in this limit, the coupling of the detector to the field is essentially zero at all points.

    Furthermore, since the units of the coupling strength $\lambda$ depend on the dimensions of spacetime one may wonder if the vanishing response of the detector in the monochromatic limit (even though the energy content of the monochromatic wavepacket is finite) is a consequence of failing to capture the scaling behaviour of the coupling strength. To see that this is not the case, suppose that we allow the coupling strength $\lambda$, which has units of $[\text{Length}]^\frac{n-3}{2}$ for linear coupling, to run with the wavepacket width $\sigma$. That is, we define a dimensionless coupling constant $\gamma \coloneqq \lambda \sigma^{\frac{n-3}{2}}$, so that we can rewrite $P^\phi$ in \eqref{eq: prob-linear-gaussian-final} as
    \begin{align}
        P^\phi &= \gamma^2  \frac{2\pi ^{\frac{4-n}{2}}}{|\bk_0|^{n-2}} \frac{\Omega ^{n-1}}{\sigma} e^{-\frac{|\bk_0|^2+\Omega ^2}{\sigma ^2}} I^2_{\frac{n-2}{2}}\left(\frac{|\bk_0| \Omega }{\sigma ^2}\right)\,.
    \end{align}
    This corresponds to having the coupling weaken (for $n>3$) or strengthen (for $n<3$) as we  decrease the wavepacket width. As it turns out, letting the coupling constant $\gamma$ run yields the universal result
    \begin{align}
        \lim_{\sigma \to 0}P^\phi(\Omega=|\bk_0|) = 0
    \end{align}
    for all $n\geq 1$, which is that the detector becomes transparent when the wavepacket is strictly monochromatic. This can be understood from the fact that in $(3+1)$ dimensions the coupling constant $\lambda$ is dimensionless, thus the variation of the probability as $\sigma$ is varied will be qualitatively similar to the $(3+1)$ dimensional case. As such, the cancellation of the response of the detector when driven by a quasi-monochromatic wavepacket at resonance is not due to the scaling of the coupling strength in different dimensions.

    \subsubsection*{Comparison with the standard intuition from optical cavities}
    \label{subsec: cavity-calculation}
    
    
    \begin{figure*}[htp]
        \centering
        \includegraphics[scale=0.95]{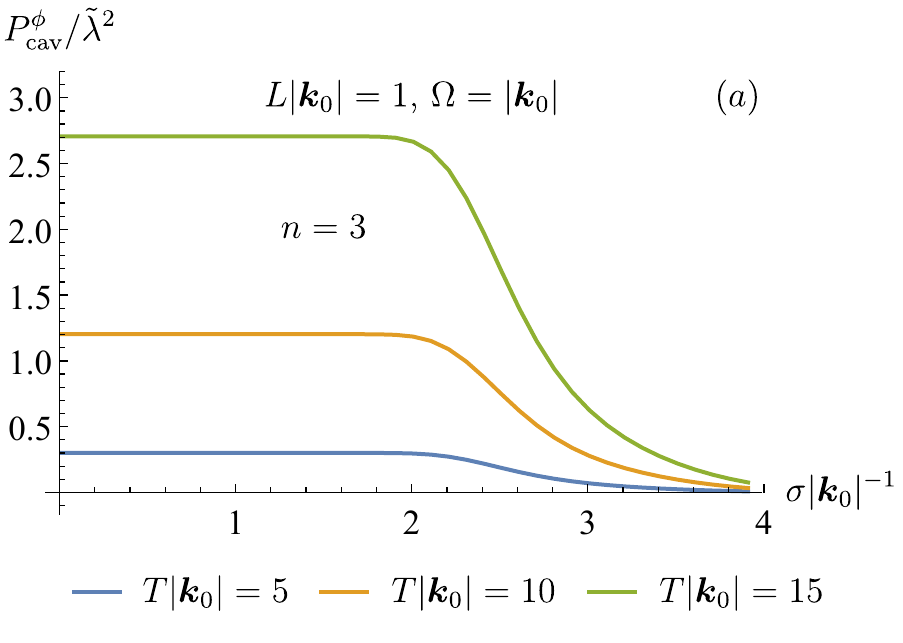}
        \includegraphics[scale=0.95]{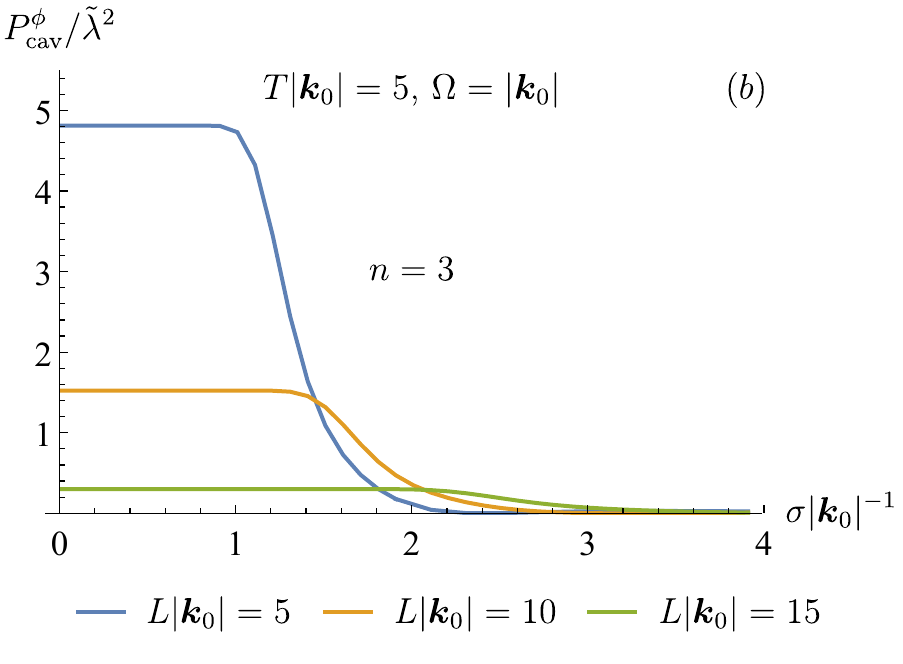}
        \caption{Excitation probability when $\Omega=|\bk_0|$ as a function of the wavepacket width $\sigma$ in a (3+1)-dimensional Dirichlet cavity. {(a)} different curves refer to different interaction time $T$ (in units of $|\bk_0|^{-1}$). The excitation probability increases with longer interaction time. {(b)} different curves refer to different cavity size $L$ (in units of $|\bk_0|^{-1}$) and the detector is always at the centre of the cavity. Note that the size of the plateau as $\sigma\to 0$ increases with cavity size, while the excitation probability decreases with size. In both cases, however, unlike the free-space setting the probability is always \textit{maximized} in the monochromatic limit. We have chosen $T$ to be large enough so that the result is within the long-time regime: the total probability $P_{\text{cav}}^\phi$ is dominated by the co-rotating contribution, while the counter-rotating and vacuum contributions are negligible. }
        \label{fig: resonant-cavity-peak-limit}
    \end{figure*}

    The result above is (in the authors' opinion) an intuition-defying one: in the monochromatic limit the detector will not be excited despite the exact frequency match between the energy gap and energy of the field quantum.  To understand this better, we will now discuss how these results compare with the (perhaps) more common expectation coming from the light-matter interaction in optical cavities. Namely, the fact that for detector-field resonance the monochromatic resonant limit (\mbox{$\sigma|\bk_0|^{-1}\to 0$}, $\Omega=|\bk_0|$) should have the largest chance of exciting the detector.

    
    The main difference between the free space (with no cavity walls imposing boundary conditions) and the cavity case is that the exact monochromatic states of the form $\hat a^\dagger_\bk\ket{0}$ are not normalizable in free space as $\braket{0|\hat a_\bk^\pdag\hat a_{\bk'}^\dagger|0} = \delta^{(n)}(\bk-\bk')$. This is unlike the situation in cavities, where the field has discrete momenta and the exact monochromatic Fock state $\hat a_{\bk}^\dagger\ket{0}$ is normalizable since $\braket{0|\hat a^{\pdag}_\bk\hat a_{\bk'}^\dagger|0} = \delta_{\bk\bk'}$, where $\delta_{\bk\bk'}$ is equal to 1 when $\bk=\bk'$ and zero otherwise. 


    We will now see how and why fields in cavities do not suffer from the probability decrease in the monochromatic resonant limit  when the energy gap matches the peak frequency of the wavepacket: instead, as intuition suggests, the excitation probability is \textit{maximized} when we take the monochromatic limit for any number of spatial dimensions in cavities.
     
    Let us consider a massless scalar field in $(n+1)$ dimensions confined to an $n$-dimensional Dirichlet cavity of dimension $L\times \cdots \times L$. The field $\phi$ satisfies Dirichlet boundary conditions whenever $x^i = 0$ and $x^i=L$ for all $i=1,...,n$, i.e. $\hat\phi(t,x^i=0) = \hat\phi(t,x^i=L)  =0$. It follows that the mode decomposition of the field is given by
    \begin{align}
        \hat \phi(t,\bx) = \sum_{I} \hat a^{\pdag}_I u^{\pdag}_I(t,\bx)+  a_I^\dagger u_I^*(t,\bx)\,, 
        \label{eq: cavity-mode-decomposition}
    \end{align}
    where $\hat a_{I}\equiv \hat a_{\bk_I}$ and $I$ is a multi-index which labels discrete momenta
    \begin{align}
        \bk_I\coloneqq (k_1,...,k_n) =  \frac{\pi}{L}(j_1,...,j_n)\,.
    \end{align}
    The summation over $I$ in the mode decomposition \eqref{eq: cavity-mode-decomposition} is a shorthand for $n$-dimensional summation over each \mbox{$j_i\in \mathbb{N}$}. Each mode with momentum $\bk_I$ is given by
    \begin{align}
        u^\pdag_I(t,\bx) &= v^\pdag_I(\bx)e^{-\ii|\bk_I|t}\,,
    \end{align}
    where
    \begin{align}
        v^\pdag_I(\bx) &\coloneqq \frac{1}{\sqrt{2|\bk_I|}}\left({\frac{L}{2}}\right)^{\!\frac{n}{2}}\prod_{i=1}^n\sin\rr{\frac{j_i\pi x^i}{L}}\,.
    \end{align}

    Now consider the cavity-field state analogous to the one-particle Fock wavepacket we considered in the continuum:
    \begin{align}
        \ket{1_f}\coloneqq \sum_{I} f_{\bk_0,\sigma}(\bk_I)\hat a_I^\dagger\ket{0}\,, 
    \end{align}
    where $f_{\bk_0,\sigma}$ is a single-peaked real-valued function  with dominant momentum $\bk_0$ (such as Gaussian) satisfying that
    \begin{align}
        \braket{1_f|1_f} = 1\Longrightarrow \sum_{I}|f_{\bk_0,\sigma}(\bk_I)|^2 = 1\,.
        \label{eq: L2-norm-cavity}
    \end{align}
    This is the discrete version of $L^2$-normalization in Section~\ref{sec: oneparticle}. The Wightman two-point function~\eqref{eq: wightman-linear-one} is given by
    \begin{align}
        W^\phi(\sx,\sx') &= \mathcal{W}^\phi_{\text{vac}}(\sx,\sx') + \rr{\mathcal{K}^*_{\bk_0}(\sx)\mathcal{K}_{\bk_0}(\sx') + \text{c.c}}\,,
        \label{eq: Wightman-linear-one-cavity}
    \end{align}
    where
    \begin{align}
        \mathcal{W}^\phi_{\text{vac}}(\sx,\sx') &= \sum_I u^\pdag_I(\sx)u_I^*(\sx')\,,\\
        \mathcal{K}_{\bk_0}(\sx) &= \sum_I f_{\bk_0,\sigma}(\bk_I)u_I^*(\sx)\,,
    \end{align}
    which are analogous to the free space counterparts $W^\phi_{\vac}$ and $K_{\bk_0}$ respectively.
    
    The excitation probability can be calculated using Eq.~\eqref{eq: general-prob-linear}, and the Wightman function defined in Eq.~\eqref{eq: Wightman-linear-one-cavity}. 
    In the adiabatic long-time regime, the vacuum contribution and the counter-rotating term can be neglected. Therefore, the excitation probability of a static detector located at $\bx=\bx_d$ is dominated by the non-vacuum co-rotating contribution, which reads
    \begin{align}
        P^\phi &= \lambda^2 \left|\sum_{I}f_{\bk_0,\sigma}(\bk_I)\tilde\chi(\Omega-|\bk_I|)v_I(\bx_d)\right|^2\,.
    \end{align}

    Let us study concretely the monochromatic limit \mbox{$\sigma |\bk_0|^{-1} \to 0$} (keeping $\bk_0$ fixed). For Gaussian spectrum, we have
    \begin{align}
        f_{\bk_0,\sigma}(\bk_I) = \NN_{\sigma}e^{-\frac{|\bk_I-\bk_0|^2}{2\sigma^2}}\,,
    \end{align}
    with $\NN_\sigma$ the normalization constant to be determined. Using \eqref{eq: L2-norm-cavity}, and writing $\bk_I = (\pi/L)(j_1,j_2,...,j_n)$ and $\bk_0 = (\pi/L)(j_1^0,...,j_n^0)$, we get
    \begin{align}
        \NN_\sigma &= \left(\prod_{i=1}^{n} \left[\sum_{m=0}^{j^0_i-1} e^{ -\alpha^2 m^2}+\frac{1}{2} \left(\vartheta _3(0,e^{-\alpha^2})-1\right)\right]\right)^{\!\!-\frac{1}{2}}\!\!\!\,,
    \end{align}
    where $\alpha = \pi/(\sigma L)$ and $\vartheta_a(u,q)$ is the Jacobi theta function \cite{NIST:DLMF}. 
    
    We remark that the crucial property of the wavepacket in the cavity scenario is that (unlike in free space) we have 
    \begin{align}
        \lim_{\sigma \to 0} \NN_\sigma &= 1\,,\hspace{0.5cm}
        \lim_{\sigma\to 0}e^{-\frac{|\bk_I-\bk_0|^2}{2\sigma^2}} = \delta_{\bk_I\bk_0}\,.
    \end{align}
    Consequently, so long as $|\bk_0|$ matches one of the frequencies of the field modes, we will have
    \begin{align}
        \lim_{\sigma\to 0}\ket{1_f} = \hat a_{\bk_0}^\dagger \ket{0}\,,
        \label{eq: monochromatic-limit-state-cavity}
    \end{align}
    which is a physically well-defined (i.e., normalizable) exact monochromatic Fock state. In this monochromatic limit, the detector excitation probability reduces to
    \begin{align}
        \lim_{\sigma\to 0} P^\phi & = \lambda^2 \bigr|\tilde\chi(\Omega-|\bk_0|)v_{\bk_0}(\bx_d)\bigr|^2
        \label{eq: monochromatic-limit-state-cavity-matching}
    \end{align}
    for any number of spatial dimensions. 
    
    We can now see that for the cavity scenario Eq.~\eqref{eq: monochromatic-limit-state-cavity-matching} shows that the probability is strongly enhanced when the energy gap matches the Fock state frequency ($\Omega\approx |\bk_0|$) and highly suppressed when it is far from resonance. The fact that the excitation probability at resonance converges to a maximum value is shown in Figure~\ref{fig: resonant-cavity-peak-limit}, where we consider a Gaussian switching given by $\chi(t) = e^{-t^2/T^2}$ and set $T|\bk_0|\gg 1$. 
    
    {In short, unlike the continuum case, the detector in a cavity can resonate with the field's quantum and the excitation probability is \textit{maximized} when the quantum frequency matches exactly with the detector gap in the monochromatic limit. Notice that in cavity, a peaked momentum wavepacket cannot be infinitely delocalized since the cavity length is finite, {thus the energy density of the wavepacket is non-zero in the monochromatic limit}. This is in stark contrast to the continuum case and can explain why the wavepacket does not become transparent for the detector in this case.}


    \subsection{Quadratic coupling: transition probability in arbitrary dimensions}
    
    We move now to the non-linear coupling between the detector and the field. Here we calculate the excitation probability of a detector interacting quadratically with a massless scalar field in analogous fashion as the previous subsection on linear coupling. 
    
    We begin by substituting our definition of a Fock state~\eqref{eq: one-particle-state} into the Wightman two-point function~\eqref{eq: wightman-quadratic-gen}, then apply the field expansion~\eqref{eq:field} and the canonical commutation relations \eqref{eq:canon}. The Wightman two-point function reads
    \begin{equation}
        \begin{split}
        &W^{\phi^2}(\sx,\sx') = \bra{1_f}\normal{\hat{\phi}^2(\sx)}\normal{ \hat{\phi}^2(\sx')}\ket{1_f}\\
        &=  2 W^\phi_{\vac}(\sx,\sx')^2 + 4 W^\phi_{\vac}(\sx,\sx')\rr{K_{\bk_0}^*(\sx)K_{\bk_0}(\sx') + \text{c.c.}}\,,
        \label{eq: wightman-quad-one}
        \end{split}
    \end{equation} 
    where $K_{\bk_0}(\sx)$ is defined in Eq.~\eqref{eq: K-integral}. Notice that the first term $2W^\phi_{\vac}(\sx,\sx')^2$ is the vacuum Wightman two-point function for a quadratic interaction \cite{Allison2017a}. This means that, as before, the response function can be split into two parts, i.e.
    \begin{align}
        P^{\phi^2} = P^{\phi^2}_{\vac}+P^{\phi^2}_{K}\,.
        \label{eq: prob-quadratic-schematic}
    \end{align}
    The first term is the vacuum contribution
    \begin{equation}
    \begin{split}
        \label{Qvacuum}
        P_{\vac}^{\phi^2} &= \frac{ 2\lambda^2}{[2(2\pi)^{n}]^2} \int
        \frac{{\dd}^n\bm{k}_1}{|\bk_1|}\int\frac{{\dd}^n\bk_2}{|\bk_2|} \\
        &\phantom{=}\times \tilde{\chi}[\Omega+|\bk_1|+|\bk_2|]^2 \tilde{F}[\bk_1+\bk_2]^2\,.
    \end{split}
    \end{equation}
    The non-vacuum contribution $P^{\phi^2}_K$ can be written in compact form by defining (cf. Eq.~\eqref{eq: prob-linear-one-compactify})
    \begin{align}
        &\mathcal{J}_\pm(\bk_1; \sigma,\bk_0,\tilde{F},\tilde{\chi})\notag\\
        &\coloneqq \int \frac{\dd^n\bk}{\sqrt{2(2\pi)^n|\bk|}}f_{\bk_0,\sigma}(\bk)\tilde{F}(\bk-\bk_1)\tilde{\chi}(\Omega+|\bk_1|\pm|\bk|).
        \label{eq: prob-quad-one-compactify}
    \end{align}
    The non-vacuum contribution now reads
    \begin{align}
        P^{\phi^2}_K &= \frac{4\lambda^2}{2(2\pi)^n}\int \frac{\dd^n\bk_1}{|\bk_1|}\left(\mathcal J_+^2(\bk_1; \sigma,\bk_0,\tilde{F},\tilde{\chi})\right.\notag\\
    	&\hspace{2.7cm}+\left.\mathcal J^2_-(\bk_1; \sigma,\bk_0,\tilde{F},\tilde{\chi})\right)\,.
    	\label{eq: J-function}
    \end{align}

    \begin{figure*}[htp]
        \includegraphics[scale=0.8]{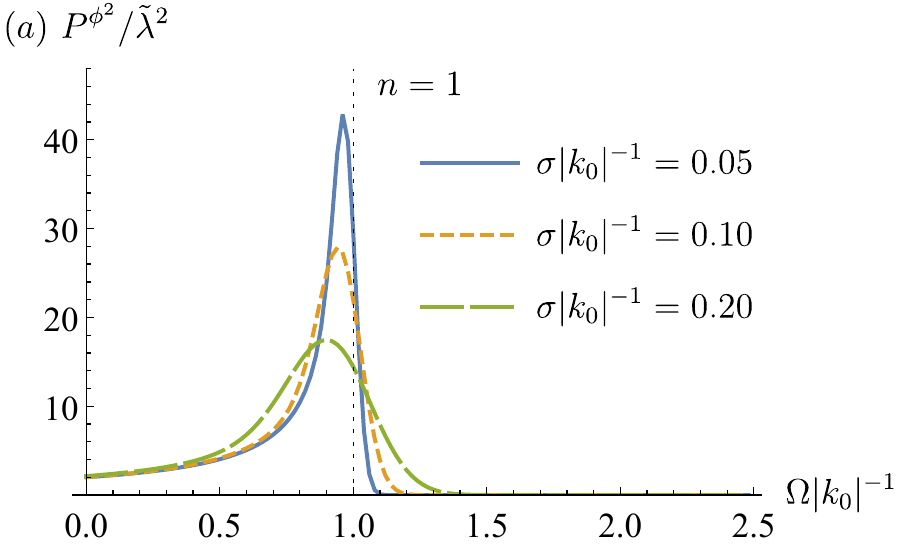}
        \includegraphics[scale=0.8]{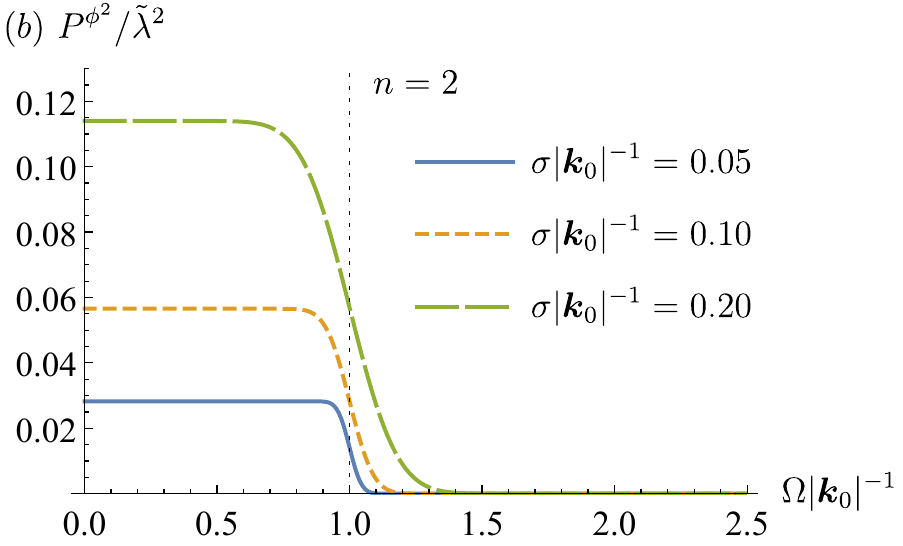}
        \includegraphics[scale=0.8]{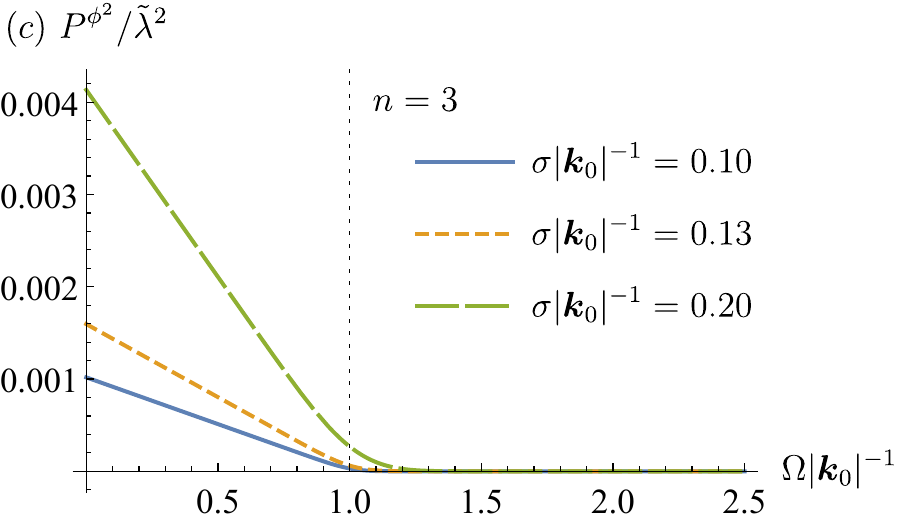}
        \includegraphics[scale=0.8]{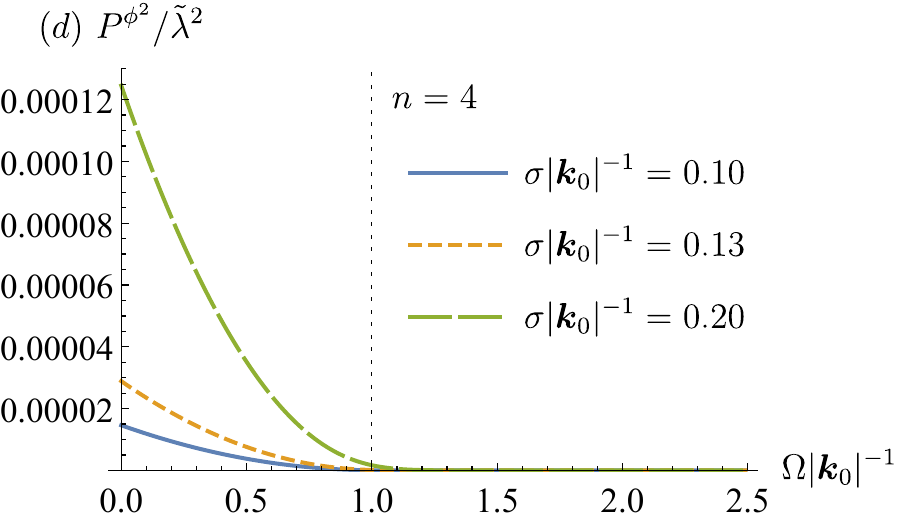}
        \caption{
        Plots of transition probability $P^\phi/\tilde{\lambda}^2$ as a function of frequency for quadratic coupling and one-particle state in various spatial dimensions $n$, as the spectral bandwidth $\sigma$. We vary $\Omega$ as we search for resonant peaks while keeping $|\bk_0|$ fixed. Here $\tilde\lambda = \lambda|\bk_0|^{(n-2)}$ is the non-dimensionalized coupling constant. The vertical lines denote the resonant frequency $|\bk_0|=\Omega$.
        }
    \label{fig: quad_resonance1}
    \end{figure*}
    	
    We now consider the effect on $P_K^{_{\phi^2}}$ of the same two limits considered in the previous case: the long switching time and pointlike regimes. In these limits, the vacuum contribution and the ``counter-rotating'' term $\mathcal{J}_+$  will vanish for a ground state detector for the similar reasons as the ones described in the linear coupling setup~\cite{Allison2017a}. With these assumptions, the full transition probability is given only in terms of the ``co-rotating'' term (cf. Eq.~\eqref{eq: final-linear-probability}):
	\begin{align}
	    P^{\phi^2} = \frac{4\lambda^2(2\pi)^2}{2(2\pi)^n}\int \frac{\dd^n\bk_1}{|\bk_1|}\mathcal{J}^2_-(\bk_1; \sigma,\bk_0)\,,
    	\label{eq: final-quad-probability}
    \end{align}
    where we short $\mathcal{J}_-(\bk_1; \sigma,\bk_0)\equiv \mathcal{J}_-(\bk_1; \sigma,\bk_0, 1,2\pi\delta )$.
    
    In order to perform explicit calculations, we need to  specify the spectrum of the one-particle wavepacket \mbox{$f_{\bk_0,\sigma}(\bk)$}. We will use the Gaussian distribution in Eq.~\eqref{eq: GaussianFrequency} and we consider two cases: $n\geq 2$ and $n=1$. For $n\geq 2$, we can simplify the expression for $\mathcal{J}_-$ using the method outlined in Appendix~\ref{appendix: exactprobability} and obtain:
    \begin{align}
        \mathcal{J}_-(\bk_1;\sigma,\bk_0)
        &=  \frac{2 \pi ^{n/2}(|\bk_1|+\Omega )^{n-\frac{3}{2}}}{\sqrt{2 (2 \pi )^n} (\pi  \sigma ^2)^{n/4}}  e^{-\frac{|\bk_0|^2+(|\bk_1|+\Omega )^2}{2 \sigma ^2}} \,\notag\\ &\hspace{0.4cm}\times\, _0\tilde{F}_1\left(\frac{n}{2};\frac{|\bk_0|^2 (|\bk_1|+\Omega )^2}{4 \sigma ^4}\right)\,,
        \label{eq: Jminus-final}
    \end{align}
    where $_0\tilde{F}_1$ is the regularized, generalized hypergeometric function \cite{NIST:DLMF}. For $n=1$, we introduce an IR cutoff $\Lambda$ and under the assumption $0<\Lambda<\Omega,|\bm k_0|$ we have by direct integration
    \begin{align}
        &\mathcal{J}_-(k_1;\sigma,k_0) \notag\\
        &= \frac{(\pi\sigma^2)^{-1/4}}{\sqrt{4\pi(|k_1|+\Omega)}}\bigg[e^{-\frac{(\Omega+|k_1|+| k_0|)^2}{2\sigma^2}} + e^{-\frac{(\Omega+|k_1|-|k_0|)^2}{2\sigma^2}}\bigg]\,.
        \label{eq: Jminus-final-1D}
    \end{align}
    Analogous to the linear results in Eqs.~\eqref{eq: prob-linear-gaussian-final}, we can show that the cutoff-free expression in Eq.~\eqref{eq: Jminus-final-1D} can be obtained by taking the limit $n\to 1$ of $\mathcal{J}_-$ in Eq.~\eqref{eq: Jminus-final}. Therefore, the expression for $\mathcal{J}_-$ in Eq.~\eqref{eq: Jminus-final} is valid in arbitrary dimensions. Unfortunately, substituting either Eq.~\eqref{eq: Jminus-final} and \eqref{eq: Jminus-final-1D} into Eq.~\eqref{eq: final-quad-probability} does not give us useful closed-form expressions, so we must proceed numerically. 
    
    We show the excitation probability for quadratic coupling for various dimensions in Figure~\ref{fig: quad_resonance1}. From these plots and  Eq.~\eqref{eq: final-quad-probability}, we can make two general observations. First, similar to the linear case, we see that the qualitative behaviour of detector-field resonances varies greatly for different spacetime dimensions, with similarities only for $n\geq 4$. Only in $n=1$ do we observe a larger transition probability as we make the wavepacket with $|\bm k_0|=\Omega$ more monochromatic; for $n\geq 2$, the transition probability \textit{decreases} as $\sigma\to 0$. Therefore, for quadratic coupling detectors are increasingly more transparent to the field excitation as the wavepacket becomes narrower. The second observation is that unlike the linear coupling where resonance peaks (a maximum in the transition probability when the Gaussian peak matches the energy gap) are always visible in any dimensions, for quadratic coupling this only occurs for $n=1$. In two or more spatial dimensions there is \textit{no resonance phenomenon} for quadratic coupling when the field is a one-particle Fock state, and the detector's response is maximized when $\Omega\ll |\bk_0|$. In some sense this result, together with the results in the linearly coupled case, highlights that the behaviour of (1+1)D detector models for massless scalar field are the exception rather than the rule.


    \section{two-particle detection}
    \label{sec:twoparticle}
    
    In this section we will investigate how a detector responds to two-particle excitations of the field. We will first define what we mean by two-particle Fock state in free space, and then we we will consider how the detector response depends on the properties of the state. Again we will see that the ability of detectors to resonate with the field quanta strongly depends on both the choice of detector-field coupling and spacetime dimensions.

    \subsection{Two-particle Fock state}
    \label{subsec: two-particle-state-def}

    Recall from Section~\ref{subsec: energy-wavepacket} that a one-particle Fock state is defined as the eigenstate of the number operator $\hat N$ with eigenvalue 1, subject to the requirement that the spectrum/wavepacket profile $f$ is $L^2$-normalizable to unity so that the state has norm 1. An important takeaway from that section is that there is no requirement on the \textit{shape} of the profile itself: in particular, it need not have, for instance, a single peak in the momentum distribution. Consequently, in general a multi-particle Fock wavepacket can also have very complicated momentum or frequency distribution. An $m$-particle Fock state need not be described by a spectrum that has $m$ peaks. The only requirement for a state to be an $m$-particle \textit{physical} Fock state is that it is a \textit{unit-norm} eigenstate of the number operator $\hat N$ with eigenvalue $m$. 
    
    Analogous to the analysis in Section~\ref{subsec: energy-wavepacket} we would like to consider a relatively simple subclass of two-particle Fock states. For instance, we would like to consider two-particle states that have two `peaks' in its frequency distributions (possibly equal). Such a choice would help in the physical interpretation of our results as the resonant peaks can be easily identified whenever they appear, and a notion analogous to monochromaticity---i.e., \textit{dichromaticity}---can be defined. 
    
    Inspired by the definition in Eq.~\eqref{eq: one-particle-state-peaked}, we can construct the following candidate for a two-particle Fock state by adding one more excitation on the one-particle Fock state $\ket{1_f}$ in Eq.~\eqref{eq: one-particle-state}, i.e.
    \begin{align}
        \ket{2_{gf}}
        &\coloneqq  \NN\int  \dd^n\bk\, g(\bk)\ann_{\bk}^\dagger \ket{1_f}\notag\\ 
        &= \NN\int  \dd^n\bk\,\dd^n\bk' g(\bk)f(\bk')\ann_{\bk}^\dagger\ann_{\bk'}^\dagger \ket{0}\,,
        \label{eq: two-particle-state}
    \end{align}
    where $f$ and $g$  have $L^2$-norm \mbox{$||f|| = ||g||=1$}. The prefactor $\NN$, which is necessary to enforce $\braket{2_{gf}|2_{gf}}=1$, is a positive normalization constant to be determined later. Applying the number operator to this wavepacket state, we get
    \begin{align}
        \hat N\ket{2_{gf}} 
        &= \int \dd^n\bm{p}\,\hat a_{\bm{p}}^\dagger\hat a_{\bm{p}}^\pdag \ket{2_{gf}} = 2\ket{2_{gf}}\,,
        \label{eq: two-particle-eigenstate}
    \end{align}
    hence it is a genuine two-particle Fock state (cf. Eq.~\eqref{eq: one-particle-eigenstate}). The state $\ket{2_{gf}}$ is a physical (normalizable) version of the na\"ive two-particle Fock state $\hat a_{\ba}^\dagger\hat a_{\bb}^\dagger\ket{0}$.
    
    The two-particle Fock state defined in Eq.~\eqref{eq: two-particle-state} is particularly useful because it allows us to introduce two peaks in the momentum distribution in a natural way. For example, we can take $g$ and $f$ to be single-peaked Gaussian functions centred at different momenta $\ba$ and $\bb$ respectively. For simplicity, we will assume that both $f$ and $g$ are given by the same single-peaked function with the same width $\sigma$ and which differ by a simple translation, namely
    \begin{align}
        g(\bk) \equiv f_{\ba,\sigma}(\bk)\,, \hspace{0.5cm}f(\bk)\equiv f_{\bb,\sigma}(\bk)\,.
    \end{align}
    Since we use the same single-peaked function $f_{\bc_j,\sigma}$ for both $f$ and $g$ which only differ by the location of their peaks at $\bc_j$, we will alleviate notation by rewriting the state as follows:
    \begin{align}
        \ket{2_{gf}}\to \ket{2_f}\,.
    \end{align}
    We will also assume that $f_{\bc_j,\sigma}$ are real-valued functions as we will be focusing on a Gaussian spectrum later.
    
    Let us now work out the normalization constant $\NN$. We first compute $\braket{2_f|2_f}$:
    \begin{align}
        \braket{2_{f}|2_{f}} &= \NN^2\!\!\int \dd^n \bk_1\,\dd^n \bk_2\,\dd^n \bk_3\,\dd^n \bk_4\braket{0|\a{\bk_1}^\pdag\a{\bk_2}^\pdag\ad{\bk_3}\ad{\bk_4}|0} \notag\\
        &\hspace{0.4cm}\times f_{\ba,\sigma}(\bk_1)f_{\bb,\sigma}(\bk_2)f_{\ba,\sigma}(\bk_3)f_{\bb,\sigma}(\bk_4)\,.
    \end{align}
    Using the canonical commutation relations and demanding that $\braket{2_f|2_f}=1$, the expression reduces to
    \begin{align}
        1 &= \NN^2\!\!\int \dd^n \bk_1\dd^n \bk_2 \bigg[f^2_{\ba,\sigma}(\bk_1)f^2_{\bb,\sigma}(\bk_1)\notag \\
        &\hspace{0.5cm} + \,f_{\ba,\sigma}(\bk_1)f_{\bb,\sigma}(\bk_1)
        f_{\ba,\sigma}(\bk_2)f_{\bb,\sigma}(\bk_2)\bigg]\,.
    \end{align}
    Using the fact that $\norm{f_{\ba,\sigma}}=\norm{f_{\bb,\sigma}}=1$, the normalization $\NN$ is given by
    \begin{align}
        \mathcal{N} &= \frac{1}{\sqrt{1 +C_{\ba\bb}^2}}\,,
        \label{eq: normalization-general-twoparticle}\\
        \spec &\coloneqq \int\dd^n\bk\,f_{\ba,\sigma}(\bk)f_{\bb,\sigma}(\bk)\,.
        \label{eq: GaussianSpectral}
    \end{align}
    
    Let us check that when the wavepacket is dichromatic then the energy expectation will be $\hbar (|\ba|+|\bb|)$. The general expression reads
    \begin{align}
        \braket{2_f|\hat H_{0,\phi}|2_f} 
        &= \NN^2\!\!\int \dd^n\bk |\bk|\,\left(|f_{\ba,\sigma}(\bk)|^2+ |f_{\bb,\sigma}(\bk)|^2\right)\notag\\
        &+ 2\NN^2\spec\!\!\int\dd^n\bk|\bk|\,f_{\ba,\sigma}(\bk)f_{\bb,\sigma}(\bk)\,.
        \label{eq: energy-expectation-two-particle}
    \end{align}
    Since, as discussed around Eq.~\eqref{eq: energy-average-nascent},  $|f_{\bc_j,\sigma}|^2$ gives rise to a family of nascent-delta functions, in the dichromatic limit $\sigma\to 0$ we have that $|f_{\bc_j,\sigma}|^2\to\delta^{(n)}(\bk-\bc_j)$
    \begin{align}
        \lim_{\sigma\to 0}\braket{2_f|\hat H_{0,\phi}|2_f} &= |\ba|+|\bb|\,,
    \end{align}
    where the limit is understood in the distributional sense. When $\sigma\to 0$, the last term of Eq.~\eqref{eq: energy-expectation-two-particle} vanishes. Furthermore, the normalization $\mathcal{N}\to 1$ because $\spec\to 0$: any nascent delta functions centred at different points are $L^2$-orthogonal in that limit. 
    

    \subsection{Linear coupling: transition probability in arbitrary dimensions}
    \label{sec:twoparticleLINEAR}
        
    We will now evaluate the linearly coupled detector excitation probability, Eq.~\eqref{eq: general-prob-linear}, and see how the detector responds to two-particle excitations in the field. 
    
    First, the Wightman two-point function we need to calculate is
    \begin{align}
        W_{\ba\bb}^{\phi}(\sx,\sx'):=\bra{2_{f}}\hat{\phi}(\sx)\hat{\phi}(\sx')\ket{2_{f}}\,,
    \end{align}
    where the subscripts $\ba,\bb$ denote the peak momenta of the two field excitations in $\ket{2_{f}}$. Direct computation yields
    \begin{align}
        &W_{\ba\bb}^\phi(\sx,\sx') = W^\phi_{\vac}(\sx,\sx')\notag\\
        &+{\NN}^2\spec\bigg[K_\ba(\sx)K^*_\bb(\sx') + K_\bb(\sx)K^*_\ba(\sx') + \text{c.c.}\bigg]\notag\\
        &+{\NN^2}\bigg[K_\ba(\sx)K^*_\ba(\sx') + K_\bb(\sx)K^*_\bb(\sx') + \text{c.c.} \bigg]  \,,
        \label{eq: linear-two-wightman}
    \end{align}
    where  $K_{\bm{\eta}_j}(\sx)$ ($j=1,2$) is defined according to Eq.~\eqref{eq: K-integral}.

    Now we can calculate the detector response to a two-particle excitation: substituting Eq.~\eqref{eq: linear-two-wightman} into Eq.~\eqref{eq: general-prob-linear}, we can write the full transition probability as the sum of four contributions:
    \begin{equation}
        \begin{split}
            \label{eq: genlinearprob2}
            P_{\ba\bb}^\phi &= P_{\vac}^\phi + P_{K,\bc_1\bc_1}^\phi+P_{K,\bc_2\bc_2}^\phi+ 2P_{K,\bc_1\bc_2}^\phi \,. 
        \end{split}
    \end{equation}
    The first term $P_{\vac}^\phi$ is the vacuum contribution, which vanishes in the adiabatic limit. We can simplify the non-vacuum contribution by defining the following integral:
    \begin{align}
        &\mathcal{M}_\pm(\sigma,\bc_j,\tilde{F},\tilde{\chi})\notag\\
        &\coloneqq {\NN}\int\frac{{\dd}^n \bm{k}}{\sqrt{2(2\pi)^n|\bm{k}|}}f_{\bc_j,\sigma}(\bm{k}) \tilde{F}[\bm{k}]\tilde{\chi}(\Omega\pm|\bm{k}|)\,.
    \end{align}
    The expression $\mathcal{M}_\pm(\sigma,\bc_j,\tilde{F},\tilde{\chi})$ enables all non-vacuum contributions in Eq.~\eqref{eq: genlinearprob2} to be written concisely:
    \begin{align}
        P_{K,\bc_i\bc_i}^\phi &= \lambda^2
        \left[\mathcal{M}^2_+(\sigma,\bm{\eta}_i,\tilde{F},\tilde{\chi})+
        \mathcal{M}^2_-(\sigma,\bm{\eta}_i,\tilde{F},\tilde{\chi})\right]\,,
        \label{eq: nonvactwoparticle1}\\
        P_{K,\ba\bb}^\phi &= \lambda^2 C_{\ba\bb}
        \left[\mathcal{M}_+(\sigma,\ba,\tilde{F},\tilde{\chi})\mathcal{M}_+(\sigma,\bb,\tilde{F},\tilde{\chi}) \right.\notag\\
        &\left.+\mathcal{M}_-(\sigma,\ba,\tilde{F},\tilde{\chi})\mathcal{M}_-(\sigma,\bb,\tilde{F},\tilde{\chi})\right]\,.
        \label{eq: nonvactwoparticle2}
    \end{align}

    \begin{figure*}[tp]
        \centering
        \includegraphics[scale=0.825]{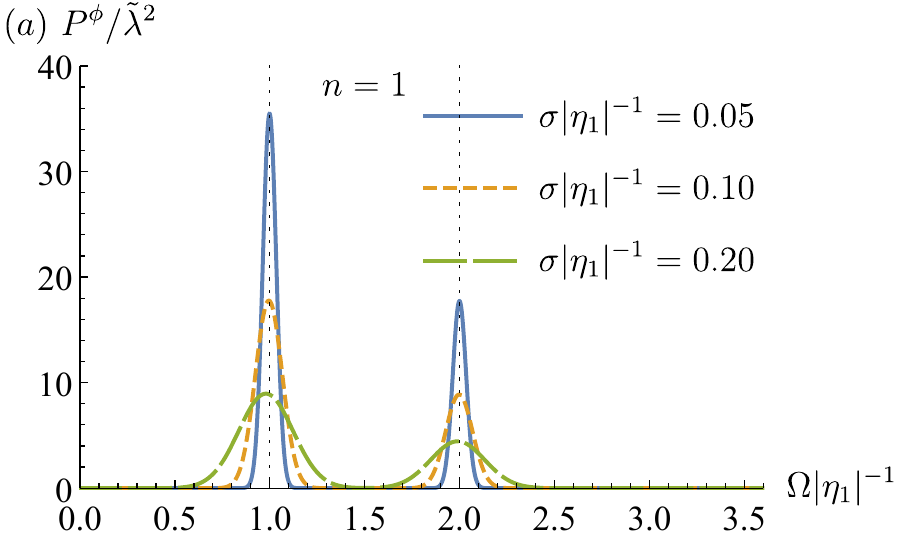}
        \includegraphics[scale=0.825]{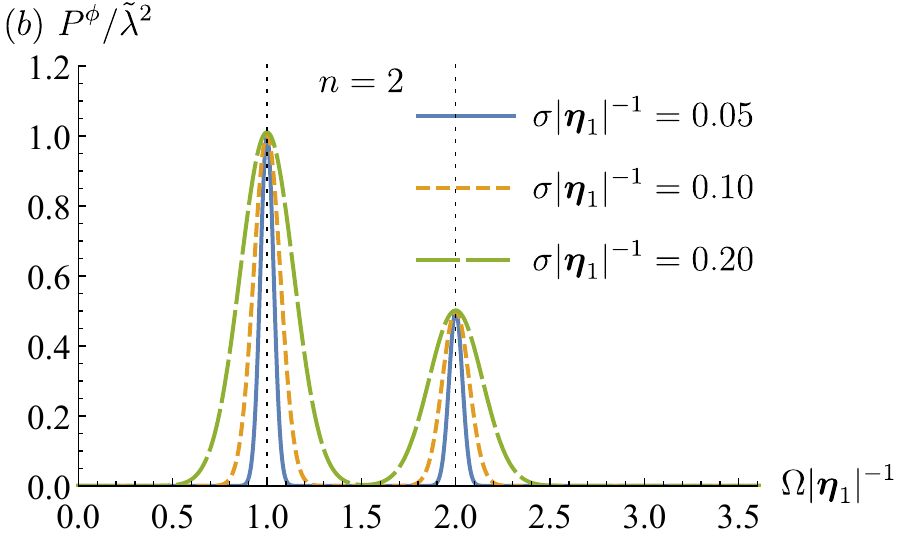}
        \includegraphics[scale=0.825]{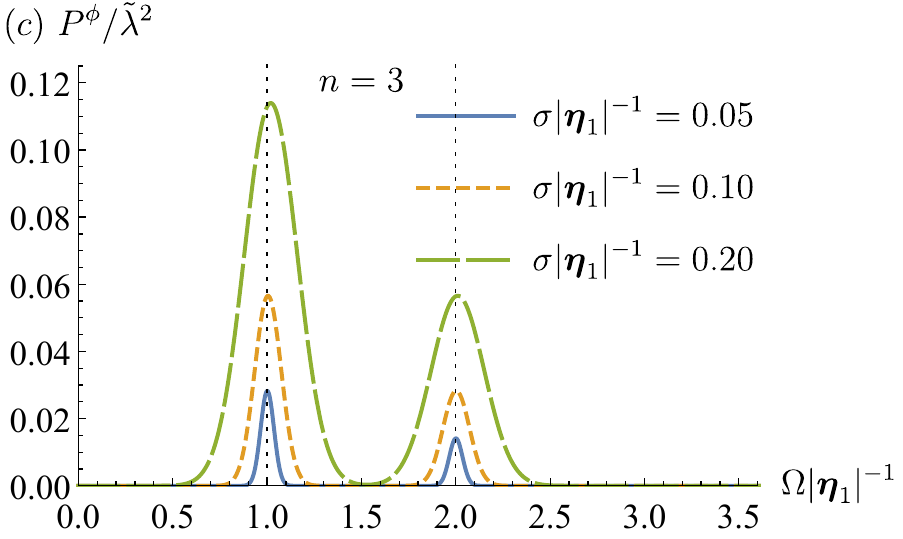}
        \includegraphics[scale=0.825]{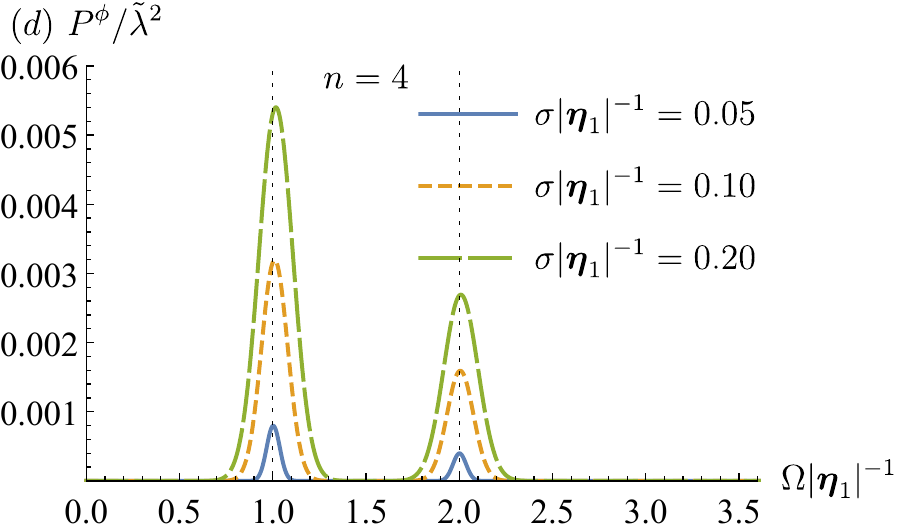}
        \caption{Plots of transition probability $P^{\phi}/\tilde{\lambda}^2$ as a function of frequency for linear coupling and two-particle state for $n=1$ and $n=3$, where $n$ is the number of spatial dimensions. We vary $\Omega$ as we search for resonant peaks while keeping $|\bk_0|$ fixed. Here $\tilde\lambda = \lambda |\bk_0|^{(n-3)/2}$ is the dimensionless coupling constant. The vertical lines denote the resonant frequencies corresponding to peak frequencies $\Omega = |\bm{\eta}_1|$ and $\Omega = |\bm{\eta}_2|=2|\bm{\eta}_1|$. As in the one-particle case, in the monochromatic limit $\sigma\to 0$ the $n=3$ the peak \textit{diminishes} in amplitude, while for $n=1$ the peak increases as $\sigma\to 0$. 
        }
        \label{fig: linear_resonance2}
    \end{figure*}
    
    Following Section~\ref{sec: oneparticle}, we will focus on the long time and pointlike limits, which simplify \eqref{eq: nonvactwoparticle1} and \eqref{eq: nonvactwoparticle2} to
    \begin{align}
        P_{K,\bc_i\bc_i}^\phi & =\lambda^2 \mathcal{M}_-(\sigma,\bm{\eta}_i)\mathcal{M}_-(\sigma,\bm{\eta}_i)\,,\\
        P_{K,\ba\bb}^\phi & =\lambda^2 C_{\ba\bb} \mathcal{M}_-(\sigma,\ba)\mathcal{M}_-(\sigma,\bb)\,,
    \end{align}
    where $\mathcal{M}_-(\sigma,\bm{\eta}_i) \equiv \mathcal{M}_-(\sigma,1,2\pi\delta,\bm{\eta}_i)$.
    The full transition probability in Eq.~\eqref{eq: genlinearprob2} is therefore given by only the co-rotating term:
    \begin{align}
        P_{\ba\bb}^\phi &= \lambda^2 \bigg[\mathcal{M}^2_-(\sigma,\ba)+\mathcal{M}^2_-(\sigma,\bb)\notag\\
        &+ 2 \spec \mathcal{M}_-(\sigma,\ba)\mathcal{M}_-(\sigma,\bb)\bigg]\,.
        \label{eq: prob-linear-two-general}
    \end{align}
        
    In order to make progress beyond this point, let us now particularize to the case when $f_{\bc_j,\sigma}$ is a Gaussian centred at $\bc_j$, i.e.
    \begin{align}
        f_{\bc_j,\sigma}(\bk) = \frac{1}{(\pi\sigma^2)^{n/4}}e^{-\frac{(\bk-\bc_j)^2}{2\sigma^2}}\,,\hspace{0.5cm}j=1,2\,.
    \end{align}
    With this choice, it follows that $C_{\ba\bb}$ is given by
    \begin{align}
        \spec = \exp\rr{-\frac{|\ba-\bb|^2}{4\sigma^2}}\,,
    \end{align}
    and the normalization constant $\NN$ for $\ket{2_f}$ reads
    \begin{align}
        \NN = \frac{1}{\sqrt{1+e^{-\frac{|\ba-\bb|^2}{2\sigma^2}}}} \in \left[\frac{1}{\sqrt{2}},1\right)\,.
        \label{eq: normalization-Gaussian}
    \end{align}    
    This range for $\mathcal{N}$ makes sense from an intuitive point of view: the two extremes correspond to the fully monochromatic limit in which (in the case of a discrete number of modes) the creation operators yield a factor 1 when creating excitations of different frequencies and a factor $\sqrt{2}$ when exciting the same mode, i.e.:
    \begin{align}
        \hat a^\dagger_{\ba}\hat a^\dagger_{\bb}\ket 0 &= 
        \begin{cases}
        \ket{1_\ba1_\bb}  \hspace{0.5cm}\ba\neq\bb\\[3mm]
        \sqrt{2}\ket{2_\ba} \hspace{0.5cm}\ba=\bb
        \end{cases}.
    \end{align}
    
    We can evaluate Eq.~\eqref{eq: prob-linear-two-general} in closed form. Notice that $\mathcal{M}_-$ has the same form as $\mathcal{I}_-$ in Eq.~\eqref{eq: prob-linear-one-compactify} except with the replacement $f_{\bk_0,\sigma}\to \NN f_{\bm{\eta}_j,\sigma}$. Therefore, it follows that for $n\geq 2$, we have
    \begin{align}
        &\mathcal{M}_-(\bm{\eta}_j,\sigma) \notag\\
        &=  \NN \sqrt{\frac{2\pi ^{2-\frac{n}{2}}}{|\bm{\eta}_j|^{n-2}}\frac{\Omega^{n-1}}{\sigma^{4-n}} } e^{-\frac{|\bm{\eta}_j|^2+\Omega ^2}{2\sigma ^2}} I_{\frac{n-2}{2}}\left(\frac{|\bm{\eta}_j| \Omega }{\sigma ^2}\right)\,.
        \label{eq:linear-twoparticle}
    \end{align}
    Substituting this into Eq.~\eqref{eq: prob-linear-two-general} gives the required closed form expression for $P_{\ba\bb}^\phi$ for $n\geq 2$. For $n=1$, we need to evaluate $\mathcal{M}_-$ with an IR cutoff $\Lambda$, which gives
    \begin{align}
        \mathcal{M}_-(\eta_j,\sigma) &=
        {\NN}\pi^{1/4} \left(\frac{e^{-\frac{(\Omega-\eta_j)^2}{2\sigma^2}}+e^{-\frac{(\Omega+\eta_j)^2}{2\sigma^2}}}{ \sqrt{\Omega \sigma} 
        }\right)\,,
        \label{eq:linear-twoparticle-1D}
    \end{align}
    where we implicitly demand that $\Omega, |\eta_j|\geq \Lambda>0$. This  expression can also be obtained by taking the limit $n\to 1$ of $\mathcal{M}_-$ in Eq.~\eqref{eq:linear-twoparticle}, thus $\mathcal{M}_-$ in Eq.~\eqref{eq:linear-twoparticle} is valid for all $n\geq 1$ as long as all relevant frequencies are above the IR cutoff.

    Fig.~\ref{fig: linear_resonance2} shows the transition probability of the detector for the choice $|\bb| = 2|\ba|$. As we vary $\Omega$, we can search for the values of the detector gap for which the detector will resonate with the two-particle excitations. We can make three observations here. First, as expected the resonance occurs around $\Omega = |\ba|$ and $\Omega=|\bb|$, with the peak aligning more closely to $\Omega$ as the wavepacket becomes more monochromatic ($\sigma\to 0$). Second, the peaks are not at equal height: the higher frequency peak is smaller than the lower frequency one, thus it is less likely for a detector to respond to higher frequency excitation of the field even in resonance. Third, we again see the dimension-dependence of the resonance peaks: for $n=1$, the peak is greater (higher transition probability) when the wavepacket is narrower, while for $n=2$ they approach constant values. For $n\geq 3$, the transition probability near resonance \textit{diminishes} with more monochromaticity, thus a detector is becoming more transparent to sharper wavepackets, similar to what happened with the one-particle case.
    
    Up to this point, the phenomenology of two-particle detection is not very different from the one-particle scenario, with the exception that there are two frequencies around which the detector can resonate. We will see in the next subsection that in addition to these phenomena, detector-field resonance for two-particle Fock state has much richer physics when non-linear coupling is considered. Some of the non-linear optical phenomena known collectively as \textit{harmonic generation} naturally arise within the quadratically coupled detector model. 
    

    \subsection{Quadratic coupling: transition probability in arbitrary dimensions}
    \label{subsec: quadratic-two-particle}
        
    Let us now study how a quadratically coupled detector responds to a two-particle Fock wavepacket. The two-point function reads
    \begin{align}
        W^{\phi^2}_{\ba\bb}(\sx,\sx'):=\braket{2_f|\normal{\phi^2(\sx)}\normal{\phi^2(\sx')}|2_f}\,,
    \end{align}
    where $\ba,\bb$ denote the peaks of the momentum distribution for the two-particle Fock wavepacket. The details of this evaluation is given in Appendix~\ref{appendix: Wightman-quad-twoparticle-proof}, and the resulting closed-form expression is
    \begin{align}
        &W_{\ba\bb}^{\phi^2}(\sx,\sx') \,\notag\\
        &= 4{\NN}^2\bigg[W_{\vac}^\phi(\sx,\sx')\rr{K_\ba {K}^{*'}_\ba + K_\bb {K}^{*'}_\bb +\text{c.c.}}  \notag\\
        &+4W_{\vac}^\phi(\sx,\sx')\spec\rr{K_\ba {K}^{*'}_\bb + K_\bb {K}^{*'}_\ba + \text{c.c.}}\, \notag \\
        &+ 4\rr{K^*_\ba K^*_\bb K'_\ba K'_\bb +K_\ba K_\bb^*{K_\ba^{*'}} K_\bb' + \text{c.c.} }  \notag\\
        &+ 4|K_\bb|^2|K_\ba'|^2 + 4|K_\ba|^2|K_\bb'|^2\bigg]+2W^\phi_{\vac}(\sx,\sx')^2\,,
       \label{eq: Wightman-quadratic-two}
    \end{align}
    where $K_{\bc_j}\equiv K_{\bc_j}(\sx)$ is defined in Eq.~\eqref{eq: linear-nonvacuum-probability-general} and we have used the shorthand $K'_{\bc_j}\equiv K_{\bc_j}(\sx')$ and $K^{*'}_{\bc_j}\equiv K_{\bc_j}^*(\sx')$. 
    
    Substituting the Wightman function \eqref{eq: Wightman-quadratic-two} into Eq.~\eqref{eq: general-prob-quad}, the transition probability can be written as
    \begin{align}
        P^{\phi^2} &= 
        P_{\vac}^{\phi^2} + ^{^{K^2}}\!\!\!\!P^{\phi^2}_{\bc_1\bc_1}+^{^{K^2}}\!\!\!\!P^{\phi^2}_{\bc_2\bc_2} + 2\big(\phantom{ } ^{^{K^2}}\!\!\!\!P^{\phi^2}_{\bc_1\bc_2}\big) + ^{^{K^4}}\!\!\!\!P^{\phi^2}_{\bc_1\bc_2}\,,
        \label{eq: probability-quadratic-two}
    \end{align}
    where, for clarity, we added the left-superindex $[K^2]$ as a shorthand notation referring to terms that depend on only products of two $K_{\bc_j}$'s in the Wightman function, and the $[K^4]$ left-superindex is notation for terms that depend on  products of four $K_{\bc_j}$'s in the Wightman two-point function. 
    
    In order to express the transition probability \eqref{eq: probability-quadratic-two} in a notationally manageable manner, we will define some functions analogous to $\mathcal{I}_\pm$, $\mathcal{J}_\pm$ and $\mathcal{M}_\pm$ in the previous subsections, namely
    \begin{align}
        &\mathcal{Q}_\pm(\bk;\sigma,\bc_j, \tilde{F},\tilde{\chi})\notag\\&
        \coloneqq 
        {\NN}\int\frac{{\dd}^n \bm{k'}}{\sqrt{2(2\pi)^n|\bm{k'}|}}\tilde{F}[\bk \pm \bk'] f_{\bc_j,\sigma}(\bm{k'})\tilde{\chi}(\Omega+|\bm{k}|\pm|\bk'|)\,,
        \label{eq: Qpm}\\
        &\mathcal{R}_\pm(\sigma,\bc_i,\bc_j, \tilde{F},\tilde{\chi}) \notag\\&
        \coloneqq 
        {\NN}\int\frac{{\dd}^n\bk\,{\dd}^n\bk'}{2(2\pi)^n\sqrt{|\bk||\bk'|}}\tilde{F}[\bk + \bk'] f_{\bc_i,\sigma}(\bk)f_{\bc_j,\sigma}(\bk')\notag\\
        &\hspace{3.5cm} \times \tilde{\chi}(\Omega\pm|\bm{k}|\pm|\bk'|)\,,
        \label{eq: Rpm}\\
        &\mathcal{S}_\pm(\sigma,\bc_i,\bc_j, \tilde{F},\tilde{\chi})\notag\\&
        \coloneqq 
        {\NN}\int\frac{{\dd}^n\bk\,{\dd}^n\bk'}{2(2\pi)^n\sqrt{|\bk||\bk'|}} \tilde{F}[\bk - \bk']f_{\bc_i,\sigma}(\bk)f_{\bc_j,\sigma}(\bk')\notag\\
        &\hspace{3.5cm} \times \tilde{\chi}(\Omega\mp |\bm{k}|\pm|\bk'|)\,.
        \label{eq: Spm}
    \end{align}
    These are defined based on the signs that appear in the argument of $\tilde\chi$ and $\Tilde{F}$. They, along with the symmetry exhibited by $S_{\pm}$ under exchange of $\bc_1$ and $\bc_2$, allow us to express the different terms in Eq.~\eqref{eq: probability-quadratic-two} as:
    \begin{align}
            ^{^{K^2}}\!\!\!\!P_{\bm{\eta}_i\bm{\eta}_i}^{\phi^2}&=                4\lambda^2\int\frac{\dd^n\bk}{2(2\pi)^n|\bk|}
            \bigg[
            \mathcal{Q}_+(\bk;\sigma,\bc_i, \tilde{F},\tilde{\chi})^2 \notag\\
            &\hspace{3cm}+
            \mathcal{Q}_-(\bk;\sigma,\bc_i, \tilde{F},\tilde{\chi})^2
            \bigg]\,, \label{eq:second-twoparticle-quadratic}\\
            ^{^{K^2}}\!\!\!\!P_{\bm{\eta}_1\bm{\eta}_2}^{\phi^2} &=                4\lambda^2\int\frac{\dd^n\bk\,{\spec}}{2(2\pi)^n|\bk|}
            \bigg[
            \mathcal{Q}_+(\bk;\sigma,\bc_1, \tilde{F},\tilde{\chi})\notag\\
            &\hspace{0.5cm}\times
            \mathcal{Q}_+(\bk;\sigma,\bc_2, \tilde{F},\tilde{\chi})+
            \mathcal{Q}_-(\bk;\sigma,\bc_1, \tilde{F},\tilde{\chi})\notag\\
            &\hspace{0.5cm}\times
            \mathcal{Q}_-(\bk;\sigma,\bc_2, \tilde{F},\tilde{\chi})
            \bigg]\,,\label{eq:third-twoparticle-quadratic}
            \\
            ^{^{K^4}}\!\!\!\!P_{\ba\bb}^{\phi^2} &= 4\lambda^2\bigg[
            \mathcal{R}_+^2(\sigma,\ba,\bb, \tilde{F},\tilde{\chi})+
            \mathcal{R}_-^2(\sigma,\ba,\bb, \tilde{F},\tilde{\chi})\notag\\
            &\hspace{0.5cm}+
            \mathcal{S}_-^2(\sigma,\ba,\bb, \tilde{F},\tilde{\chi})+
            \mathcal{S}_-^2(\sigma,\bb,\ba, \tilde{F},\tilde{\chi})\notag\\
            &\hspace{0.5cm}+
            2\mathcal{S}_-(\sigma,\bc_1,\bc_1, \tilde{F},\tilde{\chi})
            \mathcal{S}_-(\sigma,\bc_2,\bc_2, \tilde{F},\tilde{\chi})
            \bigg]\,.\label{eq:fourth-twoparticle-quadratic}
    \end{align}

    \begin{figure*}[tp]
        \centering
        \includegraphics[scale=0.64]{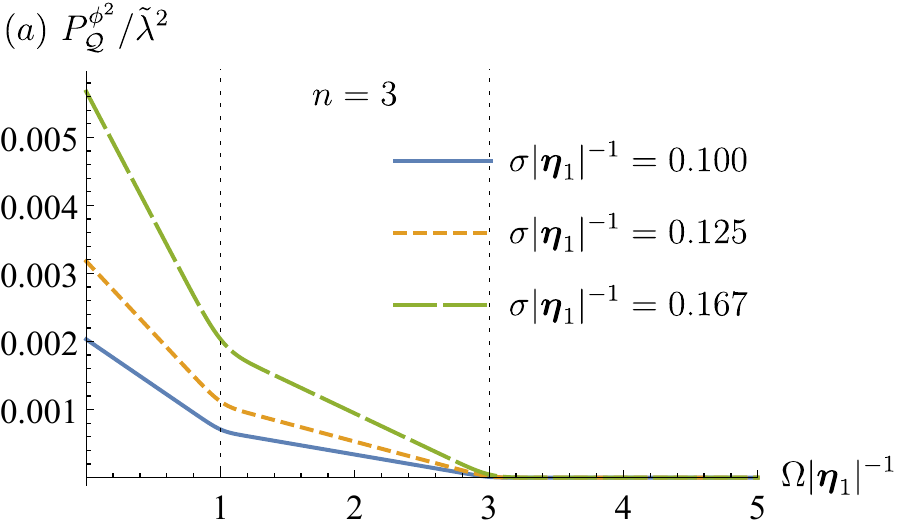}
        \includegraphics[scale=0.64]{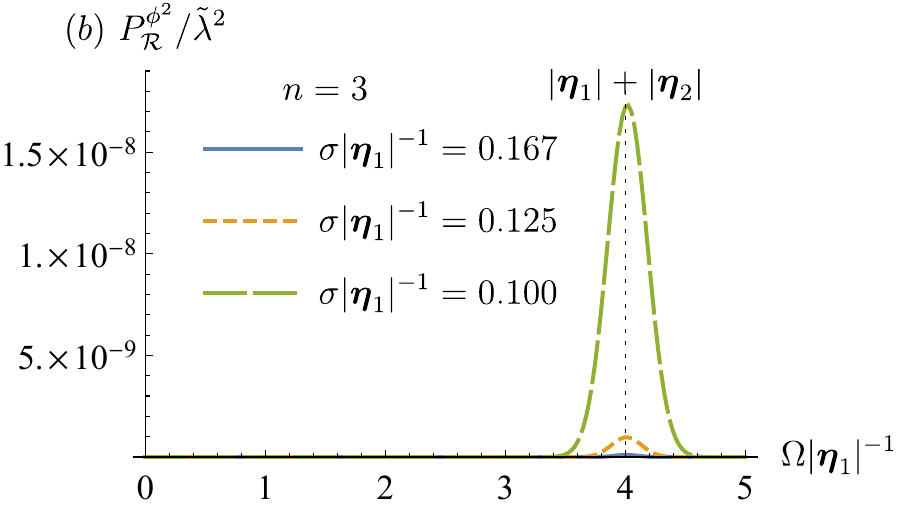}
        \includegraphics[scale=0.64]{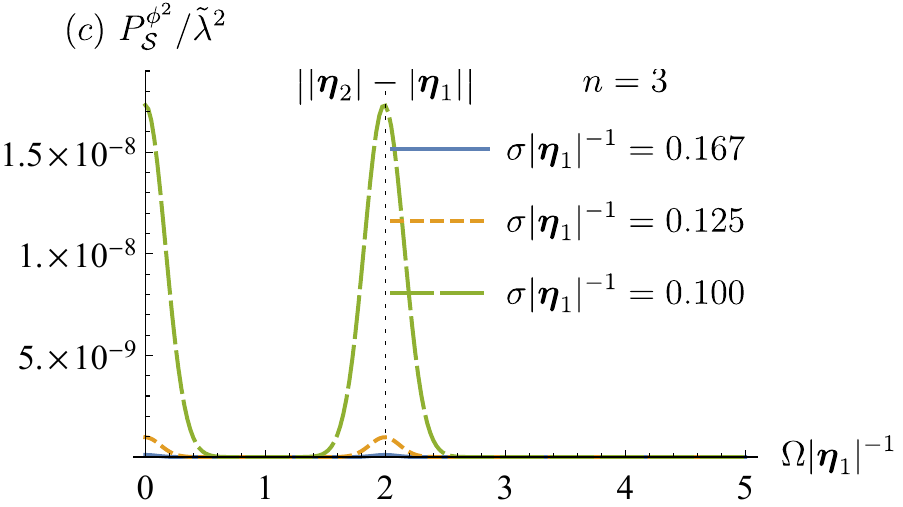}
        \caption{Various components of transition probability $P^{\phi^2}$ for detector interacting with two-particle Fock state in (3+1) dimensions. Here we fix $|\bb|=3|\ba|$. We vary $\Omega$ as we search for resonant peaks. (a) The dominant part exhibits no resonant peaks. (b) sum-frequency generation (SFG). (c) difference-frequency generation (DFG).}
        \label{fig: quad-resonance-QRS-3D}
    \end{figure*}    
 
  Finally, by considering the detector to be in its ground state ($\Omega>0$) and taking the long time and pointlike limit as per previous sections, we can simplify the transition probability \eqref{eq: probability-quadratic-two} to some extent. The vacuum contribution $P^{\phi^2}_{\vac}$ and the ``fully counter-rotating'' terms $\mathcal{Q}_+$ and $\mathcal{R}_+$ will then vanish. 
  Dropping the $\tilde F,\tilde \chi$ from the arguments,  the non-vanishing terms in Eqs~\eqref{eq: Qpm}-\eqref{eq: Spm} can then be written as
        \begin{align}
            &\mathcal{Q}_-(\bk;\sigma,\bc_j) =  \NN\frac{(|\bk|+\Omega)^{n-\frac{3}{2}}}{{2^{\frac{n-3}{2}}\pi^{\frac{n}{4}-1}\sigma^{\frac{n}{2}}}} e^{-\frac{|\bc_j|^2+(|\bk|+\Omega )^2}{2 \sigma ^2}} \notag\\
            &\hspace{2cm}\times {_0\tilde{F}_1}\left(\frac{n}{2};\frac{|\bc_j|^2 (|\bk|+\Omega )^2}{4 \sigma ^4}\right)\,,\\
            &\mathcal{R}_-(\sigma,\ba,\bb) \notag\\&= 
            \frac{ \NN}{{2^{n-2}\pi^{\frac{n}{2}-1}\sigma^{n}}}
            \int_0^\Omega \!\!\dd|\bk|\,\left(|\bk| (\Omega-|\bk| )\right)^{n-\frac{3}{2}}  e^{-\frac{ |\bk|^2+(\Omega-|\bk|)^2}{ 2\sigma ^2}} \notag\\
            &\times {_0\tilde{F}_1}\left(\frac{n}{2};\frac{|\ba|^2|\bk| ^2}{4 \sigma ^4}\right)\,
            {_0\tilde{F}_1}\left(\frac{n}{2};\frac{ |\bb|^2 (\Omega-|\bk|)^2}{4 \sigma ^4}\right)\,,\\
            &\mathcal{S}_-(\sigma,\ba,\bb)\notag\\
            &= \frac{ \NN}{{2^{n-2}\pi^{\frac{n}{2}-1}\sigma^{n}}}
            \int_0^\infty \!\!\!\dd|\bk|\,\left(|\bk| (\Omega+|\bk| )\right)^{n-\frac{3}{2}} 
            e^{-\frac{ |\bk|^2+(\Omega+|\bk|)^2}{ 2\sigma ^2}} \notag\\ 
            &\times{_0\tilde{F}_1}\left(\frac{n}{2};\frac{ |\ba|^2|\bk|^2}{4 \sigma ^4}\right)\,
            {_0\tilde{F}_1}\left(\frac{n}{2};\frac{|\bb|^2 (\Omega+|\bk|)^2}{4 \sigma ^4}\right)\,,
        \end{align}
    where ${_0\Tilde{F}_1}$ is the regularized hypergeometric function. These expressions are valid for all $n\geq 1$, noting that for $n=1$ all the energy scales have to be larger than the IR cutoff (analogous to the situation in Eqs.~\eqref{eq:linear-twoparticle}).
        
    Due to the simplifications above, we can write the full transition probability as $P^{\phi^2} = P^{\phi^2}_{\mathcal{Q}} +P^{\phi^2}_{\mathcal{R}} +P^{\phi^2}_{\mathcal{S}}$,
    where\footnote{Note that for $n=1$ we need to include the IR cutoff $\Lambda$ for the computation of $P^{\phi^2}_\mathcal{Q}$.}
    \begin{align}
        P^{\phi^2}_{\mathcal{Q}} &=                 4\lambda^2\int\frac{\dd^n\bk}{2(2\pi)^n|\bk|}
        \mathcal{Q}_-(\bk;\sigma,\ba)^2+\mathcal{Q}_-(\bk;\sigma,\bb)^2\notag\\ 
        & + {8}\lambda^2\int\frac{\dd^n\bk\,{\spec}}{2(2\pi)^n|\bk|}
        \mathcal{Q}_-(\bk;\sigma,\bc_1)
        \mathcal{Q}_-(\bk;\sigma,\bc_2)\,,\label{eq: ProbQ}
        \\
        P^{\phi^2}_{\mathcal{R}} &= 4\lambda^2
        \mathcal{R}_-^2(\sigma,\ba,\bb)\,,\label{eq: ProbR}\\
        P^{\phi^2}_{\mathcal{S}} &= 4\lambda^2\bigg[
        \mathcal{S}_-^2(\sigma,\ba,\bb)+
        \mathcal{S}_-^2(\sigma,\bb,\ba)\notag\\
        &\hspace{0.8cm}+
        2\mathcal{S}_-(\sigma,\bc_1,\bc_1)
        \mathcal{S}_-(\sigma,\bc_2,\bc_2)\bigg]\,.\label{eq: ProbS}
    \end{align}
    
    \begin{figure*}[tp]
        \centering
        \includegraphics[scale=0.64]{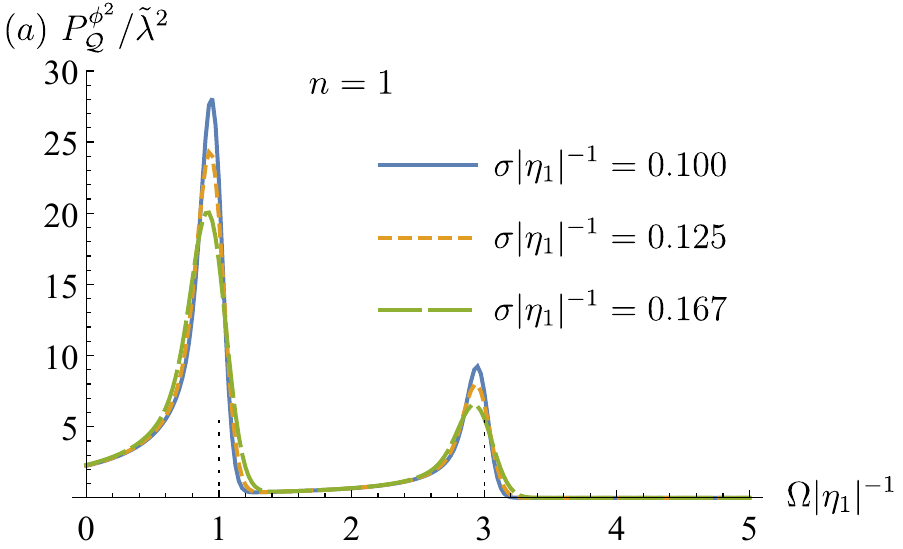}
        \includegraphics[scale=0.64]{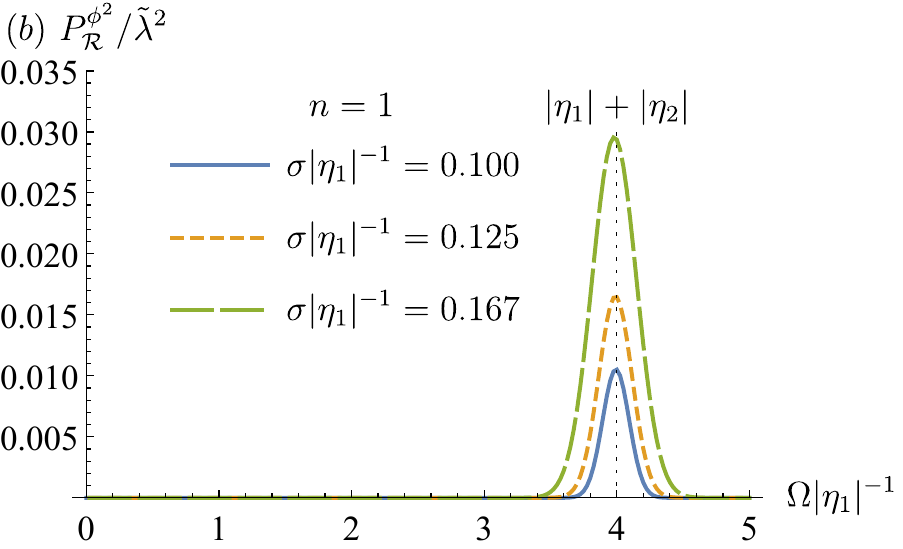}
        \includegraphics[scale=0.64]{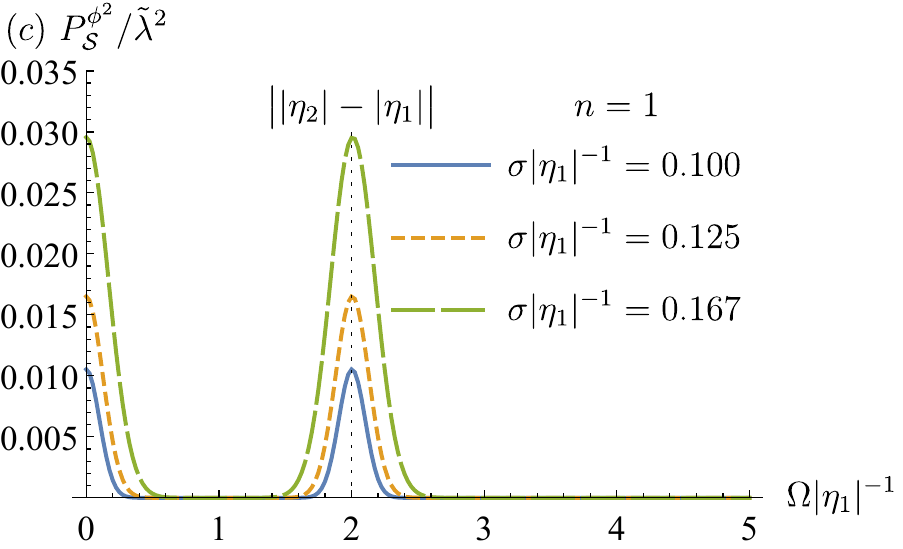}
        \includegraphics[scale=0.64]{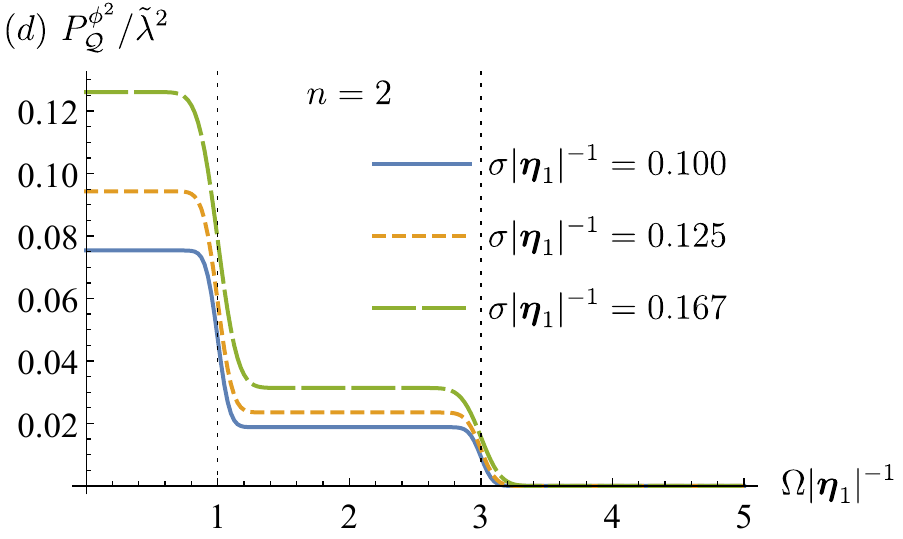}
        \includegraphics[scale=0.64]{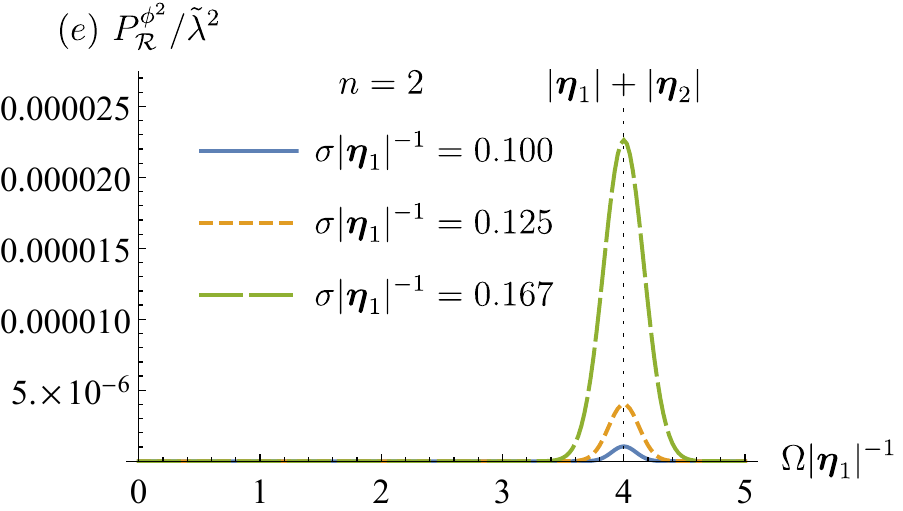}
        \includegraphics[scale=0.64]{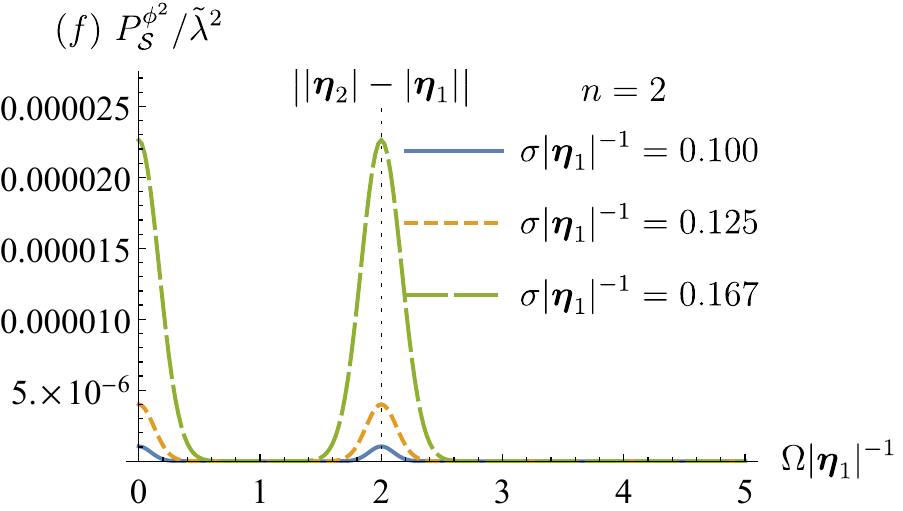}
        \includegraphics[scale=0.64]{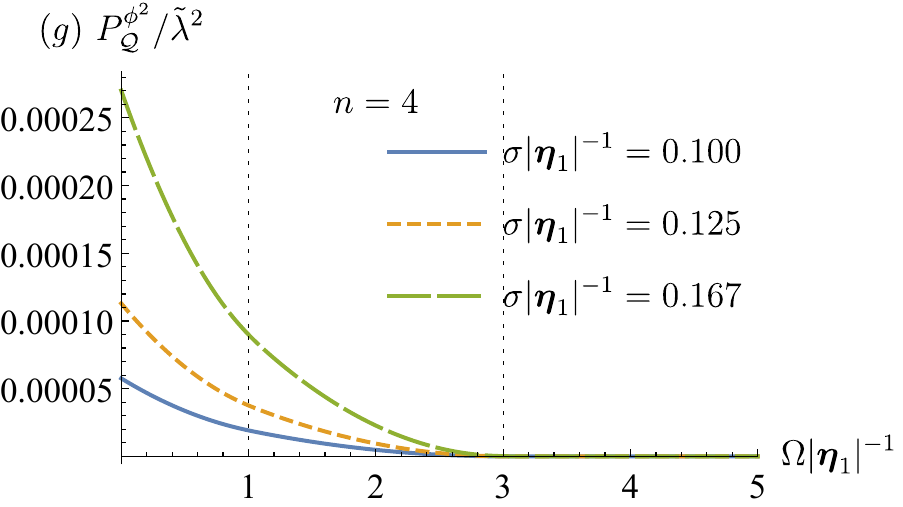}
        \includegraphics[scale=0.64]{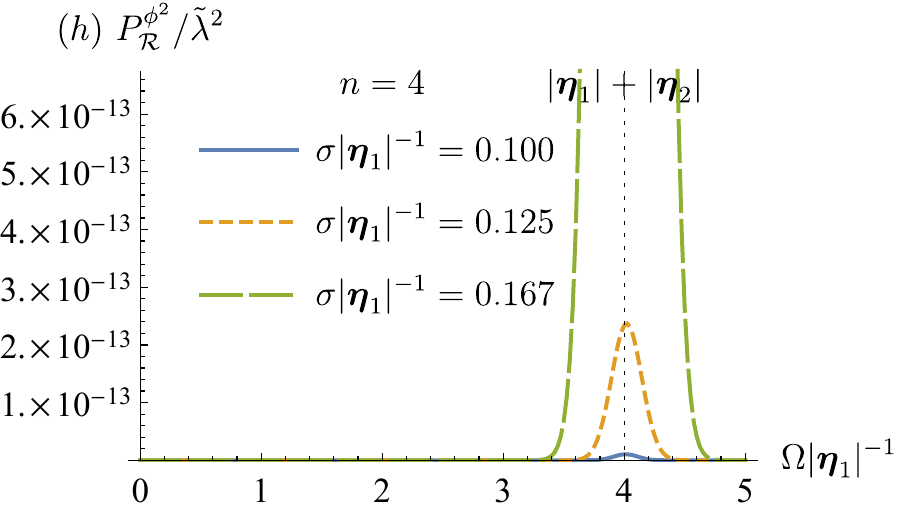}
        \includegraphics[scale=0.64]{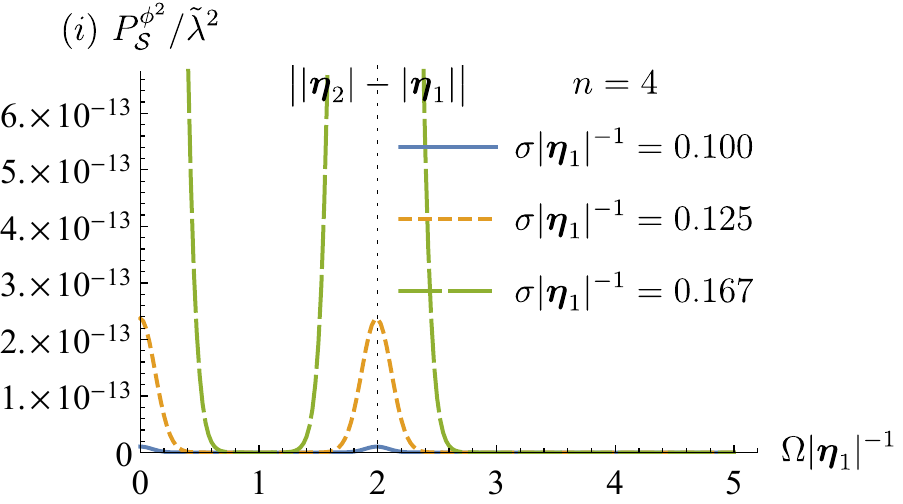}
        \caption{Various components of transition probability $P^{\phi^2}$ for detector interacting with two-particle Fock state in various dimensions. The case for $n=4$ is qualitatively representative of the higher-dimensional counterparts ($n\geq 5$). Here we set $|\bb|=3|\ba|$ for concreteness and we vary $\Omega$ in order to search for resonant-like phenomena. In all plots, we see that $P^{\phi^2}_\mathcal{Q}$ exhibits no resonant peaks for $n\geq 2$, $P^{\phi^2}_\mathcal{R}$ contains sum-frequency generation (SFG) and $P^{\phi^2}_\mathcal{S}$ contains difference-frequency generation (DFG).}
        \label{fig: quad-resonance-QRS-therest}
    \end{figure*}

    Before we study the dependence of the transition probability on the number of spacetime dimensions, we first plot in Figure~\ref{fig: quad-resonance-QRS-3D} the separate components of the transition probability in \eqref{eq: ProbQ}-\eqref{eq: ProbS} to better understand each of the terms that make up the total probability. Let us choose as a particular case study $n=3$. In this case, there are three interesting observations we can make from Figure~\ref{fig: quad-resonance-QRS-3D}. 
    
    First, the dominant contribution comes from the ``non-resonant'' piece (Eq.~\eqref{eq: ProbQ}) which  does not peak around resonance. Furthermore, from Figure~\ref{fig: quad-resonance-QRS-3D}(a) we see that the peak frequencies $\omega_1=|\ba| $ and $ \omega_2 = |\bb|$ of the two-particle Fock wavepacket delineate the different regimes where the slope of this dominant contribution changes. Second, from Figure~\ref{fig: quad-resonance-QRS-3D}(b) we observe that the term in Eq.~\eqref{eq: ProbR} contributes to a resonant peak at the \textit{sum} of the peak frequencies $\omega_1+\omega_2$ of the two-particle Fock wavepacket. This is a nonlinear optical effect which would correspond to \textit{sum-frequency generation} (SFG) in quantum optics literature \cite{boyd2008nonlinear}. Third, in Figure~\ref{fig: quad-resonance-QRS-3D}(c) we see that the contribution from the term in Eq.~\eqref{eq: ProbS} accounts for two maxima, one not associated to resonance (near zero gap), and another one corresponding to a resonant peak at the \textit{difference} of the two frequencies $\left|\omega_1-\omega_2\right|$. This is another nonlinear optical effect which would correspond to \textit{difference-frequency generation} (DFG) in quantum optics literature \cite{boyd2008nonlinear}. 
    The authors find it satisfying that a relativistic particle detector model is able reproduce two well-known nonlinear optical phenomena (SFG and DFG) in a unified manner.
    
    In Figure~\ref{fig: quad-resonance-QRS-therest} we consider the separate contributions from $P^{\phi^2}_{\mathcal{Q}},P^{\phi^2}_{\mathcal{R}},P^{\phi^2}_{\mathcal{S}}$ for different number of spatial dimensions. The results for $n=4$ are qualitatively similar to $n\geq 5$, so we take $n = 4$ to represent the higher-dimensional cases. We can make three important observations regarding the dimension dependence of the transition probability for the quadratic model interacting with the two-particle Fock state.
    
    First, note that in all dimensions, the $\mathcal{Q}$-dependent contribution (Eq.~\eqref{eq: ProbQ}) dominates compared to the $\mathcal{R}$-dependent contribution from Eq.~\eqref{eq: ProbR} associated to SFG and the $\mathcal{S}$-dependent contribution from Eq.~\eqref{eq: ProbS} associated to DFG. However, when $n\geq 2$, there is no resonant peak at the Fock wavepacket peak frequencies $\omega_1=|\ba|$ and $\omega_2=|\bb|$. Only in $n=1$ do the detectors have significant resonance aligned with the peak frequencies of the wavepacket. Similar to the results in the previous subsections, only for $n=1$ do we see that \eqref{eq: ProbQ} increases in the ``dichromatic'' limit (decreasing bandwidth $\sigma$), while for $n\geq 2$ we see that \eqref{eq: ProbQ} decreases as $\sigma$ decreases. We also see that the only contribution which is qualitatively different in different dimensions is the $\mathcal{Q}$-dependent one.
    
    
    The second observation is that in all dimensions, the nonlinear optical phenomena (SFG and DFG) persist, but the rate at which the magnitude of the peaks diminishes in the dichromatic limit $\sigma\to 0$ differs for different $n$. In higher dimensions the SFG and DFG peaks decrease as $\sigma\to 0$, and appears to decrease faster the larger the spatial dimensions. The third observation is that---similar to the one-particle scenario---the transition probability for quadratic coupling decreases for all $n\geq 2$ when $\sigma\to 0$: this is in contrast to the linear coupling model, where the transition probability only decreases when $n\geq 3$, approaches a constant value as $n=2$ and increases when $n=1$.



    \section{Energy deposited in the field}
    \label{sec:energydeposit}

    In this section we compare the converse scenario, where an excited detector interacts with the vacuum state of the field. This will provide a complementary picture on the light-matter interaction by studying how energy is transferred from an excited detector to the field's vacuum depending on how the detector is coupled to the field. 
    
    We are interested in the expectation values of the number of excitations in each frequency mode of the cavity after the interaction with an excited detector and how it varies with duration of the interaction. More specifically, we consider the global initial state
    \begin{align}
        \hat\rho_0 = \ket{0}\!\bra{0}\otimes\ket{e}\!\bra{e}\,,
    \end{align}
    where $\ket{0}$ is the field's vacuum state and $\ket{e}$ is the detector's excited state, which after interaction yields the final global state $\hat\rho = \hat U\hat\rho_0\hat U^\dagger$. In order to know the energy distribution of the field on each of the field modes, we can compute the number expectation value $N_j$ on each mode labelled by the positive integer $j$, defined by
    \begin{align}
        N_{j} \coloneqq \tr\left[\hat\rho \hat N_j\right]\,,
    \end{align}
    where $\hat N_j=\hat a_j^\dagger \hat a_j^{\pdag}$ is the number operator associated to mode $j$. Since the field is in initially in its vacuum state and we are interested in how a detector deposits its energy in it, we only need to consider the contribution coming from $\rho^{(1,1)}$, i.e.
    \begin{align}
        N_j &=  \tr\left[\hat\rho_{\phi}^{(1,1)}\hat N_j\right] + O(\lambda^3)\,,
        \label{eq: number-expectation-cavity}\\
        \hat\rho_{\phi}^{(1,1)} &\coloneqq \tr_{\text{d}}\left(\hat U^{(1)}\hat\rho_0 \hat U^{(1)\dagger}\right)\,,
    \end{align}
    where $\hat\rho_\phi^{(1,1)}$ denotes the leading order reduced density matrix of the field after interaction that account for detector's de-excitation to its ground state. Therefore, it suffices to find the expression for the first-order term $\hat U^{(1)}$ in the Dyson expansion of the full time evolution operator $\hat U$. The form of $\hat U^{(1)}$ depends on the choice of detector-field coupling (linear vs quadratic), and is given as an integral over the interaction Hamiltonian:
    \begin{align}
        \hat U^{(1)} = -\ii\int\dd t\,\hat H_I(t)\,,
    \end{align}
    where $\hat H_I(t)$ is given by either \eqref{eq: linear-hamiltonian} or \eqref{eq: quadratic-hamiltonian}. 
    
    In this section, we will also focus on the scenario where we have a massless scalar field confined to a (1+1)-dimensional Dirichlet cavity. While a (1+1) dimensional cavity is very different from (and certainly not a good model for) a `thin' (3+1) dimensional cavity (see e.g. \cite{Lopp:2018cavity}), it is a good enough testbed to understand the differences in resonant behaviour between linear and quadratic couplings. Indeed, the resonant behaviour for a field in an $(n+1)$-dimensional cavity is qualitatively similar to the $(1+1)$-dimensional case, as we discussed in Section~\ref{subsec: cavity-calculation}.
    
    We consider a Dirichlet cavity of length $L$ in the field's quantization frame with coordinates $(t,x)$. We impose the Dirichlet boundary condition 
    \begin{align}
        \hat\phi(t,0) = \hat\phi(t,L) = 0\,.    
    \end{align}
    It follows that the mode decomposition of the massless scalar field in  the $(1+1)$-dimensional cavity takes the form
    \begin{align}
        \hat\phi(t,x) = \sum_{n=1}^\infty \frac{1}{\sqrt{n\pi}}\sin\omega_n x\left[\hat a_{n}e^{\ii\omega_n t} + \hat a^\dagger_{n}e^{-\ii\omega_n t}\right]
        \label{eq: mode-decomposition-cavity-1D}
    \end{align}
    where $\omega_n = n\pi /L$ and $n\in \mathbb{N}$. 
    
    In what follows, we will restrict our attention to the special case where the detector is pointlike and comoving in the quantization frame, i.e. $F(\bx)=\delta(x-x_\text{d})$, where $x_\text{d}\in (0,L)$. This will simplify the calculations considerably, especially for the quadratic coupling, and it corresponds to the regime where the cavity is much larger that the size of the detector. We set the switching function to be a Gaussian \begin{equation}
    \chi(t)=e^{-\frac{t^2}{T^2}}\notag
    \end{equation} where $T$ prescribes the effective duration of the interaction. This allows us to study how the energy distribution changes with the duration of interaction between the short-time and long-time regimes.

    \subsection{Linear coupling}

    In this subsection we consider the energy distribution left by an excited detector in Dirichlet cavity when the detector is linearly coupled to the field. The relevant $(1,1)$ component of the leading order reduced density matrix of the field reads
    \begin{align}
        \hat\rho_\phi^{(1,1)}
        &= \lambda^2\int\dd t\,\dd t'\, e^{\ii\Omega(t-t')}\chi(t)\chi(t')\!\!\sum_{i,j=1}^\infty\!\! u_i^{*'}u_j^{\pdag}\ket{1_i}\!\bra{1_j}\,,
        \label{eq: field-state-cavity-linear}
    \end{align}
    where $u_j\equiv u_j(t,x_\text{d})$ is the eigenmode of the scalar field evaluated along the detector's trajectory,
    \begin{equation}
        u_j\equiv u_j(t,x_{\text{d}}) = \frac{1}{\sqrt{j\pi}}\sin(\omega_jx_\text{d})e^{-\ii\omega_j t}\,,
        \label{eq: eigenmodes-1D-cavity}
    \end{equation}
    and $u_j'\equiv u_j(t',x_{\text{d}})$. 
    
    \begin{figure*}[tp]
        \includegraphics[scale=0.5]{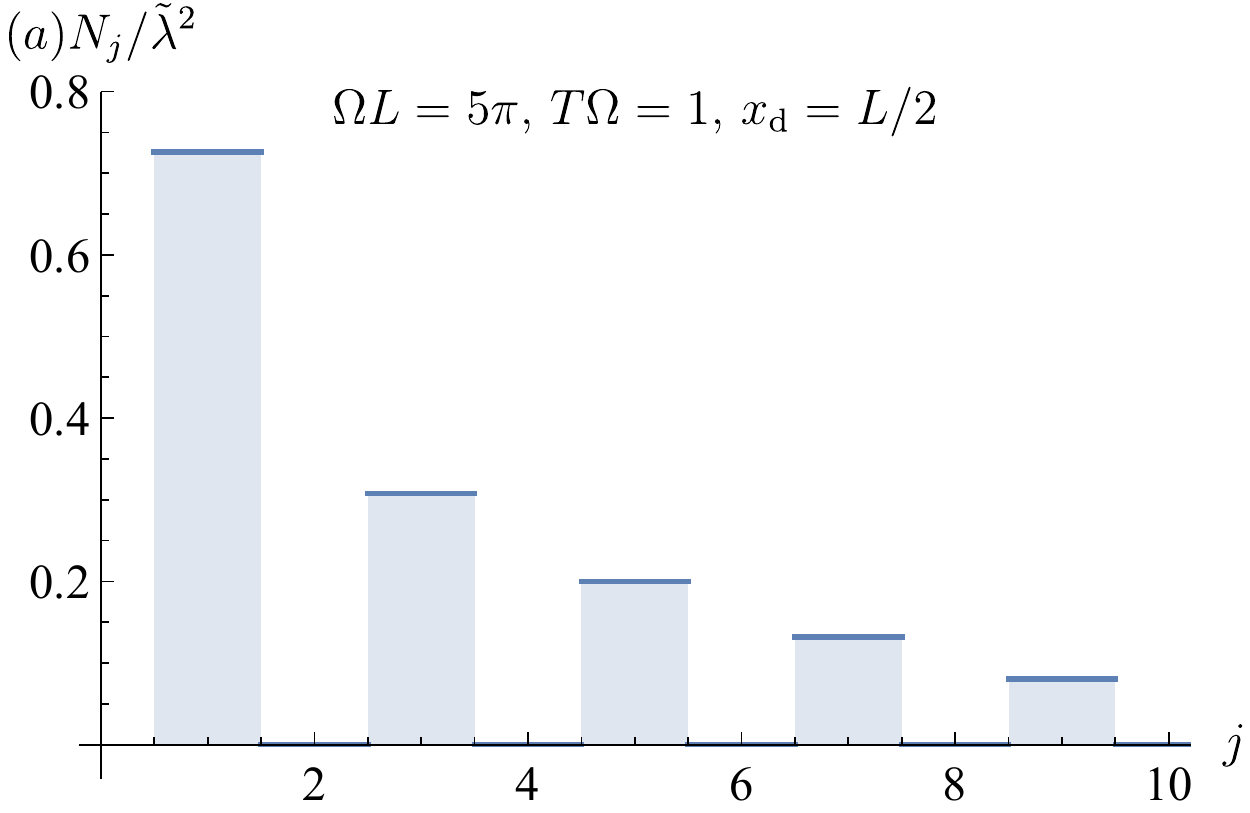}
        \includegraphics[scale=0.5]{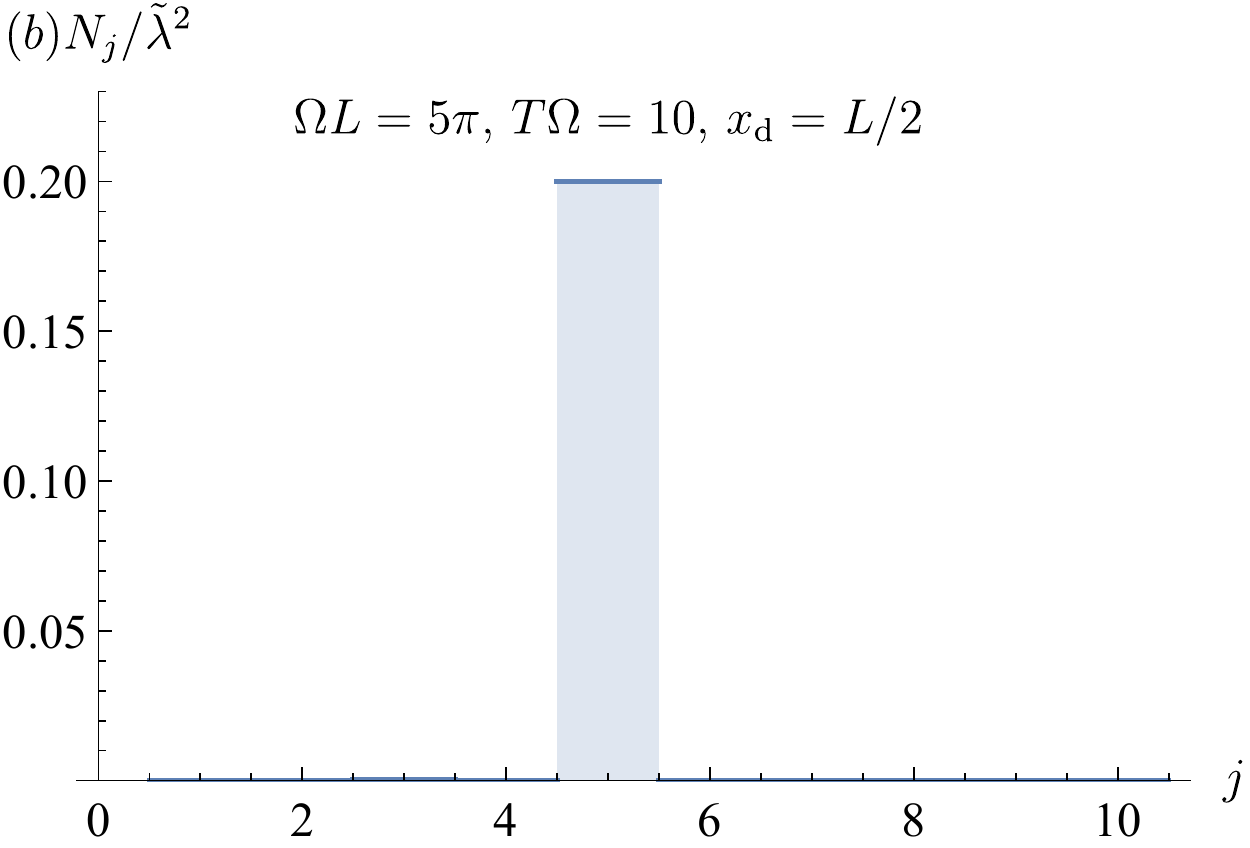}
        \includegraphics[scale=0.5]{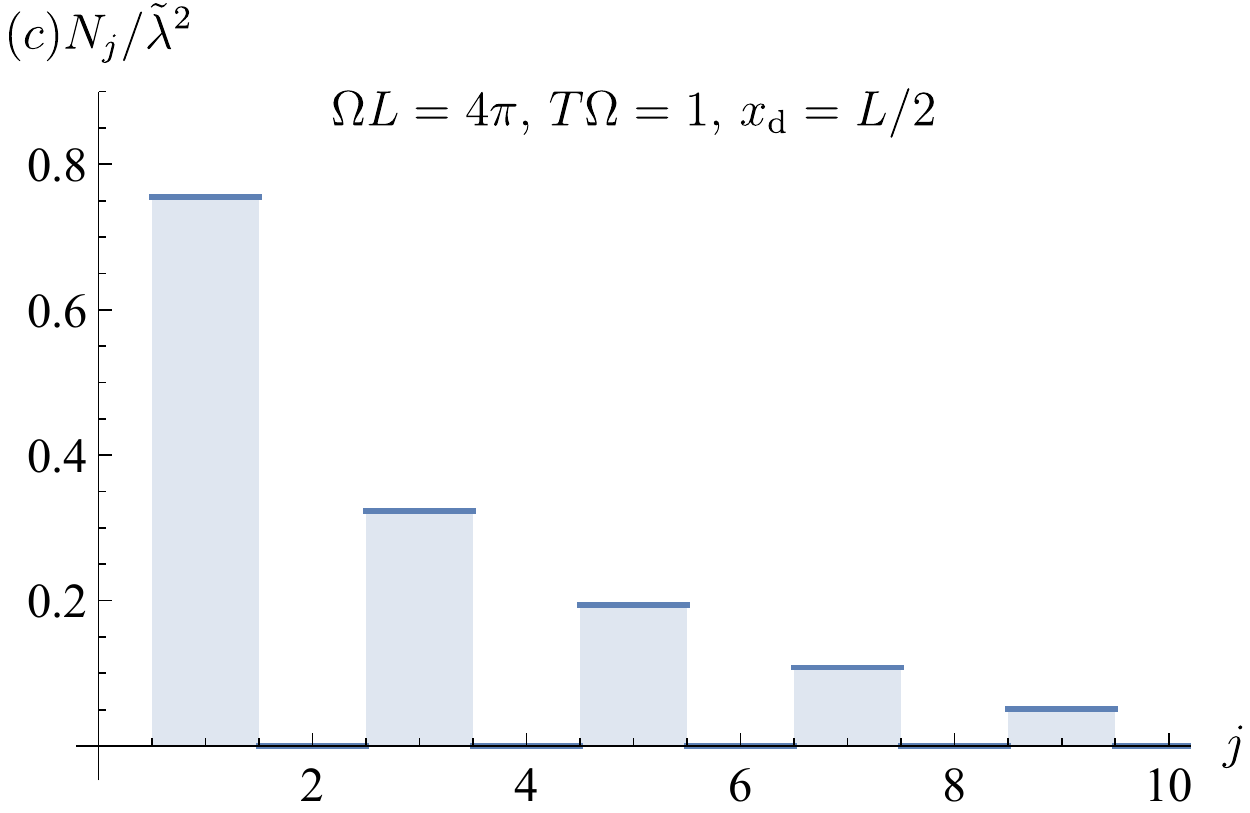}
        \includegraphics[scale=0.5]{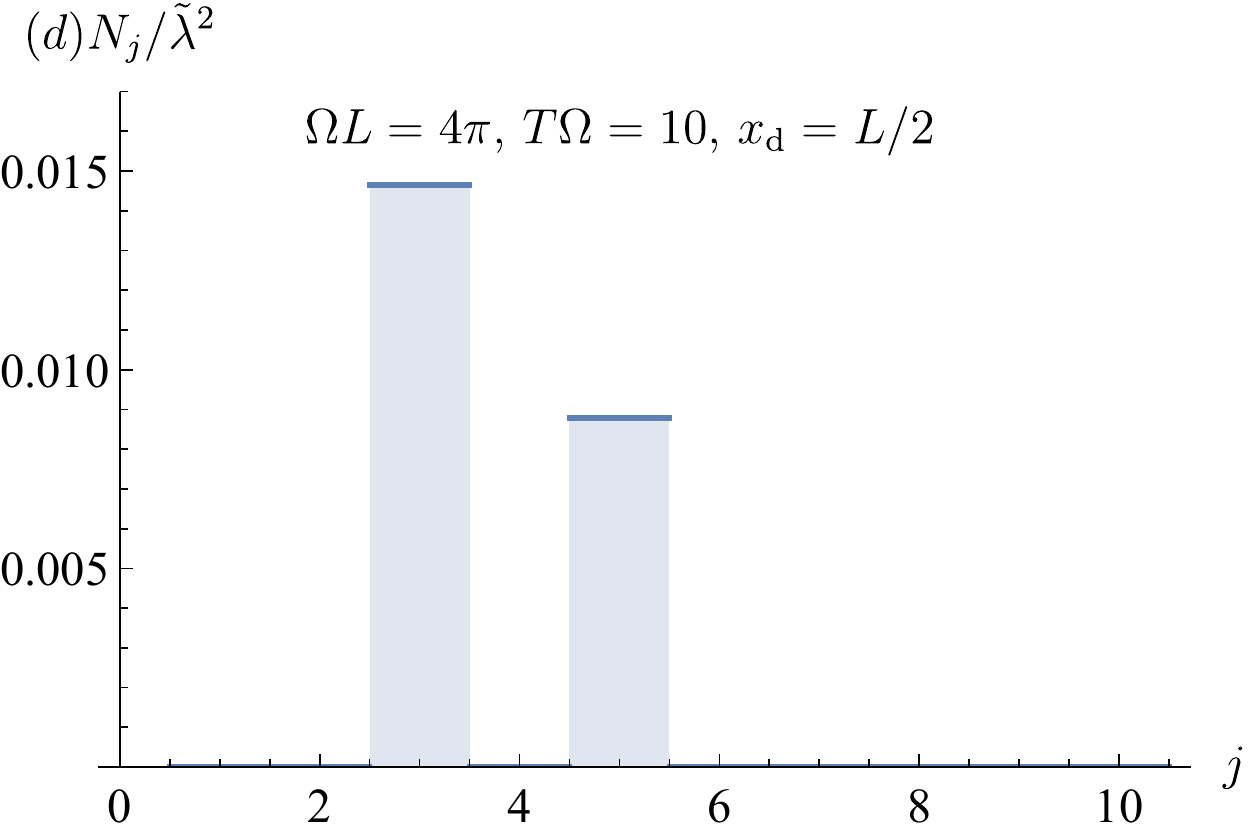}
        \includegraphics[scale=0.5]{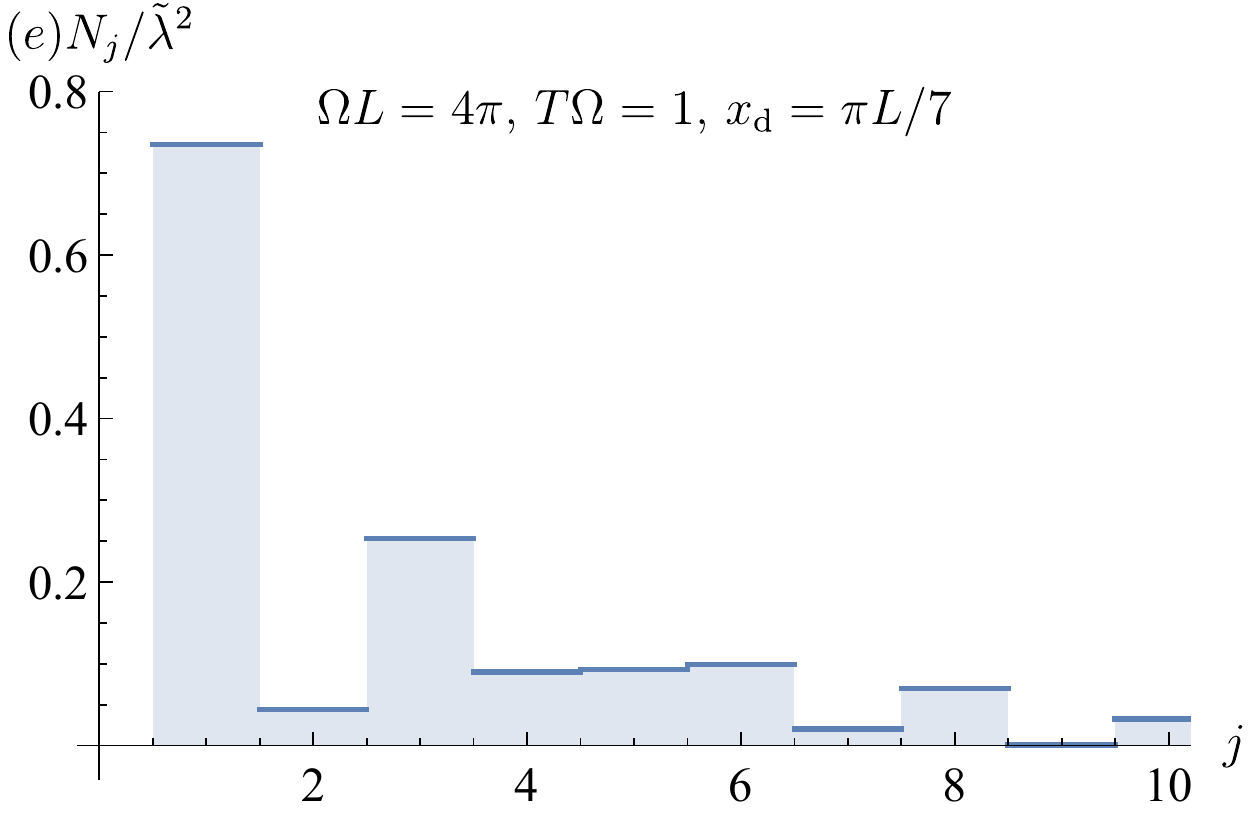}
        \includegraphics[scale=0.5]{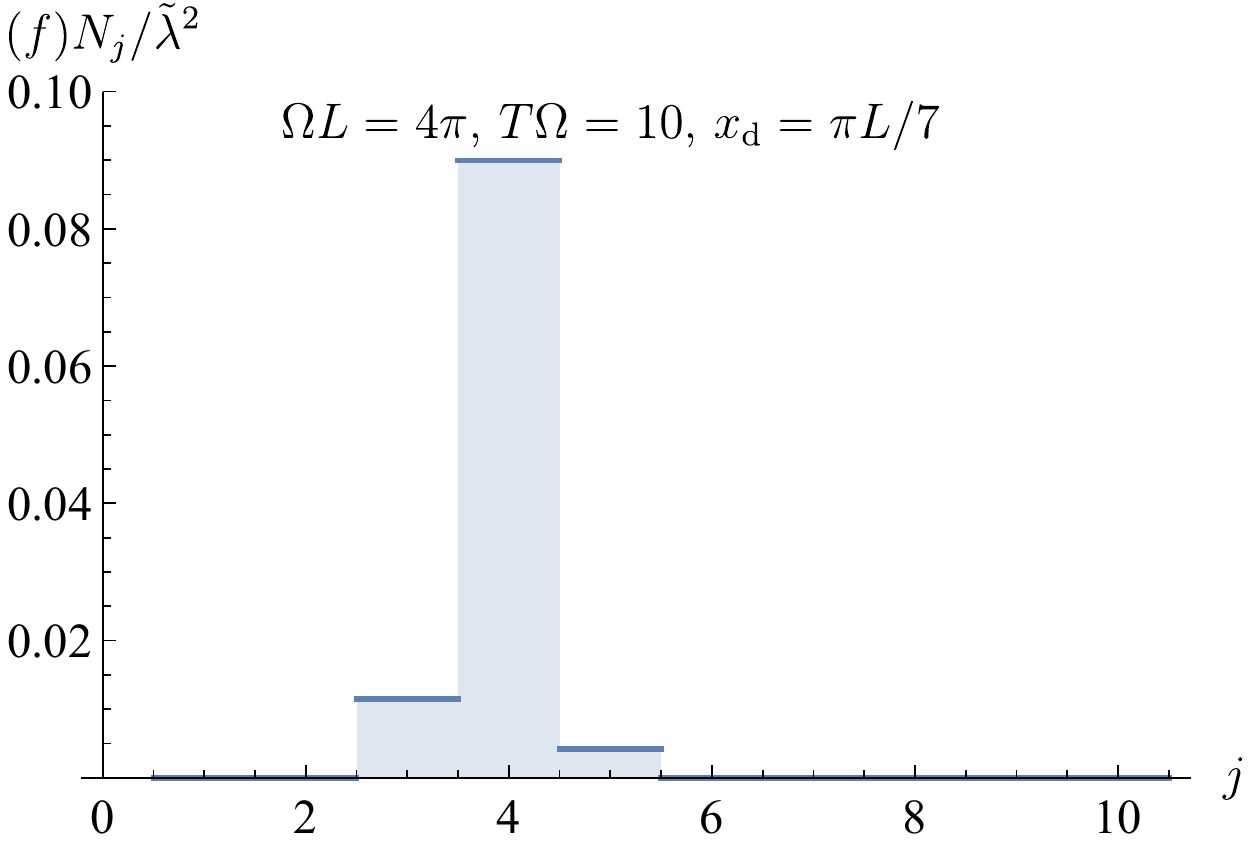}
        \caption{Number expectation value for the $n$-th harmonics as a function of $n$ for linearly coupled detector-field system.} 
        \label{fig: lineardeposit1}
    \end{figure*}
    
    We are now ready to compute the number expectation $N_j$. Substituting $\rho^{(1,1)}_\phi$ into the expression for the number expectation \eqref{eq: number-expectation-cavity} it follows that 
    \begin{equation}
        N_j = \langle \hat a_j^\dagger \hat a_j^{\vphantom{\dagger}} \rangle 
        = \lambda^2\left|\tilde\chi(\Omega-\omega_j)\frac{1}{\sqrt{j\pi}}\sin (\omega_jx_\text{d}) \right|^2.
        \label{eq: Nj}
    \end{equation}
    For a Gaussian switching function, this expression reads
    \begin{align}
         N_j = {\frac{\tilde\lambda^2}{j}}e^{-\frac{1}{2} T^2 (\Omega-\omega_j)^2}\sin^2(\omega_jx_\text{d})\,,
         \label{eq: Nj-Gaussian}
    \end{align}
    where we pull out the factor $T^2$ from the Fourier transform of the switching function to make the dimensionless coupling constant\footnote{Recall that in natural units $[T]=[L]$ and  $[\lambda]=[L]^{\frac{n-3}{2}}$ for the linear coupling where $n$ is the number of spatial dimensions.} $\widetilde\lambda=\lambda T$.
    
    We plot $N_j$ as a function of $j$ to aid visualization in Figure~\ref{fig: lineardeposit1}. We can make several observations on the behaviour of the number expection $N_j$ based on the expression in Eq.~\eqref{eq: Nj-Gaussian}.  We choose $\Omega$ to be an integer multiple of $\pi/L$ in order to make the resonance with field modes exact so that $\Omega$ is equal to $\omega_j$. We consider how $N_j$ varies as a function of the duration of the interaction $T$ and the detector's position $x_d$, keeping the cavity size fixed.
    
    First, in the short-time regime (say, $T\Omega \lesssim 1$) $N_j$ is large for small $j$ and decreases with increasing $j$. In this regime the Gaussian does not impose any effective frequency cutoff on the interaction  since $e^{-x^2}\approx 1+O(x^2$). From Figure~\ref{fig: lineardeposit1}(a),(c),(e), we see that most of the energy is deposited into the field mode with largest wavelength $j=1$, and decreases as $1/j$. Note that because $x_d$ appears as an argument of $\sin(\omega_jx_d)$, when $x_d = L/2$, there is no energy deposited when $j$ is an even number, as we show in Figure~\ref{fig: lineardeposit1}(a) and (c). In particular, which modes are accessible for the detector to dump its energy depends on the zeros of $\sin(\omega_j x_d)$. For generic $x_d\in (0,L)$ the behaviour is closer to that in Figure~\ref{fig: lineardeposit1} where all the modes are accessible (because $\omega_jx_d\neq 0$ for all $j$ when $x_d = \pi/7$). 
    
    Second, in the long-time regime ($T\Omega\gg 1$), most of the energy is dumped in a single resonant mode $\omega_j = \Omega$, as shown clearly in Figure~\ref{fig: lineardeposit1}(b) and (f). If the resonant mode cannot be exactly obtained, the energy will be dumped mainly on the nearest-neighbouring modes. An example of this is shown in Figure~\ref{fig: lineardeposit1}(d): when $x_d=L/2$, the detector is in a node of the the resonant mode $\omega_4$. When this happens, the energy is deposited in the nearest neighbour modes $\omega_3$ and $\omega_5$. It also follows that in this long time regime, $N_5<N_3$ because $N_j$ scales with $1/j$. In the case when every mode is accessible, such as when $x_d = \pi L/7$ (Figure~\ref{fig: lineardeposit1}(f)), the dominant mode where most of the energy is deposited will be the resonant mode $\omega_j=\Omega$.

    We summarize our results as follows: we see that for linearly coupled detector, the detector preferentially deposits energy to the mode(s) closest to the energy gap $\Omega$ when the interaction time is large due to resonant effect. Conversely, in the short interaction regime the detector preferentially deposits its energy to the lowest cavity mode due to the $1/j$ modulation in $N_j$. The results of this subsection are indeed not very surprising in the context of quantum optical intuition, but it is nice to have as a consistency check for the model as well as for completeness.

    \subsection{Quadratic interaction}
    
    \begin{figure*}[tp]
        \includegraphics[scale=0.5]{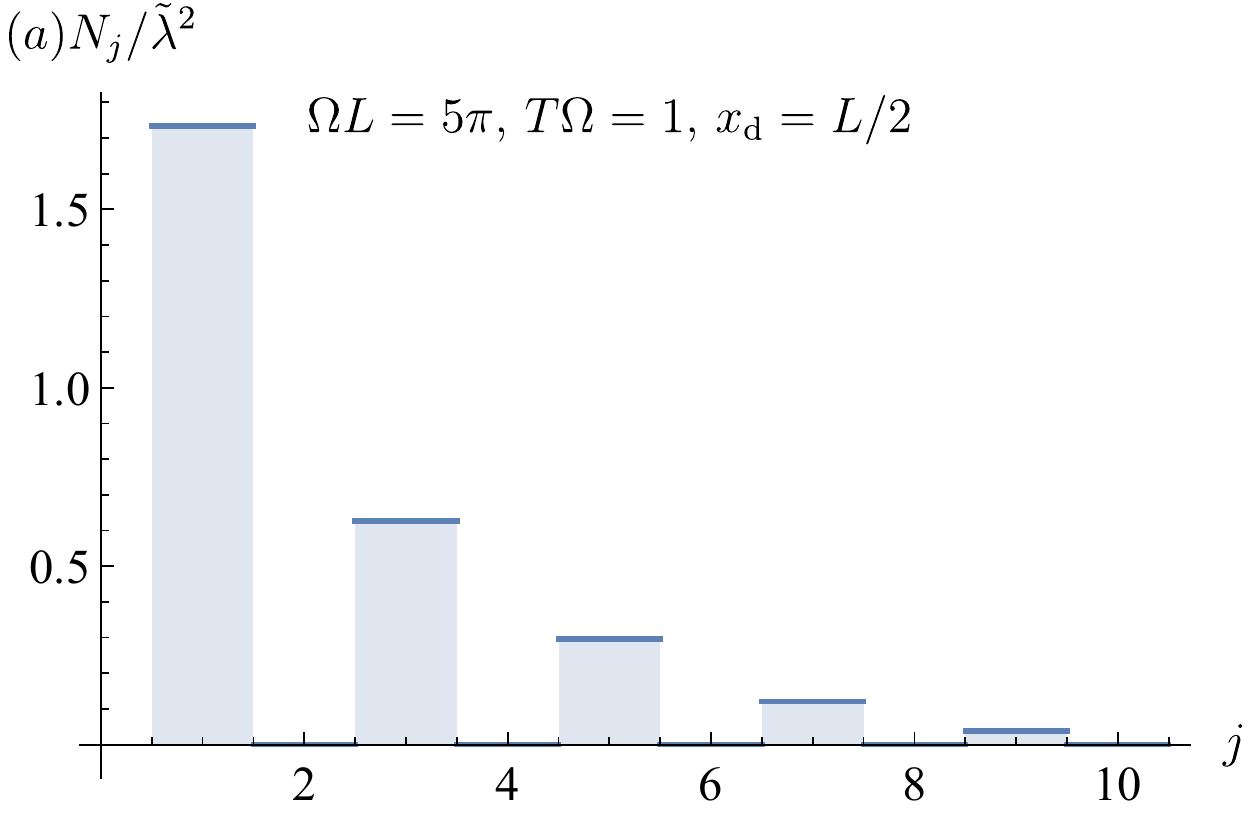}
        \includegraphics[scale=0.5]{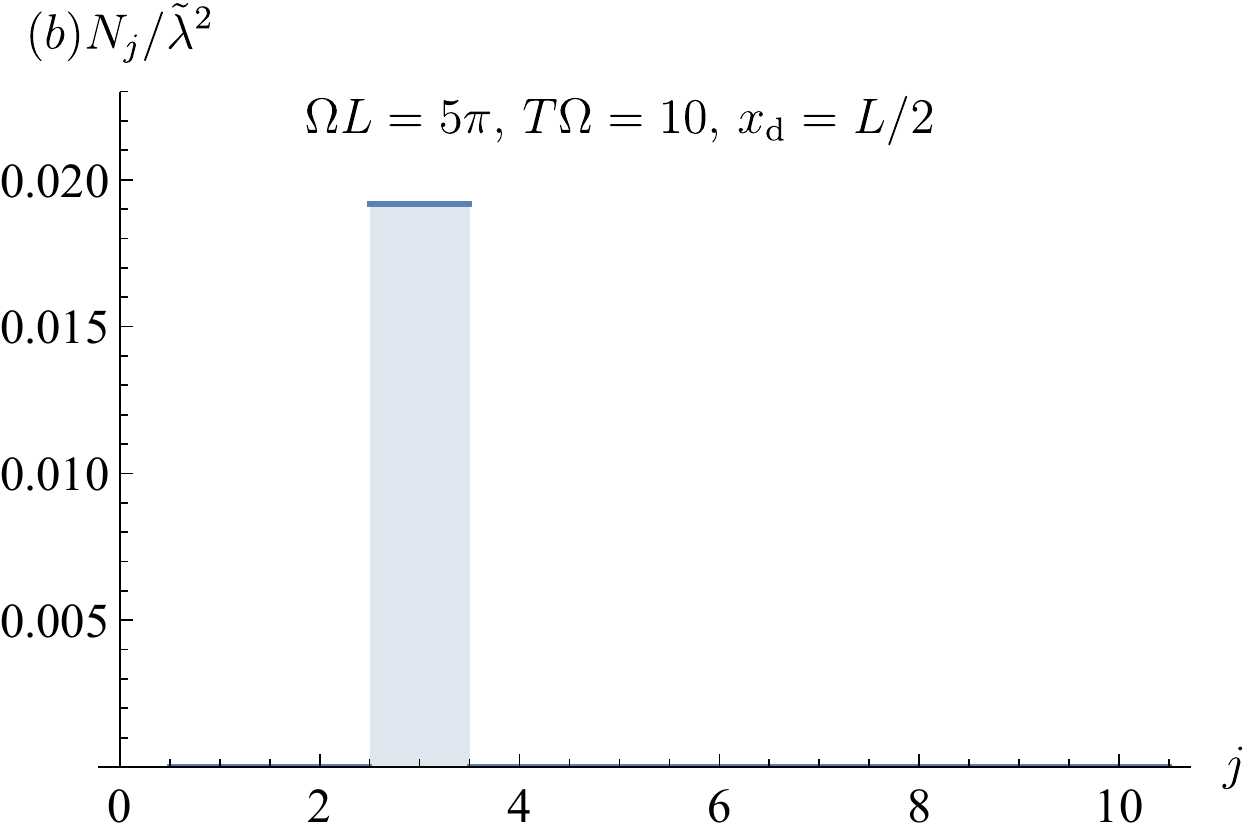}
        \includegraphics[scale=0.5]{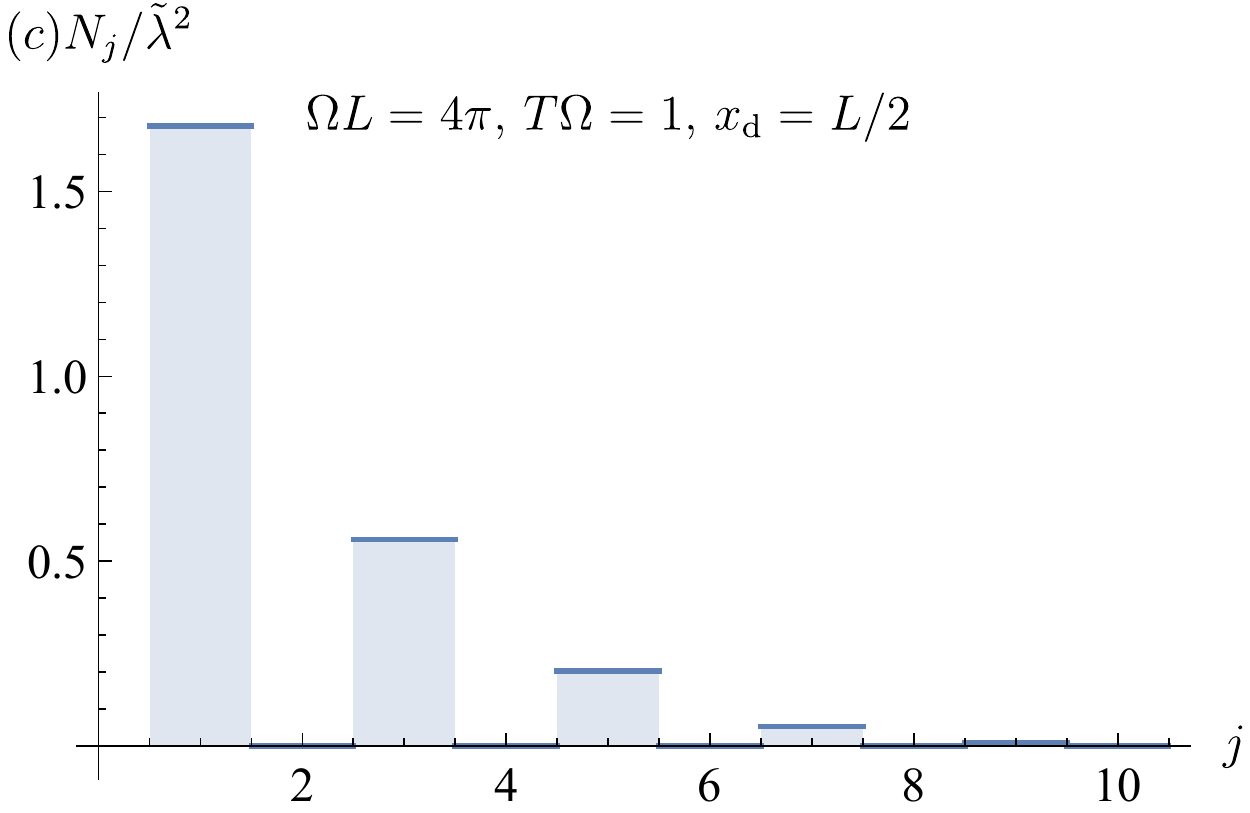}
        \includegraphics[scale=0.5]{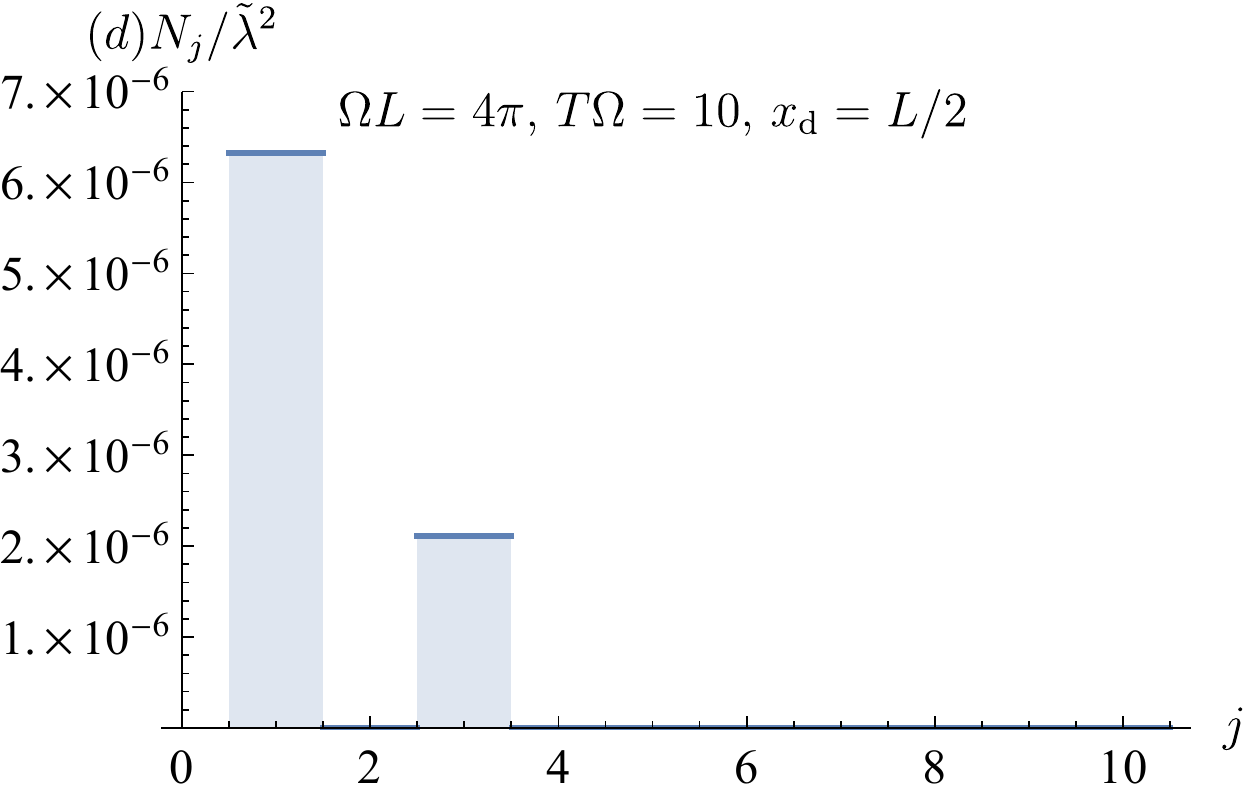}
        \includegraphics[scale=0.5]{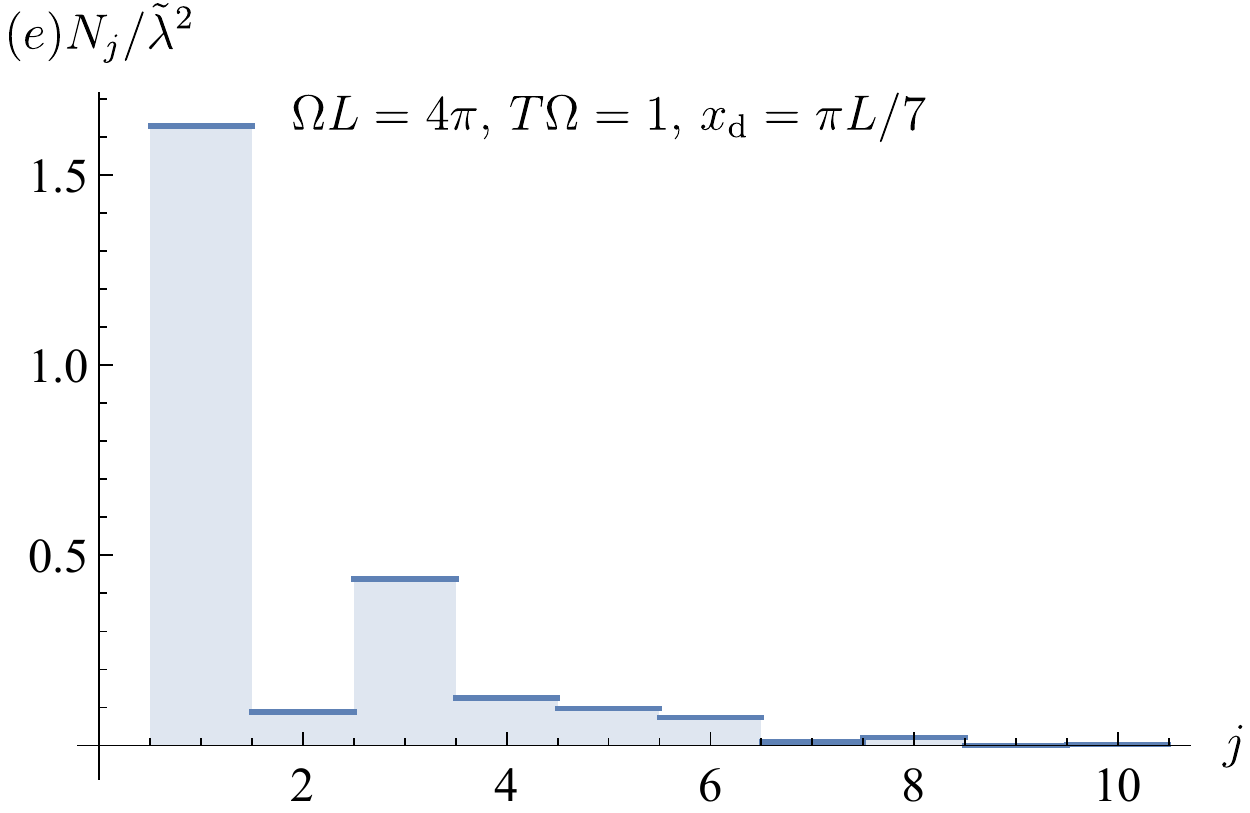}
        \includegraphics[scale=0.5]{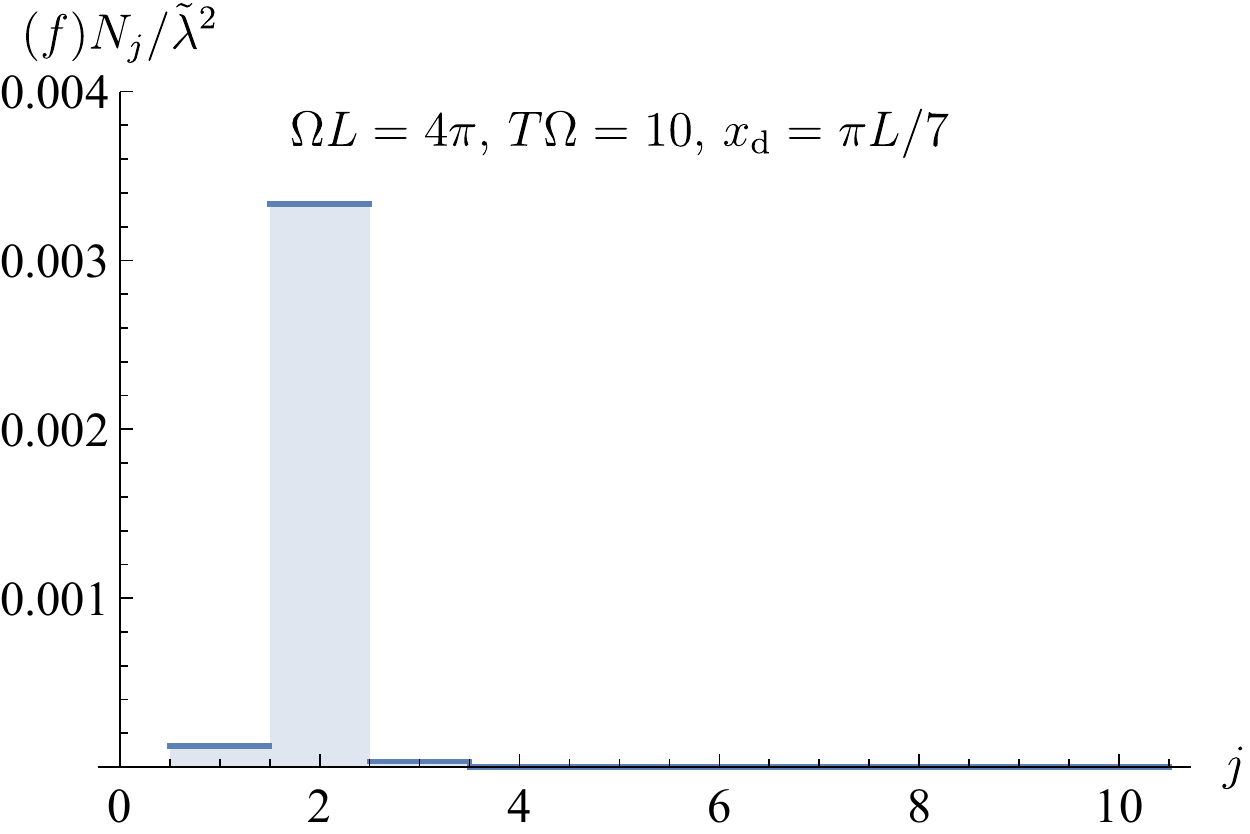} \caption{Number expectation value for the $n$-th harmonics as a function of $n$ for quadratically coupled detector-field system. }
        \label{fig: quadraticdeposit1}
    \end{figure*}

    In this subsection we consider the energy distribution when the detector is quadratically coupled to the field. The relevant $(1,1)$ contribution to the leading order reduced density matrix of the field after interaction reads
    \begin{align}
        \rho_\phi^{(1,1)}
        &=\lambda^2\sum_{j,k,l,m}\int\dd t\,\dd t'\,\chi(t)\chi(t') e^{i\Omega(t-t')} u_j^{*'}u_{k}^{*'}u_lu_m\notag\\
        &\hspace{0.5cm}\times \left(1+\delta_{jk}(\sqrt{2}-1)\right)\left(1+\delta_{lm}(\sqrt{2}-1)\right)\notag\\
        &\hspace{0.5cm}\times\ket {1_j1_k}\!\bra {1_l1_m}\,,
        \label{eq: field-state-cavity-quadratic}
    \end{align}
    where $u_j\equiv u_j(t,x_\text{d})$ is the eigenmode of the scalar field evaluated along the detector's trajectory given in Eq.~\eqref{eq: eigenmodes-1D-cavity}. In obtaining \eqref{eq: field-state-cavity-quadratic} we have used the fact that
    \begin{align}
        \hat a^\dagger_j \hat a^\dagger_k \ket{0} =  \left(1+\delta_{jk}(\sqrt{2}-1)\right)\ket{1_j1_k}\,.
    \end{align}
        
    We are now ready to compute the number expectation $N_j$. Substituting the reduced density matrix $\rho^{(1,1)}_\phi$ into the definition of $N_j$ in \eqref{eq: number-expectation-cavity}, we get
    \begin{align}
        N_{j} 
        &= \lambda^2\sum_{k=1}^\infty\left| \frac{4}{jk\pi} \tilde\chi(\Omega-2\omega_j)\tilde\chi(\Omega-2\omega_k)\right|\notag\\
        &\hspace{0.5cm}\times \sin^2(\omega_jx_\text{d})\sin^2(\omega_kx_\text{d})\,.
        \label{eq: Njk}
    \end{align}
    For Gaussian switching, this reads
    \begin{align}
        N_{j} 
        &=  \tilde\lambda^2\sum_{k=1}^\infty\frac{4}{jk\pi} e^{-\frac{T^2}{4}(\Omega-2\omega_j)^2}e^{-\frac{T^2}{4}(\Omega-2\omega_k)^2}\notag\\
        &\hspace{0.5cm}\times \sin^2(\omega_jx_\text{d})\sin^2(\omega_kx_\text{d})\,,
        \label{eq: Njk-Gaussian}
    \end{align}
    where $\tilde\lambda=\lambda T$ is dimensionless\footnote{For quadratic coupling, in natural units $\lambda$ has dimension \mbox{$[\lambda] = [L]^{n-2}$} in $(n+1)$ spacetime dimensions.}. Note that we can write $2\omega_j = \omega_{2j}$.
        
    Let us analyse the results for $N_j$ with the aid of Figure~\ref{fig: quadraticdeposit1}. First, analogous to the linear coupling case,  in the short time regime $T\Omega\lesssim 1$ the lower cavity modes are preferred due to the factor $j^{-1}$ in the expression of $N_j$. Second,  Eq.~\eqref{eq: Njk-Gaussian} shows that for quadratic coupling, the detector preferentially deposits its energy to modes whose frequency is \textit{half} the frequency of the energy gap $\Omega=2\omega_j \equiv \omega_{2j}$ in the long time regime $T\Omega\gg 1$. Since $N_j$ is modulated by $\sin(\omega_jx_\text{d})\sin(\omega_k x_\text{d})$, the zeros of the sine functions may render certain modes to be inaccessible: for example, by choosing $x_\text{d}=L/2$ $N_j$ is only nonzero for odd $j$. When this occurs, the energy will be deposited in the nearest neighbouring mode. We show this in Figure~\ref{fig: quadraticdeposit1}(b), where in this case given $\Omega = 5\pi/L$ the detector will deposit most of its energy to $\omega_3$ because it is closest to $\Omega/2 = 2.5\pi/L$ (note that $\omega_2$ is inaccessible). Similar to the linearly coupled case in Figure~\ref{fig: lineardeposit1}, if we choose $x_\text{d}$ such that every mode is accessible ($\omega_jx_{\text{d}}$ is not a zero of the the sine function for all $j$), then the detector will always dump its energy on the mode $\omega_j = \Omega/2$ in the long time regime.

    We summarize our results as follows. We see that a quadratically coupled detector preferentially deposits energy to the mode(s) closest to \textit{half} the energy gap $\Omega/2$ when the interaction time is large due to resonance effects. In the short interaction regime the detector preferentially deposits its energy to the lower cavity mode due to the $1/j$ modulation in $N_j$. Finally, we note that this splitting of energy into two parts in the long time regime has a correspondence in standard quantum optics, an effect known as half-harmonic generation.

    \section{Conclusion}
    \label{sec:conclusions}

    In this paper we focused on understanding the differences between linear and quadratic couplings between light (modelled by a scalar field) and matter (modelled by a particle detector) and how to interpret the phenomenology of detector excitations in both scenarios.
    
    More specifically, we study how a linearly coupled Unruh-DeWitt detector resonates with one-particle and two-particle Fock states of the field and how they differ from the quadratically coupled variant of the detector model. We explore the effects of spacetime dimension and the width of the Fock wavepacket (bandwidth) on the detector's responses to the field's excitations. We also complement our study with the converse scenario where an excited detector deposits its energy to the field in its vacuum state through their interaction and explore how linearly and quadratically coupled detectors differ in this regard.
    
    We present three main results. First, we show that generically in free space (in absence of boundary conditions) where the field has a continuous spectrum, the detector becomes more transparent to a Fock wavepacket as it becomes more monochromatic, even if it is in resonance with the detector. This happens despite the fact that the energy of the wavepacket in the monochromatic resonant limit is the expected $\hbar \Omega$. In other words, shining more monochromatic light on a detector in free space will make the detector click less and not more, contradicting intuition from results in optical cavities. Indeed, in the cavity  scenario the excitation probability near resonance is always amplified when the field's state has tighter frequency range around the energy gap of the detector. More specifically, for a linearly coupled detector, this \textit{transparency at resonance} for detectors in free-space happens for $(n+1)$-dimensional spacetimes with $n\geq 3$. For a quadratically coupled detector this happens for $n\geq 2$. Only in the $(1+1)$-dimensional setting do we have larger transition probability as the wavepacket bandwidth more closely matches the resonant frequency in free-space.
    
    Second, we show that for quadratically coupled detectors, non-linear optical phenomena known as sum-frequency generation (SFG) and difference-frequency generation (DFG) naturally arise within a relativistic particle-detector model formalism. Finally, we show that an excited linearly coupled detector deposits its energy in the field differently from the quadratically coupled detector. The quadratically coupled detector preferentially deposits its energy in the field modes with a frequency of {half} the detector's energy gap, while a linearly coupled detector preferentially deposits its energy in field modes with frequency equal to the detector's energy gap.


    The main takeaway of our study is that when it comes to light-matter interactions and Fock states, there are distinctions between  free space and a very large cavity. This is particularly relevant because a very large cavity is often used to extrapolate arguments about the physics of quantum fields in free space. Our results emphasize the point that coupling the detector and then taking the large cavity limit is not the same as coupling detectors to a field in free space. The reason is fundamentally tied to the discrete vs continuous spectrum of the field (cavity vs free space) and how this affects the definition of physically meaningful Fock states.
    
    {Since a peaked wavepacket in momentum corresponds to a very delocalized wavepacket in space, these results could be read as the detector becoming insensitive to a very delocalized wavepacket in free space due to the fact it couples locally to the field. This means that even when the energy of the wavepacket is localized around its resonance frequency, the spatial spread of the state makes the detector insensitive to it. This reasoning does not apply in cavity settings since the energy of the wavepacket cannot be infinitely spread in space.}
    
   {Our results can also be interpreted as the detector becoming insensitive to a wavepacket whose energy density approaches zero in the monochromatic limit. A localized wavepacket in momentum space corresponds to a delocalized wavepacket in position space. However, the energy content of the wavepacket approaches a finite value in the monochromatic limit (see section \ref{sec: oneparticle}) and its energy density approaches zero (see Appendix~\ref{appendix: energyexpectation}), and the wavepacket becomes completely transparent to a detector. This contrasts to the cavity setting, where the the wavepacket is spread over a finite volume and the energy density converges to a finite value in the monochromatic limit.}
    

    \acknowledgements
    The authors thank Tales R. Perche, Luis J. Garay, Bruno S. L. Torres and  and Emma McKay for useful discussions. E. T. acknowledges the support of Mike-Ophelia Lazaridis Fellowship. E. M-M. is funded by the NSERC Discovery program as well as his Ontario Early Researcher Award.

    \appendix

    \section{Exact transition probability for one-particle states}
    \label{appendix: exactprobability}
    Here we compute the exact expression for the transition probability of a detector reacting to a one-particle Fock states when the detector-field coupling is linear. First, using the Gaussian spectrum in \eqref{eq: GaussianFrequency} let us rewrite Eq.~\eqref{eq: final-linear-probability} into a more useful form:
    \begin{align}
	    P^{\phi} &= \frac{\lambda^2(2\pi)^2}{(\pi\sigma^2)^{n/2}}\frac{e^{-\frac{|\bk_0|^2}{\sigma^2}}}{2(2\pi)^n}\abs{G_n}^2\,,
	\end{align}
	where we define
	\begin{align}
	    G_n 
	    &\coloneqq \int\frac{\dd^n \bm{k}}{\sqrt{|\bk|}}e^{-\frac{|\bk|^2}{2\sigma^2}}e^{\frac{|\bk||\bk_0|\cos\theta}{\sigma^2}}\delta(\Omega-|\bk|)\notag\\
	    &= \int_0^\infty \!\!\dd|\bk|\dd\Omega_{n-1}\, |\bk|^{n-\frac{3}{2}} e^{-\frac{|\bk|^2}{2\sigma^2}} e^{\frac{|\bk_0||\bk|\cos\theta}{\sigma^2}}\delta(\Omega-|\bk|)\,.
	    \label{eq: Gn}
	\end{align}
	We would like to obtain closed-form expressions for $G_n$. We consider two distinct cases, namely $n\geq 2$ and $n=1$. This is because for $n=1$ the integral over the momentum has no angular part and we require an IR cutoff. For convenience, in this Appendix we will write the transition probability as $P^\phi_n$ with the subscript $n$ labelling the number of spatial dimensions.
	
	\textbf{Case 1:} suppose $n\geq 2$. The trick is to recognize that we can write
	\begin{align}
	    &\int \dd\Omega_{n-1}e^{\frac{|\bk_0||\bk|\cos\theta}{\sigma^2}} \notag\\
	    &= \int \dd\mu_{n-2}\int_0^\pi \dd\theta\,(\sin\theta)^{n-2}e^{\frac{|\bk_0||\bk|\cos\theta}{\sigma^2}}\,,
	\end{align}
	where $\dd\Omega_{n-1}$ is the area element of the unit sphere $S^{n-1}$
	\begin{align}
	    \dd\Omega_{n-1} = \dd\theta (\sin\theta)^{n-2}\prod_{i=1}^{n-2}\dd\varphi_i\,(\sin\varphi_i)^{n-2-i}\,,
	\end{align}
	and $ \dd\mu_{n-2}$ is the area element without the $(\sin\theta)^{n-2}\dd\theta$. First let us deal with the $\dd\mu_{n-2}$ part. Note that
	\begin{align}
	    \int \dd\Omega_{n-1} &= \frac{2 \pi ^{n/2}}{\Gamma \left(\frac{n}{2}\right)}\,,\\
	    \int_0^\pi\dd\theta (\sin\theta)^{n-2}  &= \frac{\sqrt{\pi } \Gamma \left(\frac{n-1}{2}\right)}{\Gamma \left(\frac{n}{2}\right)}\,,
	\end{align}
    hence we can write
    \begin{align}
        \int \dd\mu_{n-2} &= \frac{\int \dd\Omega_{n-1}}{\int_0^\pi\dd\theta (\sin\theta)^{n-2} } 
        = \frac{2 \pi ^{\frac{n-1}{2}}}{\Gamma \left(\frac{n-1}{2}\right)}\,.
    \end{align}
    
    Next, the integral over $\theta$ can be solved analytically and reads
    \begin{align}
        &\int_0^\pi \dd\theta\,(\sin\theta)^{n-2}e^{\frac{|\bk_0||\bk|\cos\theta}{\sigma^2}}\notag\\
        &= \sqrt{\pi } \Gamma \left(\frac{n-1}{2}\right) \, _0\tilde{F}_1\left(\frac{n}{2};\frac{|\bk|^2 |\bk_0|^2}{4 \sigma ^4}\right)\,,
        \label{eq: reg-hyper}
    \end{align}
    where $_p\tilde{F}_q$ is the regularized generalized hypergeometric function \cite{NIST:DLMF}.
    
    Putting everything together into $G_n$ and integrating over $|\bk|$, we obtain after some algebraic manipulation and simplification the expression of $P^\phi$ for $n\geq 2$:
    \begin{align}
        P_n^\phi &= \lambda^2\frac{2 \pi ^{2-\frac{n}{2}}}{|\bk_0|^{n-2}} \frac{\Omega ^{n-1}}{\sigma^{4-n}} e^{-\frac{|\bk_0|^2+\Omega ^2}{\sigma ^2}} I_{\frac{n-2}{2}}\left(\frac{|\bk_0| \Omega }{\sigma ^2}\right)^2\,,
        \label{eq: appendix-prob-linear-one}
    \end{align}
    where $I_\alpha(z)$ is the modified Bessel function of the first kind of order $\alpha$. Note that from this expression, we can read off $\mathcal{I}_-(\sigma,\bk_0)$ that appear in Eq.~\eqref{eq: final-linear-probability}.
    
	\textbf{Case 2:} suppose $n=1$. Since there is no angular part, we have
	\begin{align}
	    G_1 &=  \int\frac{\dd k }{\sqrt{|k|}}e^{-\frac{|k|^2}{2\sigma^2}}e^{\frac{kk_0}{\sigma^2}}\delta(\Omega-|k|)\,.
	\end{align}
	This integral is divergent at $k=0$, so we need an IR cutoff $\Lambda>0$ for the integral. In other words, we should replace $G_1$ with IR-regulated version, namely 
	\begin{align}
	    G_1^\Lambda &\coloneqq  \int_{-\infty}^{-\Lambda}+\int_\Lambda^\infty\frac{\dd k }{\sqrt{|k|}}e^{-\frac{|k|^2}{2\sigma^2}}e^{\frac{kk_0}{\sigma^2}}\delta(\Omega-|k|)\,,
	\end{align}
	and we require that $\Omega,|\bk_0|>\Lambda$. Under this constraint, the integral over $k$ can be performed and we get
	\begin{align}
	    G^\Lambda_1\Bigr|_{\Omega,|\bk_0|\geq \Lambda} = \frac{e^{-\frac{\Omega ^2}{2 \sigma ^2}} \left(e^{-\frac{|k_0| \Omega }{\sigma ^2}}+e^{\frac{|k_0| \Omega }{\sigma ^2}}\right)}{\sqrt{\Omega }}
	\end{align}
	Putting everything together, we obtain for the transition probability for $n=1$:
    \begin{align}
        P^\phi_{1} &= 
	    \lambda^2\frac{\sqrt{\pi }}{\sigma  \Omega } e^{-\frac{(|k_0|+\Omega )^2}{\sigma ^2}} \left(e^{\frac{2 |k_0| \Omega }{\sigma ^2}}+1\right)^2\,.
    \end{align}
    We emphasize that although this expression does not explicitly depend on $\Lambda$, it has an implicit dependence on the IR cutoff, since the Dirac delta function that appears in $G_1^\Lambda$ must be evaluated for $\Omega >\Lambda$ and we also need for consistency that all length scales in the problem (such as $|k_0|$) is larger than $\Lambda$. However, once these are satisfied, the final expression is free from any IR cutoff dependence.
    
    {Finally, we note a remarkable fact: we can also obtain the result for $n=1$ (\textit{after} the IR cutoff requirement has been implemented) by taking the limit $n\to 1$ of $P^\phi_n$ in Eq.~\eqref{eq: appendix-prob-linear-one}:
	\begin{align}
	    \lim_{n\to 1} P^\phi_n = P^\phi_{1}\,.
	\end{align}
	Therefore, the result obtained by manipulating the angular part of the integral for $n\geq 2$ can be ``analytically continued'' to $n=1$, but only \textit{after} the IR cutoff constraint is satisfied ($\Omega,|k_0|\geq \Lambda$) since $P_n^\phi$ does not depend on $\Lambda$ from the outset.}

\section{Energy expectation value of one-particle state}
\label{appendix: energyexpectation}
	
	Here we compute the energy expectation value for the one-particle Fock wavepacket. Although, as discussed in the main text,  the limit is the same for any spectrum whose modulus-squared is a nascent delta in the monochromatic limit, here we show the explicit evaluation for a Gaussian spectral function. Using the Hamiltonian \eqref{eq:free-field-hamiltonian} we get
	\begin{align}
	    &\braket{1_f|\hat H_{0,\phi}|1_f}\notag\\
	    &= \int \dd^n\bk \,|\bk|\,\left|f_{\sigma,\bk_0}(\bk)\right|^2\notag\\
	    &= \frac{1}{\sqrt{\pi\sigma^2}^{n}}\int\dd^n\bk |\bk|e^{-\frac{(\bk-\bk_0)^2}{\sigma^2}}\notag\\
	    &= \frac{e^{-\frac{|\bk_0|^2}{\sigma^2}}}{\sqrt{\pi\sigma^2}^{n}}\int \dd|\bk|\dd\Omega_{n-1}\,|\bk|^ne^{-\frac{|\bk|^2}{\sigma^2}}e^{\frac{2|\bk||\bk_0|\cos\theta}{\sigma^2}}\,.
	    \label{eq: energy-expectation-temp}
	\end{align}
	We can use the same trick in Appendix~\ref{appendix: exactprobability}: the most important step is to first write
	\begin{align}
	    \dd\Omega_{n-1} = \dd\mu_{n-2}\dd\theta\,(\sin\theta)^{n-2}\,,
	\end{align}
	and the only non-triviality is the integral over the angular variable $\theta$ (cf. Eq.~\eqref{eq: reg-hyper}):
	\begin{align}
	    &\int_0^\pi \dd\theta\,(\sin\theta)^{n-2}e^{\frac{2|\bk_0||\bk|\cos\theta}{\sigma^2}}\notag\\
        &= \sqrt{\pi } \Gamma \left(\frac{n-1}{2}\right) \, _0\tilde{F}_1\left(\frac{n}{2};\frac{|\bk|^2 |\bk_0|^2}{\sigma ^4}\right)\,,
	\end{align}
	where $_p\tilde{F}_q$ is the regularized generalized hypergeometric function \cite{NIST:DLMF}. This is precisely the same as Eq.~\eqref{eq: reg-hyper} except we replace $|\bk_0|\to 2|\bk_0|$. Substituting this into Eq.~\eqref{eq: energy-expectation-temp}, we get the energy expectation value for $n\geq 2$:
	\begin{align}
	    \braket{1_f|\hat H_{0,\phi}|1_f}
	    &= \sigma  \Gamma \left(\frac{n+1}{2}\right) \, _1\tilde{F}_1\left(-\frac{1}{2};\frac{n}{2};-\frac{|\bk_0|^2}{\sigma ^2}\right)\,.
	    \label{eq: energy-expectation-dimtwo}
	\end{align}
    In the monochromatic limit, we have for $n\geq 2$
	\begin{align}
	    \lim_{\sigma\to 0}\braket{1_f|\hat H_{0,\phi}|1_f}  = |\bk_0|\,,
	\end{align}
    as expected. 
    
    For $n=1$ we can simply perform direct integration and we obtain
    \begin{align}
        &\braket{1_f|\hat H_{0,\phi}|1_f}_{n=1}\notag  \\
        &=\frac{1}{2\sqrt{\pi}} \left[\sqrt{\pi } |k_0| \left(\text{erf}\left(\frac{\Lambda +|k_0|}{\sigma }\right)-\text{erf}\left(\frac{\Lambda -|k_0|}{\sigma }\right)\right)+\right.\notag \\
        &\hspace{1.5cm} \left. \sigma  \left(e^{-\frac{(\Lambda -|k_0|)^2}{\sigma ^2}}+e^{-\frac{(\Lambda +|k_0|)^2}{\sigma ^2}}\right)\right]\,.
    \end{align}
    At this point, the reader would likely be less surprised by the still remarkable existence of the exact monochromatic limit  $\sigma\to 0$ at the same time that we lift the IR cutoff: 
    \begin{align}
        \lim_{\sigma\to 0}\lim_{\Lambda\to 0}\braket{1_f|\hat H_{0,\phi}|1_f}_{n=1}
        &= |k_0| \,.
    \end{align}
    Equivalently, analogous to Appendix~\ref{appendix: exactprobability}, we can obtain this result by taking the limit $n\to 1$ for the energy expectation value \eqref{eq: energy-expectation-dimtwo}:
    \begin{align}
        \lim_{n\to 1}\braket{1_f|\hat H_{0,\phi}|1_f} = \lim_{\Lambda\to 0}\braket{1_f|\hat H_{0,\phi}|1_f}_{n=1}\,.
    \end{align}
    Therefore, we showed explicitly how the Gaussian wavepacket indeed goes to the expected energy expectation $\hbar|\bk_0|$ in the monochromatic limit in all dimensions, as we expected from the nascent delta argument in the main text.
    
    For completeness, we include here the energy density of the field which can be obtained from the $tt$-component from the renormalized stress-energy tensor for the massless scalar field. The renormalized $tt$-component of the stress-energy tensor is precisely the Hamiltonian density, which reads
    \begin{align}
        &\braket{1_f|:\hat T_{tt}(\sx):|1_f} \notag\\
        &= \int\frac{\dd^n\bk\, \dd^n\bk'}{2(2\pi)^n\sqrt{|\bk||\bk'|}}(|\bk||\bk'|+\bk\cdot\bk')f_{\bk_0,\sigma}(\bk)f_{\bk_0,\sigma}(\bk')\notag\\
        &\hspace{3cm}\times\cos\left[(k_\mu-k_\mu')x^\mu\right]\,, 
    \end{align}
    where $k_\mu x^\mu = -|\bk|t+\bk\cdot \bx$ and $:\hat T_{tt}(\sx):$  is the normal ordered $\hat T_{tt}(\sx)$ operator. It is straightforward to check that we recover the energy expectation $\eqref{eq: energy-average-before-nascent}$ by performing spatial integral:
    \begin{align}
        \braket{1_f|\hat H_{0,\phi}|1_f} &= \int \dd^n\bx \braket{1_f|:\hat T_{tt}(\sx):|1_f}\,.
    \end{align}
    Note that $f_{\bk_0,\sigma}(\bk)$ does not define a nascent delta function because $f_{\bk_0,\sigma}$ is $L^2$-normalized to unity. We can make it into a nascent delta function by multiplying it by the right power of $\sigma$. For example, in the case of a Gaussian spectrum \eqref{eq: GaussianFrequency} we can write 
    \begin{align}
        f_{\bk_0,\sigma}(\bk)={(4\pi\sigma^2)^{n/4}}{\mathfrak{f}_{\bk_0,\sigma}(\bk)}\,,
    \end{align}
    where
    \begin{align}
        \mathfrak{f}_{\bk_0,\sigma}(\bk) &= \frac{1}{(2\pi\sigma^2)^{n/2}}e^{-\frac{(\bk-\bk_0)^2}{2\sigma^2}}\,.
    \end{align}
    $\mathfrak{f}_{\bk_0,\sigma}$ defines a family of nascent delta function since 
    \begin{align}
        \int \dd^n\bk\, \mathfrak{f}_{\bk_0,\sigma}(\bk)=1
    \end{align}
    even in the limit as $\sigma\to 0$. We can then write
    \begin{align}
        \lim_{\sigma\to 0}\frac{1}{\sigma^n}\braket{1_f|:\hat T_{tt}(\sx):|1_f} &=  \frac{|\bk_0|}{\pi^{n/2}}\,.
    \end{align}
    From this, it follows that
    \begin{align}
        \braket{1_f|:\hat T_{tt}(\sx):|1_f}\sim |\bk_0|\sigma^n
        \label{eq: RSET-scaling}
    \end{align}
    and hence, in the monochromatic limit, the energy density of the wavepacket goes to zero with $\sigma^n$ (which is the inverse of the spatial volume scale of the wavepacket). 
    
    Note that this will be true for any choice of $L^2$-normalizable spectrum $f$, since it can always be made into a nascent delta function multiplied by some geometric factor and $\sigma^{n/2}$. This proves that for any choice of spectrum the energy density goes to zero as the wavepacket becomes infinitely delocalized in the monochromatic limit.



    \section{Computation of the two-point function for the quadratic model and the one-particle Fock wavepacket}
    \label{appendix: Wightman-quad-oneparticle-proof}

    Here we prove that the two-point function for a one-particle Fock state 
    \mbox{$W^{\phi^2}(\sx,\sx')\coloneqq \bra{1_f}\normal{\hat{\phi}^2(\sx)}\normal{ \hat{\phi}^2(\sx')}\ket{1_f}$} is given by Eq.~\eqref{eq: wightman-quad-one}. First, we split $ W^{\phi^2}(\sx,\sx')$ into two parts using properties of normal ordering~\cite{Allison2017a}:
    \begin{align}
        W^{\phi^2}(\sx,\sx') = W^{\phi^2}_{\text{I}} (\sx,\sx')+ W_{\text{II}}^{\phi^2}(\sx,\sx')\,,
    \end{align}
    where
    \begin{align}
        W^{\phi^2}_{\text{I}}(\sx,\sx')&= \braket{1_f|\hat\phi^2(\sx)\hat\phi^2(\sx')|1_f}\,,\\
        W^{\phi^2}_{\text{II}}(\sx,\sx')&=- \braket{1_f|\hat\phi^2(\sx)|1_f}\braket{0|\hat\phi^2(\sx')|0}\notag\\
        &\hspace{0.5cm}- \braket{0|\hat\phi^2(\sx)|0}\braket{1_f|\hat\phi^2(\sx')|1_f}\notag\\
        &\hspace{0.5cm}+ \braket{0|\hat\phi^2(\sx)|0}\braket{0|\hat\phi^2(\sx')|0}\braket{1_f|1_f}\,.\\
        \notag
    \end{align}
    By Wick's theorem, only terms with equal number of annihilation and creation operators can contribute, thus $W_\text{I}^{\phi^2}$ yields the following integral:
    \begin{widetext}
    \begin{align}
        W_\text{I}^{\phi^2}(\sx,\sx') &= \int \frac{\prod_{j=1}^6\dd^n\bk_j\,f_{\bk_0,\sigma}(\bk_1)f_{\bk_0,\sigma}(\bk_6)}{[2(2\pi)^n)]^2\sqrt{|\bk_2||\bk_3||\bk_4||\bk_5|}} \left[\bra{0}\a{\bk_1}\a{\bk_2}\a{\bk_3}\ad{\bk_4}\ad{\bk_5}\ad{\bk_6}\ket{0}
        e^{-\ii k_2^\mu x_\mu -\ii k_3^\mu x_\mu+\ii k_4^\mu x'_\mu+\ii k_5^\mu x'_\mu}\right.\notag\\
        &\hspace{5.25cm}+\bra{0}\a{\bk_1}\a{\bk_2}\ad{\bk_3}\a{\bk_4}\ad{\bk_5}\ad{\bk_6}\ket{0}
        e^{-\ii k_2^\mu x_\mu +\ii k_3^\mu x_\mu-\ii k_4^\mu x'_\mu+\ii k_5^\mu x'_\mu}\notag\\
        &\hspace{5.25cm}+\bra{0}\a{\bk_1}\a{\bk_2}\ad{\bk_3}\ad{\bk_4}\a{\bk_5}\ad{\bk_6}\ket{0}
        e^{-\ii k_2^\mu x_\mu +\ii k_3^\mu x_\mu+\ii k_4^\mu x'_\mu-\ii k_5^\mu x'_\mu}\notag\\
        &\hspace{5.25cm}+\bra{0}\a{\bk_1}\ad{\bk_2}\ad{\bk_3}\a{\bk_4}\a{\bk_5}\ad{\bk_6}\ket{0}
        e^{\ii k_2^\mu x_\mu +\ii k_3^\mu x_\mu-\ii k_4^\mu x'_\mu-\ii k_5^\mu x'_\mu}\notag\\
        &\hspace{5.25cm}+\bra{0}\a{\bk_1}\ad{\bk_2}\a{\bk_3}\a{\bk_4}\ad{\bk_5}\ad{\bk_6}\ket{0}
        e^{\ii k_2^\mu x_\mu -\ii k_3^\mu x_\mu-\ii k_4^\mu x'_\mu+\ii k_5^\mu x'_\mu}\notag\\
        &\left.\hspace{5.25cm}+\bra{0}\a{\bk_1}\ad{\bk_2}\a{\bk_3}\ad{\bk_4}\a{\bk_5}\ad{\bk_6}\ket{0}
        e^{\ii k_2^\mu x_\mu -\ii k_3^\mu x_\mu+\ii k_4^\mu x'_\mu-\ii k_5^\mu x'_\mu}\right]\,.
        \label{eq: Wightman-quad-oneparticle-unordered}
    \end{align}
    \end{widetext}
    {In the above expression we have used the shorthand \mbox{$k_j^\mu x_\mu = |\bk_j|t-\bk_j\cdot \bx$} to reduce notational clutter.} 
    
    For brevity, we write $W_{\text{I}}^{\phi^2}$ as a sum of six integrals
    \begin{equation}
        W_{\text{I}}^{\phi^2}  =  {A_1+A_2+A_3+A_4+A_5+A_6}\,,
    \end{equation}
    where each $A_j$ corresponds to each vacuum expectation value of the ladder operators in \eqref{eq: Wightman-quad-oneparticle-unordered}. It will be very convenient for us to construct a compact notation for the vacuum expectation values over these ladder operators. First, we define 
    \begin{align}
        \delta([abc][def]) \coloneqq \delta(\bk_a-\bk_d)\delta(\bk_b-\bk_e)\delta(\bk_c-\bk_f)\,,
        \label{eq: delta-permutation-def}
    \end{align}
    where the index $[def]$ in the second square bracket will be \textit{fixed}, and we will only vary the indices in the first bracket to avoid double-counting. We also use the shorthand $\delta(\bk)\equiv\delta^{(n)}(\bk)$ for the RHS of \eqref{eq: delta-permutation-def}. Next, we define 
    \begin{align}
        \delta([abc+a'b'c'][def]) &\coloneqq \delta([abc][def])+\delta([a'b'c'][def])\,.
    \end{align}
    Finally, we define $\pi[a\hat bcd...]$ to mean \textit{summation} over permutation of strings $abcd...$ but excluding all permutations involving $b$ on that specific position. For example, $\pi[1\hat 2 3]$ means we exclude cases when $2$ is in the second position (namely [123] and [321]). We will list the permutation explicitly when this notation is not useful.

    Let us illustrate our convention with three examples. First, we have
    \begin{align}
        \delta([123][456]) =  \delta(\bk_1-\bk_4)\delta(\bk_2-\bk_5)\delta(\bk_3-\bk_6)\,.
        \label{eq: delta-permutation-example}
    \end{align}
    Second, when we have $\pi[123]$, we sum over all possible combinations coming from permutations of $[123]$:
    \begin{align}
        \delta(\pi[123][456])  &=\delta(\bk_1-\bk_4)\delta(\bk_2-\bk_5)\delta(\bk_3-\bk_6)  \notag\\
        &+\delta(\bk_1-\bk_4)\delta(\bk_3-\bk_5)\delta(\bk_2-\bk_6) +  \dots \notag\\
        &+\delta(\bk_3-\bk_4)\delta(\bk_2-\bk_5)\delta(\bk_1-\bk_6)\,,
    \end{align}
    where we recall that in this convention the positions of $\bk_4,\bk_5,\bk_6$ are held \textit{fixed} while $\bk_1,\bk_2,\bk_3$ are permuted and summed over. Finally, we have for instance
    \begin{align}
        \delta(\pi[\hat 412]&[356]) \notag\\ &=\delta(\bk_1-\bk_3)\delta(\bk_2-\bk_5)\delta(\bk_4-\bk_6) + \dots \notag\\
        &+\delta(\bk_2-\bk_3)\delta(\bk_4-\bk_5)\delta(\bk_1-\bk_6)\,, 
    \end{align}
    where terms involving $\delta(\bk_4-\bk_3)$ are excluded because we exclude all cases when index `4' is in the first position. 
    
    Using this notation, the vacuum expectation values now read
    \begin{align}
        \label{eq:VEVoneparticleQuad}
        \bra{0}\a{\bk_1}\a{\bk_2}\a{\bk_3}\ad{\bk_4}\ad{\bk_5}\ad{\bk_6}\ket{0} &= \delta(\pi[123][456])\,,\notag\\
        \bra{0}\a{\bk_1}\a{\bk_2}\ad{\bk_3}\a{\bk_4}\ad{\bk_5}\ad{\bk_6}\ket{0}&= \delta(\pi[\hat 412][356])\,,\notag\\
        \bra{0}\a{\bk_1}\a{\bk_2}\ad{\bk_3}\ad{\bk_4}\a{\bk_5}\ad{\bk_6}\ket{0}&= \delta([125+215][346])\,,\notag\\
        \bra{0}\a{\bk_1}\ad{\bk_2}\ad{\bk_3}\a{\bk_4}\a{\bk_5}\ad{\bk_6}\ket{0}&=0\,,\notag\\
        \bra{0}\a{\bk_1}\ad{\bk_2}\a{\bk_3}\a{\bk_4}\ad{\bk_5}\ad{\bk_6}\ket{0}&= \delta([134+143][256])\,,\notag\\
        \bra{0}\a{\bk_1}\ad{\bk_2}\a{\bk_3}\ad{\bk_4}\a{\bk_5}\ad{\bk_6}\ket{0}&=\delta([135][246])\,.
    \end{align}
    Substituting Eq.\eqref{eq:VEVoneparticleQuad} into Eq.~\eqref{eq: Wightman-quad-oneparticle-unordered}, we can readily obtain the expressions for each $A_j$:
    \begin{align}
        A_1 &= 2W^\phi_{\vac}(\sx,\sx')^2+4W^\phi_{\vac}(\sx,\sx')K_{\bk_0}^*(\sx)K_{\bk_0}(\sx')\,,\\
        A_2 &= W^\phi_{\vac}(\sx,\sx)|K_{\bk_0}(\sx')|^2+W^\phi_{\vac}(\sx',\sx')|K_{\bk_0}(\sx)|^2 \notag\\
        &+ W^\phi_{\vac}(\sx,\sx)W_{\vac}(\sx',\sx')\,,\\
        A_3 &= W^\phi_{\vac}(\sx,\sx')K_{\bk_0}(\sx)K_{\bk_0}^*(\sx') + W^\phi_{\vac}(\sx,\sx)|K_{\bk_0}(\sx')|^2\,,\\
        A_4 &= 0\,,\\
        A_5 &= W^\phi_{\vac}(\sx,\sx')K_{\bk_0}(\sx)K_{\bk_0}^*(\sx') + W^\phi_{\vac}(\sx',\sx')|K_{\bk_0}(\sx)|^2\,,\\
        A_6 &= W^\phi_{\vac}(\sx,\sx')K_{\bk_0}(\sx)K_{\bk_0}^*(\sx')\,.
    \end{align}
    where $K_{\bk_0}(\sx)$ and $W^\phi_{\vac}(\sx,\sx')$ are defined in Section~\ref{sec: oneparticle}. Note that $A_2,A_3,A_5$ are singular because they involve the coincidence limit of the vacuum two-point function. 
    
    Putting all the expressions for $A_j$ together, we obtain
    \begin{align}
        W_\text{I}^{\phi^2} &= 4W^\phi_{\vac}(\sx,\sx')\left(K_{\bk_0}(\sx)K_{\bk_0}^*(\sx') + \text{c.c.}\right) \notag\\
        &+2W^\phi_{\vac}(\sx,\sx)|K_{\bk_0}(\sx')|^2+2W_{\vac}(\sx',\sx')|K_{\bk_0}(\sx)|^2\notag\\
        &+2W_{\vac}^\phi(\sx,\sx')^2+W^\phi_{\vac}(\sx,\sx)W_{\vac}^\phi(\sx',\sx')\,.\\\notag
    \end{align}
    The $W_{\text{II}}^{\phi^2}$ can be readily obtained by taking the coincidence limit of the linearly coupled Wightman two-point functions $W^\phi$, which gives
    \begin{align}
        &\braket{1_f|\hat\phi(\sx)^2|1_f} \braket{0|\hat\phi(\sx')^2|0} \notag \\
        &\hspace{1cm}= W_{\vac}^\phi(\sx',\sx')\left[W_{\vac}^\phi(\sx,\sx)+2|K_{\bk_0}(\sx)|^2\right]\,,\\
        &\braket{1_f|\hat\phi(\sx')^2|1_f} \braket{0|\hat\phi(\sx)^2|0} \notag \\
        &\hspace{1cm}= W_{\vac}^\phi(\sx,\sx)\left[W_{\vac}^\phi(\sx',\sx')+2|K_{\bk_0}(\sx')|^2\right]\,,\\
        &\braket{0|\hat\phi(\sx)^2|0} \braket{0|\hat\phi(\sx')^2|0} = W_{\vac}^\phi(\sx,\sx)W_{\vac}^\phi(\sx',\sx')\,.
    \end{align}
    Adding these together gives us
    \begin{align}
        W_{\text{II}}^{\phi^2} &= -W_{\vac}^\phi(\sx,\sx)W_{\vac}^\phi(\sx',\sx')-2W_{\vac}^\phi(\sx,\sx)|K_{\bk_0}(\sx')|^2\notag \\
        &-2W_{\vac}^\phi(\sx',\sx')|K_{\bk_0}(\sx)|^2\,.
    \end{align}
    Putting all these together, the full two-point function $W^{\phi^2} = W^{\phi^2}_\text{I}+W^{\phi^2}_{\text{II}}$ for the one-particle Fock state now reads
    \begin{align}
        W^{\phi^2}(\sx,\sx') 
        &= 4 W^\phi_{\vac}(\sx,\sx')\rr{K_{\bk_0}^*(\sx)K_{\bk_0}(\sx')+ \text{c.c.}}\notag\\
        &+2W^\phi_{\vac}(\sx,\sx')^2\,,
    \end{align}
    which is precisely Eq.~\eqref{eq: wightman-quad-one}.

    \section{Computation of the two-point function for the quadratic model and the two-particle Fock wavepacket}
    \label{appendix: Wightman-quad-twoparticle-proof}
    
    We will now prove that the following two-point function for the two-particle Fock wavepacket  \mbox{$W^{\phi^2}_{\ba\bb}(\sx,\sx')\coloneqq \bra{2_f}\normal{\hat{\phi}^2(\sx)}\normal{ \hat{\phi}^2(\sx')}\ket{2_f}$} (where $\ba,\bb$ are the dominant momenta of the two-particle Fock state) is given by Eq.~\eqref{eq: Wightman-quadratic-two}. First, let us define a shorthand
    \begin{align}
        &f_{1278}^{\ba\bb}\coloneqq f_{\ba,\sigma}(\bk_1)f_{\bb,\sigma}(\bk_2)f_{\ba,\sigma}(\bk_7)f_{\bb,\sigma}(\bk_8)\,.
    \end{align}
    We can split $W^{\phi^2}_{\ba\bb}$ in two parts using the properties of normal ordering:
    \begin{align}
        W^{\phi^2}_{\ba\bb} &=  W^{\phi^2}_{\ba\bb,\text{I}}+W^{\phi^2}_{\ba\bb,\text{II}}\,,
        \label{eq: Wightman-quad-twoparticle-splitting}
    \end{align}
    where (dropping the $(\sx,\sx')$ from the LHS for brevity)
    \begin{align}
        W^{\phi^2}_{\ba\bb,\text{I}} = &\braket{2_f|\phi^2(\sx)\phi^2(\sx')|2_f}\,,\\ W^{\phi^2}_{\ba\bb,\text{II}}= 
        & -\braket{2_f|\phi^2(\sx)|2_f}\braket{0|\phi^2(\sx')|0}\notag\\
        & -\braket{0|\phi^2(\sx)|0}\braket{2_f|\phi^2(\sx')|2_f}\notag \\
        & +\braket{0|\phi^2(\sx)|0}\braket{0|\phi^2(\sx')|0}\braket{2_f|2_f}\,.
        \label{eq: W-quad-sing-two}
    \end{align}
    Again by Wick's theorem, only terms with equal number of annihilation and creation operators can contribute, thus $W_{\ba\bb,\text{I}}^{\phi^2}$ yields the following integral:
    \begin{widetext}
    \begin{equation}
        \begin{split}
        W^{\phi^2}_{\ba\bb,\text{I}} &= {\NN^2\!\!}\int\frac{\prod_{j=1}^8\dd^n \bk_j  \,\,f_{1278}^{\ba\bb}}{[2(2\pi)^n]^2\sqrt{|\bk_3||\bk_4||\bk_5||\bk_6|}} 
        \left[ 
        \braket{0|\a{k_1} \a{k_2} \a{k_3} \a{k_4} \ad{k_5} \ad{k_6}\ad{k_7} \ad{k_8}|0}e^{-\ii k_3^\mu x_\mu}e^{-\ii k_4^\mu x_\mu}e^{\ii k_5^\mu x'_\mu}e^{\ii k_6^\mu x'_\mu} \right. \\
        &\hspace{5.2cm}+
        \braket{0|\a{k_1} \a{k_2} \a{k_3} \ad{k_4} \a{k_5} \ad{k_6}\ad{k_7} \ad{k_8}|0}e^{-\ii k_3^\mu x_\mu}e^{\ii k_4^\mu x_\mu}e^{-\ii k_5^\mu x'_\mu}e^{\ii k_6^\mu x'_\mu} \\
        &\hspace{5.2cm}+
        \braket{0|\a{k_1} \a{k_2} \a{k_3} \ad{k_4} \ad{k_5} \a{k_6}\ad{k_7} \ad{k_8}|0}e^{-\ii k_3^\mu x_\mu}e^{\ii k_4^\mu x_\mu}e^{\ii k_5^\mu x'_\mu}e^{-\ii k_6^\mu x'_\mu}  \\
        &\hspace{5.2cm}+
        \braket{0|\a{k_1} \a{k_2} \ad{k_3} \a{k_4} \a{k_5} \ad{k_6}\ad{k_7} \ad{k_8}|0}e^{\ii k_3^\mu x_\mu}e^{-\ii k_4^\mu x_\mu}e^{-\ii k_5^\mu x'_\mu}e^{\ii k_6^\mu x'_\mu} \\
        &\hspace{5.2cm}+
        \braket{0|\a{k_1} \a{k_2} \ad{k_3} \a{k_4} \ad{k_5} \a{k_6}\ad{k_7} \ad{k_8}|0}e^{\ii k_3^\mu x_\mu}e^{-\ii k_4^\mu x_\mu}e^{\ii k_5^\mu x'_\mu}e^{-\ii k_6^\mu x'_\mu} \\
        &\left.\hspace{5.2cm}+
        \braket{0|\a{k_1} \a{k_2} \ad{k_3} \ad{k_4} \a{k_5} \a{k_6}\ad{k_7} \ad{k_8}|0}e^{\ii k_3^\mu x_\mu}e^{\ii k_4^\mu x_\mu}e^{-\ii k_5^\mu x'_\mu}e^{-\ii k_6^\mu x'_\mu} \right]\,,
        \end{split}
        \label{eq: Wightman-quad-twoparticle-unordered}
    \end{equation}
    \end{widetext}
    {For brevity, we will express the above integral as
    \begin{equation}
        W_{\ba\bb,\text{I}}^{\phi^2} = B_1+B_2+B_3+B_4+B_5+B_6\,,
    \end{equation}
    where $B_j$ corresponds to the integral over each vacuum expectation value of the ladder operators in Eq.~\eqref{eq: Wightman-quad-twoparticle-unordered}.}
    
    We need to work out the vacuum expectation values of the six terms in the eightfold nested $n$-dimensional integral. We will employ the permutation notation defined in Appendix~\ref{appendix: Wightman-quad-oneparticle-proof} but generalized to eight ladder operators, so we will have the delta functions over eight indices $\delta([abcd][efgh])$ instead of six indices $\delta([abc][def])$ in the previous section. In addition to the convention used there, we will have one more rule: we define $\widehat{11}234$ to mean that we are excluding cases $1abc$ \textit{and} $a1bc$ (i.e. when the index `1' is in \textit{either} the first or the second position). 
    
    We illustrate these conventions using two examples. First,  $\delta(\pi[1234][5678])$ means summing over all permutations of $[1234]$ while holding the last four indices fixed:
    \begin{align}
        &\delta(\pi[1234][5678])\notag\\
        &=\delta(\bk_1-\bk_5)\delta(\bk_2-\bk_6)\delta(\bk_3-\bk_7)\delta(\bk_4-\bk_8) + ... \notag\\
        &+\delta(\bk_4-\bk_5)\delta(\bk_3-\bk_6)\delta(\bk_2-\bk_7)\delta(\bk_1-\bk_8)\,.
    \end{align}
    Second, our new rule applied to $\delta(\pi[\widehat{66}123][4578])$ leads to the following expression
    \begin{align}
        &\delta(\pi[\widehat{66} 123][4578])\notag\\
        &=\delta(\bk_1-\bk_4)\delta(\bk_2-\bk_5)\delta(\bk_3-\bk_7)\delta(\bk_5-\bk_8) + ... \notag\\
        &+\delta(\bk_3-\bk_4)\delta(\bk_2-\bk_5)\delta(\bk_6-\bk_7)\delta(\bk_1-\bk_8)\,,
    \end{align}
    where we sum over all permutations of $[6123]$ but excluding the cases containing $\delta(\bk_6-\bk_4)$ and $\delta(\bk_6-\bk_5)$.
    
    With these conventions, we can express the vacuum expectation values in compact form as
    \begin{equation}\begin{split}
        \braket{0|\a{k_1} \a{k_2} \a{k_3} \a{k_4} \ad{k_5} \ad{k_6}\ad{k_7} \ad{k_8}|0} &= \delta(\pi[1234][5678])\,, \\
        \braket{0|\a{k_1} \a{k_2} \a{k_3} \ad{k_4} \a{k_5} \ad{k_6}\ad{k_7} \ad{k_8}|0} &= \delta(\pi[\hat 5123][4678])\,,\\
        \braket{0|\a{k_1} \a{k_2} \a{k_3} \ad{k_4} \ad{k_5} \a{k_6}\ad{k_7} \ad{k_8}|0} &= 
        \delta(\pi[\widehat{66}123][4578])\,,\\
         \braket{0|\a{k_1} \a{k_2} \ad{k_3} \a{k_4} \a{k_5} \ad{k_6}\ad{k_7} \ad{k_8}|0} &
        \notag\\
        &\hspace{-5cm}=\delta([1245+1254+2145+2154+2415+2451\notag\\
        &\hspace{-3.425cm}+2514+2541][3678])\,,\\
        \braket{0|\a{k_1} \a{k_2} \ad{k_3} \a{k_4} \ad{k_5} \a{k_6}\ad{k_7} \ad{k_8}|0} & 
        \notag\\
        &\hspace{-5cm}=\delta([1246+1264+1426+1462+2146+2164\notag\\
        &\hspace{-3.425cm}+2416+2461][3578])\,,\\
        \braket{0|\a{k_1} \a{k_2} \ad{k_3} \ad{k_4} \a{k_5} \a{k_6}\ad{k_7} \ad{k_8}|0} &  \notag\\
        &\hspace{-5cm}=\delta([1256+1265+2156+2165][3478])\,.
     \end{split}
     \end{equation}
    
    Let us now solve the six integrals over each vacuum expectation value. The first integral comes from the 24 permutations of $[1234]$, which reads
    \begin{equation}
    \begin{split}
        B_1 &= 2W^\phi_{\vac}(\sx,\sx')^2
        + 4{\NN}^2K'_\ba K'_\bb K^*_\ba K^*_\bb  \\
        &+ 4{\NN}^2W_{\vac}^\phi(\sx,\sx')\spec\rr{K_\ba'K^*_\bb +K_\bb'K^*_\ba } \\
        &+ 4{\NN}^2W_{\vac}^\phi(\sx,\sx')\rr{K_\ba'K^*_\ba +K_\bb'K^*_\bb }\,,
    \end{split}
    \end{equation}
    where we have used the shorthand $K'_{\bc_j}$ to denote $K_{\bc_j}(\sx')$ to simplify notation, where $K_{\bc_j}(\sx)$ is defined by Eq.~\eqref{eq: K-integral}.
    The second integral comes from 18 permutations after removing the $[5abc]$ terms, which reads
    \begin{equation}
        \begin{split}
        B_2 &= {\NN}^2\bigg[|K_\ba|^2|K_\bb'|^2 + |K_\bb|^2|K_\ba'|^2  \\
        &+
        K_\ba K_\bb^*K_\ba^{*'} K_\bb' +  K_\ba^* K_\bb K_\ba' K_\bb^{*'} \\
        &+
        W_{\vac}^\phi(\sx,\sx')\rr{K_\ba {K}^{*'}_\ba + K_\bb {K}^{*'}_\bb } \\
        &+
        W_{\vac}^\phi(\sx,\sx')\spec\rr{K_\ba {K}^{*'}_\bb + K_\bb {K}^{*'}_\ba }  \\
        &+
        W_{\vac}^\phi(\sx',\sx')\rr{|K_\ba|^2 + |K_\bb|^2 }  \\
        &+
        W_{\vac}^\phi(\sx',\sx')\spec\rr{K_\ba K^*_\bb + K_\bb K^{*}_\ba }  \\
        &+
        W_{\vac}^\phi(\sx,\sx)\rr{|K'_\ba|^2 + |K'_\bb|^2 }  \\
        &+
        W_{\vac}^\phi(\sx,\sx)\spec\rr{K'_\ba {K}^{*'}_\bb + K'_\bb {K}^{*'}_\ba } \bigg]\\
        &+
        W_{\vac}^\phi(\sx,\sx)W_{\vac}^\phi(\sx',\sx')\,.
        \end{split}
    \end{equation}
    Notice that this second integral contains distributional divergences due to the coincidence limit of the vacuum two-point function, and even products of two divergent two-point functions. These divergences will be cancelled exactly by normal ordering as we will see. 
    
    The third integral comes from 12 terms involving permutations of $[1236]$ but excluding $[6abc]$ and $[a6bc]$:
    \begin{equation}
        \begin{split}
            B_3 &= {\NN}^2\bigg[|K_\ba|^2|K_\bb'|^2 + |K_\bb|^2|K_\ba'|^2 \\
            &+
            K_\ba K^*_\bb {K}^{*'}_\ba K'_\bb +  K^*_\ba K_\bb K'_\ba {K}^{*'}_\bb  \\
            &+
            W^\phi_{\vac}(\sx,\sx')\rr{K_\ba K_\ba^{*'}+K_\bb {K}^{*'}_\bb} \\
            &+
            W^\phi_{\vac}(\sx,\sx')\spec\rr{K_\ba {K}^{*'}_\bb+K_\bb K_\ba^{*'}}\\
            &+
            W_{\vac}^\phi(\sx,\sx)\left(|K_\ba'|^2+|K_\bb'|^2\right)\\
            &+             W_{\vac}^\phi(\sx,\sx)\spec\left(K_\ba'K_\bb^{*'}+K_\bb'K_\ba^{*'}\right)\bigg]\,.
        \end{split}
    \end{equation}
    The fourth integral also comes from 12 permutations,
    \begin{equation}
        \begin{split}
            B_4&= {\NN}^2\bigg[|K_\ba|^2|K_\bb'|^2 + |K_\bb|^2|K_\ba'|^2 \\
            &+ K_\ba K_\bb^* K_\bb' {K}^{*'}_\ba+K_\bb K_\ba^* K_\ba' {K}^{*'}_\bb\\
            &+
            W_{\vac}(\sx,\sx')\rr{K_\ba K_\ba^{*'}+K_\bb K_\bb^{*'}}\\
            &+
            W_{\vac}(\sx,\sx')\spec\rr{K_\ba K_\bb^{*'}+K_\bb K_\ba^{*'}}\\
            &+
            W_{\vac}^\phi(\sx',\sx')\rr{|K_\ba|^2+|K_\bb|^2}\\
            &+
            W_{\vac}^\phi(\sx',\sx')C_{\ba\bb}\rr{K_\ba K^*_\bb+K_\bb K^*_\ba}\bigg]\,.
        \end{split}
    \end{equation}
    The fifth integral comes from 8 permutations,
    \begin{equation}
    \begin{split}
        B_5 &= {\NN}^2\bigg[|K_{\ba}|^2|K_{\bb}'|^2+|K_{\bb}|^2|K_{\ba}'|^2\\
        &+ K_\ba^* K_\bb K'_\ba K^{'*}_\bb + K_\ba K^*_\bb K^{'*}_\ba K^{'}_\bb  \\
        &+ W_{\vac}^\phi(\sx,\sx')\rr{K_{\ba}K_\ba^{'*}+K_{\bb}K_\bb^{'*}}  \\
        &+ W_{\vac}^\phi(\sx,\sx')\spec\rr{K_{\ba}K_\bb^{'*}+K_{\bb}K_\ba^{'*}}\bigg] \,.
        \end{split}
    \end{equation}
    Finally, the sixth integral comes from 4 permutations,
    \begin{equation}
        \begin{split}
        B_6 &= 4{\NN}^2 K_{\ba}K_\bb K_\ba^{'*} K_\bb^{'*}\,.
        \end{split}
    \end{equation}
    Overall, only $B_2,B_3,B_4,B_5$ contain divergent terms coming from the coincidence limit of the vacuum Wightman two-point function $W_{\vac}^\phi$. Next, the $W_{\ba\bb,\text{II}}^{\phi^2}$ term in Eq.~\eqref{eq: W-quad-sing-two} contains three summands which are made of the products of the following quantities:
    \begin{align}
        \braket{0|\phi^2(\sx)|0} &= W_{\vac}^\phi(\sx,\sx)
            \,,\\
        \braket{0|\phi^2(\sx')|0} &= W_{\vac}^{\phi}(\sx',\sx')
            \,,\\
        \braket{2_f|\phi^2(\sx)|2_f}&= W^\phi_{\vac}(\sx,\sx)+ {\NN}^2\bigg[2|K_\ba|^2 + 2|K_\bb|^2 \notag\\
        &+  \spec\rr{2K_\ba K^*_\bb + 2K_\bb K^*_\ba}\bigg] \,,\\
        \braket{2_f|\phi^2(\sx')|2_f}&=  W_{\vac}^\phi(\sx',\sx')+ {\NN}^2\bigg[2|K_\ba'|^2 + 2|K_\bb'|^2 \notag\\
        &+ \spec\rr{2K'_\ba {K}^{*'}_\bb + 2K'_\bb {K}^{*'}_\ba}\bigg] \,.
    \end{align}
    Putting these together, we get
    \begin{align}
        W_{\ba\bb,\text{II}}^{\phi^2} 
        &= -{\NN}^2W_{\vac}^\phi(\sx,\sx)\bigg[2|K_\ba'|^2 + 2|K_\bb'|^2 \notag\\
        &+ \spec\rr{2K'_\ba {K}^{*'}_\bb + 2K'_\bb {K}^{*'}_\ba}\bigg]\notag\\
        &- {\NN}^2 W_{\vac}^{\phi}(\sx',\sx')\bigg[2|K_\ba|^2 + 2|K_\bb|^2\notag\\
        &+ \spec\rr{2K_\ba K^*_\bb + 2K_\bb K^*_\ba} \bigg]\notag\\
        &- W_{\vac}^\phi(\sx,\sx)W_{\vac}^{\phi}(\sx',\sx')\,.
    \end{align}
    
    Finally, by comparing the six integrals $B_1$ to $B_6$ coming from $W_{\ba\bb,\text{I}}^{\phi^2}$ and $W_{\ba\bb,\text{II}}^{\phi^2}$, it can be readily checked that when adding the two terms in Eq.~\eqref{eq: Wightman-quad-twoparticle-splitting} the divergent parts  are exactly cancelled, yielding Eq.~\eqref{eq: Wightman-quadratic-two}:
    \begin{align}
        &W_{\ba\bb}^{\phi^2}(\sx,\sx') \notag\\
        &={\NN}^2\bigg[4W_{\vac}^\phi(\sx,\sx')\rr{K_\ba {K}^{*'}_\ba + K_\bb {K}^{*'}_\bb +\text{c.c.}}\notag\\
        &+4W_{\vac}^\phi(\sx,\sx')\spec\rr{K_\ba {K}^{*'}_\bb + K_\bb {K}^{*'}_\ba + \text{c.c.}}\, \notag \\
        &+ 4\rr{K^*_\ba K^*_\bb K'_\ba K'_\bb +K_\ba K_\bb^*K_\ba^{*'} K_\bb' + \text{c.c.} }\notag\\
        &+ 4|K_\bb|^2|K_\ba'|^2 + 4|K_\ba|^2|K_\bb'|^2\bigg] + 2 W^\phi_{\vac}(\sx,\sx')^2\,.
    \end{align}

\bibliography{fullref}

\end{document}